\documentstyle[11pt, aaspp]{article}

\def\deg{$^{\circ}\,$}
\def\kms{km s$^{-1}$}

\begin{document}

\title{Tides, Interactions, and Fine-Scale Substructures in Galaxy Clusters}

\author{Christopher J. Conselice$^1$ and John S. Gallagher III}

\affil{Department of Astronomy, University of Wisconsin, Madison, 475 N. 
Charter St., Madison, WI., 53706}

\altaffiltext{1}{chris@astro.wisc.edu}

\begin{abstract}

We present the results of a study on galaxy interactions, tides, and
other processes which produce luminous fine-scale substructures in
the galaxy clusters: Coma, Perseus, Abell 2199, AWM 3 and AWM 5.
All unusual structures in these clusters can be categorized in seven
morphologies: interacting galaxies, multiple galaxies
(non-interacting), distorted galaxies, tailed galaxies, line
galaxies, dwarf galaxy groups and galaxy aggregates.  The various
morphologies are described, and a catalog is presented of 248
objects in these five clusters along with color, and positional
information obtained from CCD images taken with the WIYN 3.5m
telescope in broadband B and R filters.

Distorted, interacting, and fine-scale substructures have a range of
colors extending from blue objects with B-R $\approx$ 0, to redder
colors at B-R $\approx$ 2.5.  We also find that the structures with
the most disturbed morphology have the bluest colors.  Additionally,
the relative number distributions of these structures, suggests that
two separate classes of galaxy clusters exist: one dominated by
distorted structures and the other dominated by galaxy
associations.  The Coma and Perseus clusters, respectively, are
proposed as models for these types of clusters.   These structures
avoid the deep potentials of the dominant D or cD galaxies in the
Coma and Perseus clusters, and tend to clump together.

Possible mechanisms for the production of fine-scale substructure
are reviewed and compared to observations of z $\approx$ 0.4
Butcher-Oemler clusters.  We conclude, based on color, positional,
and statistical data, that the most likely mechanism for the
creation of these structures is through an interaction with the
gravitational potential of the cluster, possibly coupled with
effects of weak interactions with large cluster ellipticals.

\end{abstract}

\newpage

\section{INTRODUCTION}

Galaxy interactions and mergers have been recognized since the work
of Toomre and Toomre (1972) as an important, if not the crucial
aspect, for understanding galaxy evolution (see Schweizer, 1986 and
Barnes \& Hernquist, 1992 for reviews).  Since very few galaxies are
isolated (Ramella et al. 1989 and references therein),  galaxy
evolution must occur for the majority of galaxies in a dense cluster,
or group environment, facilitating interactions.   Also, galaxy
clusters in general are not relaxed systems (e.g. West 1994; Girardi
et al. 1997), and by definition have an enhanced number density of
galaxies.  If clusters are not viralized, an increased number of
interactions likely will occur among member galaxies.  Furthermore,
clusters are probably in part created hierarchically though accretion
of field galaxies, possibly accounting for the large proportion of
blue galaxies in rich clusters at high redshifts (Butchter \& Oemler
1978, hereafter B-O effect); this process, or one similar to it,
might be occurring in nearby clusters at a reduced rate,  with
similar observational consequences.  The interplay between galaxy
cluster formation and evolution, substructures in clusters and the
morphological appearance of distant clusters can be further
illuminated by studying the nature of interactions, tides and
fine-scale substructures in nearby clusters.    This paper is a
morphological and physical study aimed at detecting and explaining
these structures in the five nearby galaxy clusters: Coma, Perseus,
Abell 2199, AWM 3 and AWM 5.

Distorted blue galaxies play a significant role in the evolution of
clusters seen at moderate redshifts z $\approx$ 0.4,  first noticed
in the work of Butcher and Oemler (1978, 1984).   Using the high
resolution capabilities of the Hubble Space Telescopes,  a high
fraction of the blue galaxies in distant clusters were discovered
to be either spirals, or distorted galaxies (Oemler, Dressler and
Butcher 1997; Dressler et al. 1994; Couch et al. 1994).  The origin
of these objects is still a bit of a mystery.  Many possible
scenarios for the creation of B-O galaxies in distant clusters have
been proposed: low-velocity galaxy-galaxy interactions (Icke 1985;
Lavery \& Henry 1988), interactions with a cluster gravitational
potential (Henriksen and Byrd 1996; Valluri 1993; Byrd and Valtonen
1990), ram-pressure stripping (Gunn and Gott 1972; Dressler and
Gunn 1983), and galaxy harassment (Moore, Lake and Katz, 1998).
Many of these models are based on a scenario where a field galaxy
falls into the cluster.   If clusters are formed this way,
hierarchically, then we should still see in-fall occurring,
although at a much reduced rate (Kauffmann 1995).  If we could
locate which galaxies are in-falling into nearby clusters, this
would give an excellent opportunity to test models which predict
how clusters evolve, and determine which mechanisms are responsible
for modifying field galaxies within clusters to yield the distinct
galaxy populations in clusters (e.g. Dressler et al. 1997).

Another, and possibly related process involving interactions of
galaxies in clusters, is the merger hypothesis for the creation of
elliptical galaxies (e.g. Lake \& Dressler 1986; Toomre 1977).  The
merger of galaxies has long been a popular explanation for the build
up of the large  central cD or D galaxies found at the centers of
clusters (Ostriker and Tremaine 1975; Toomre 1978).  In this model
of galactic "cannibalism", a large central galaxy accretes nearby
smaller galaxies,  building up its mass to become the giant galaxies
we observe at the centers of rich clusters.  Previous observations
of cluster centers (Hoessel 1980; Schneider et al. 1983) have shown
multiple nuclei to be common, present in at least 1/3 to 1/2 of all
clusters.  Projection effects are barely adequate to explain this
high fraction of multiple nuclei observed in the cores of first
ranked cD galaxies in clusters, and thus some are possibly remnants
of cannibalized galaxies.   There are however, some objections to
modelling the formation of the more common cluster ellipticals as
being soley a result of mergers (e.g. Gunn 1987; van den Bergh
1982).

A less studied feature of nearby cluster evolution is the
possibility of interactions, mergers and stripping occurring in
nearby galaxy clusters.   Early work on this topic included indirect
studies on the sizes of cluster galaxies, suggesting that the sizes
of the ellipticals shrink as they get closer to the core of the
cluster (Strom \& Strom, 1978).  This shrinkage was attributed to
stripping of material from the galaxies closest to the center.
Some later studies have however disputed this result (Currie 1983),
interpreting the Strom and Strom trend as an effect of luminosity
segregation in clusters.  Another form of distorted galaxy is the
radio head-tail galaxies, well known features of clusters seen at
radio wavelengths.  Radio head-tail galaxies have been observed in
several nearby clusters, including Coma and Perseus (e.g. Owen and
Eilek 1998).  The usual interpretation given for the existence of
these tails is a scenario where an in-falling galaxy is losing gas
from an interaction with the intracluster medium; however, internal
dynamical causes cannot be completely excluded.

Additionally, there are several possible interacting galaxies and
merger remnants in nearby clusters. Less rich clusters such as Virgo
and Fornax, contain possible low-velocity galaxy-galaxy
interactions, as well as distorted galaxies.  One example, NGC
4438/35 in the Virgo cluster, is a pair of galaxies displaying a
distorted morphology in both optical and radio (CO and HI) emission
(Kenney et al. 1995). Another Virgo galaxy, the Sa NGC 4424 is
probably the result of a merger (Kenney et al. 1996).   A
quantitative study of 84 Virgo disk galaxies (Koopmann and Kenney
1998) shows a systematic bias in previous morphological studies
towards early Hubble type classification, yet the physical
properties of these galaxies other than their morphological Hubble
types, suggest they are physically related to later type galaxies.
This change in morphology is possibly the result of stripping and
galaxy interactions changing slightly the morphology of the galaxies
in Virgo.  Rich clusters, which are generally thought of as being
more dynamically stable and mature systems, do not have as many
examples of luminous interacting systems.  However,  NGC 4676, ``the
Mice'' is  one of the best examples of a low-velocity galaxy-galaxy
interaction and is on the Coma cluster's outskirts.  This galaxy
pair is the result of a two disk galaxy merger (Barnes 1998).
Despite these dynamical and observational examples of interacting
and distorted galaxies in clusters, there has been no attempt to
systematically classify and study these systems in nearby clusters.

This paper presents results from a morphological study of fields in
nearby galaxy clusters based on moderately deep B and R images of
clusters taken with the WIYN 3.5m telescope.  We use these images to
systematically investigate the types, distributions, frequency of,
and physical features which could be associated with interactions,
tides or stripping in clusters.  Because such processes tend to
produce anomalous light distributions on galactic, or smaller
scales, we refer to such features by the generic term 'fine-scale
substructures'.  Since this is a morphological survey, follow-up
measurements are required to prove that any particular fine scale
substructure is in a given cluster.  We carried out our survey
looking for galaxies which have morphological properties of galaxy
interactions, or distortions, and to investigate any fine-scale
substructures in the clusters:  Abell 2199, Coma, Perseus, AWM 3 and
AWM 5.  We show that all structures found in these clusters of
various richness and morphology can be classified into just seven
different categories: Multiple Galaxies (MG), Galaxy Interactions
(IG), dwarf galaxies groups (dwG), tailed galaxies (TG), and
distorted galaxies (DG), line galaxies (LG), and galaxy aggregates
(AG).  In this paper we describe these different morphologies,
present color, magnitude, and positional information, and discuss
possible origins for the existence of these structures, including
comparisons to models.

Processes similar to those responsible for creating the large
fractions of distorted galaxies at moderate redshifts are probably
occurring in nearby clusters, albeit at lower levels.   We further
propose that the mechanism producing most of these disturbed
galaxies is via a method similar to the proposed galaxy harassment
models (Moore et al. 1998).  The drop-off in severity of distorted
morphologies in clusters from z $\approx$ 0.4 to z $\approx$ 0 can
be accounted for by the decayed rate of in-falling galaxies in
hierarchical clustering, which is theoretically predicted to  peak
at moderate redshifts (Kauffmann 1995), or from the longer dynamical
time scales associated with disturbing lower-mass (and hence lower
luminous) galaxies.

\section{OBSERVATIONS AND METHOD}

All of the observations used in this paper were obtained with the
WIYN\footnote{The WIYN Observatory is a joint facility of the
University of Wisconsin-Madison, Indiana University, Yale University,
and the National Optical Astronomy Observatories.} 3.5m, f/6.2
telescope located at Kitt Peak National Observatory.  A thinned
2048$^2$ pixel S2kB charged coupled device (CCD) was used to image the
cluster fields, mostly in or near cluster cores.  The resulting images
are at  a scale of 0.2 arcsec per pixel, and cover a field-of-view of
6.8 x 6.8 arcmin. The Coma, Abell 2199, AWM 3 and AWM 5 clusters were
imaged between the nights of May 31 to June 2 1997.  The Coma cluster
core was imaged in 6 different fields (see Table 1), straddled by the
two D galaxies NGC 4874 and NGC 4889.  Two additional fields in Coma
were taken centered on NGC 4881 and IC 4051.  One field image for Abell
2199, AWM 3 and AWM5, centered about their cD galaxies  were also
taken.  Exposure times were 900s for the B-band and 600s for the
R-band. The seeing ranged from 0.7 to 1 arcseconds full width at half
maximum.  The Perseus cluster was imaged in the fall on 1996, when the
WIYN CCD was non-linear.  We do not present photometric information for
this cluster due to the uncertainty in the performance of the CCD.  The
non-linearity does not affect the high-resolution image quality of the
cluster, and we are still able to make positive morphological
identifications of galaxies in the cluster.  We also obtained, as a
control sample for identification of background galaxies, a field
centered on the Hubble Deep Field (Williams et al. 1996).  We chose
this area for its known lack of nearby clusters, galaxies and stars, as
well as for its low Galactic extinction.

All of the images were then searched by eye, examining in detail,
every region of each image in the R and B bands.   Interesting
features, distorted galaxies, multiple galaxies, or anything
unusual was noted.   For the images centered on NGC 4881, IC 4051,
AWM 5 and AWM 3, there exists a large central galaxy. These
galaxies were removed using the IRAF routines ``isophote'' and
``bmodel'' for the less crowded fields.   Subtracting the central
galaxy yields a better view of the entire cluster, particularly the
central part of the core. In the denser fields, where bmodel and
isophote produce significant artifacts due to the presence of field
galaxies, the central galaxies were removed using a 180\deg
rotation about their centers and subtracted out in a manner similar
to that presented by Conselice (1997).  The symmetry method allows
a clearer view of the cores of these large E galaxies, nearly all
of which show some form of distortion, or have multiple nuclei.

Initially, the images were searched for unusual or outstanding
features, without reference to any distorted galaxy morphological
system.  Each object's position and a description of it were recorded.
By a closer examination of the list of structures, we created a catalog
of all the unusual objects in our cluster fields, and found that all
structures (248 total) could be classified into seven different
categories.

We then performed aperture photometry on the core of each structure,
to determine both colors and magnitudes; only the core of each
structure was measured to maximize our S/N ratio, and to to provide
a guide for future spectroscopic studies.  We use Landolt (1992)
standard stars to calibrate our images and are able to obtain
photometry with errors typically less than 0.1 magnitudes.  We do
not attempt to obtain color and magnitude information for entire
galaxies.  The tidal and distorted features are typically faint and
photometry of these features is not possible with our relatively
shallow imaging;  deeper images are necessary before we can obtain
photometry of these fainter features.   In our comparison blank field,
we performed the same kind of analysis, looking for any unusual
structures that are ubiquitous to a random area of the sky.  From this
comparison field, we are able to put constraints on the background
contamination in our cluster images.

\section{SAMPLE}

Our sample includes the Coma (Abell 1656), Perseus (Abell 0426), AWM
3, AWM 5, and Abell 2199 clusters.  What follows is a brief summary
of what is known about these clusters and their relationship to any
possible structures.

\subsection{Coma Cluster}

The Coma Cluster with a radial velocity of 6942 \kms (z $\approx$
0.023), is the prototypical rich cluster and the most exhaustively
studied rich cluster in history\footnote {It may be argued that Coma
is overtaking the less dense Virgo for the number one spot as the
most intensively studied cluster.  A search of the astro-ph preprint
server shows that as of February 1998, there are 35 Coma papers, 27
Virgo and 19 on the Perseus cluster entered in the last year.  This
may be biased by a number of papers coming from the Coma meeting in
Marseille in June, 1997, but still shows its growing popularity.},
with a BM (Bautz and Morgan 1970) richness class of II  (see Biviano
1997 for a recent review).

Coma at large scales has a compact symmetrical spherical shape (Kent
and Gunn, 1982), and until the X-ray work of Johnson et al. (1978) was
generally assumed  to be a relaxed, viralized cluster.  The amount and
intensity of X-ray inhomogeneity and substructure recognized in Coma
has increased over the past few years, particularly with ROSAT (Briel,
Henry and Bohringer 1992; White Briel and Henry 1993).  Additionally,
the observed X-ray emission cannot be completely modeled by an
isothermal distribution (Henriksen 1985).

X-ray substructure is the most obvious form, but substructure also
exists at optical wavelengths. The best example of substructure in
Coma is the existence of two D galaxies, NGC 4874 and NGC 4889,
both with similar masses.   D galaxies are the central locations of
deep potential wells (Beers and Geller 1983), and the existence of
two in Coma suggests that the cluster is not in complete viralized
equilibrium.  Two-dimension image maps, analyzed by statistical
maximum likelihood methods, also imply that the core of Coma is not
in an equilibrium distribution (Fitchett and Webster 1987).
Inspired by the amount of substructure seen in ROSAT maps, Colless
and Dunn (1996) analyzed 552 Coma galaxy redshifts, finding that
the velocity distributions can be fit by two Gaussians centered
about the D galaxies NGC 4874 and NGC 4889, an effect first hinted
at by Fitchett and Webster (1987).  The general interpretation of
this result is that the present Coma cluster formed sometime in the
past by a merger between two clusters once centered about NGC 4874
and NGC 4889.

A less studied effect in Coma, as opposed to Virgo, is the presence
of fine scale structures in its member galaxies.  Searches for this
type of feature is necessarily limited by the technology of
telescopes and imaging techniques.  In the last year or so, after the
advent of new image processing and telescopes of high resolution
capabilities, several fine scale structures have been found in Coma.
Luminous arcs of material have been seen in several different areas
of Coma (Trentham \& Mobasher 1998; Secker 1998; this paper), and
small dense 'galaxy aggregates' have been found (Conselice and
Gallagher 1998).  These arcs and aggregates are evidence for some
form of interaction between galaxies, or between galaxies and the
cluster potential.   The bulk of this paper refers to fine scale
substructures in Coma (116 in all), a significant portion of which
are probably the result of physical effects similar to those which
produced these arcs and aggregates.

\subsection{Perseus Cluster}

 The Perseus cluster (Abell 0426) is a relatively nearby cluster,
 D$\approx$75 Mpc (H$_o$ = 75 \kms Mpc$^{-1}$), with a moderate
 Galactic extinction, having a radial velocity of 5486 \kms
 (z=0.0183) and BM  class II-III.   Compared to Coma, little
 attention has been given to Perseus in the past few decades.

Perseus contains a peculiar cD galaxy, NGC 1275 (Perseus A), which is
a strong radio source.  NGC 1275 is extremely distorted with
filaments and abundant structure, and potentially has an origin that
differs from other cD galaxies.  Perseus differs from Coma in a few
important areas.  Perseus has only one large central galaxy, NGC 1275,
as opposed to the two found in the Coma cluster.  Perseus also has a
compact core R$_c$ $\approx$ 0.1 Mpc and a high velocity dispersion
$\sigma$ $\approx$ 1260 \kms, possibly related to the disturbed
morphology of NGC 1275.  Perseus also has a unimodal structure, with no
significant optical substructure, as well as a regular velocity
distribution (Girardi et al. 1997).  However, both Slezak et al. (1994)
and Mohr, Fabricant and Geller (1993) have found evidence for
substructure in X-rays in this cluster.  These X-ray studies are
however limited to the central part of the cluster, and may be
displaying effects of the central distorted galaxy NGC 1275, and the
high velocity dispersion near the cluster center.   Agreement between
the ratio of the intracluster medium temperature and the velocity
dispersion ($\beta$) for the entire cluster agrees well with
theoretical predicted values.  Therefore, for the most part, this
cluster can be considered as viralized, with a possible recent
significant in-fall event.

The Perseus cluster is also known to contain at least five examples
of radio head-tail galaxies (Sijbring \& DeBruyn 1998), including NGC
1275.  These radio head-tail galaxies are generally interpreted as
the result of an interaction between an in-falling galaxy and the
intracluster medium (Sijbring \& DeBruyn 1998).  Head-tail galaxies,
which these objects are usually referred to as, are only found in
cluster.  Perseus contains more of these objects than any other
cluster (Sijbring \& DeBruyn 1998).  These radio tailed sources may
also have a similar origin to the optical tailed galaxies as found in
this study.  In addition to having radio head-tail galaxies, Perseus
contains a significant dwarf galaxy population, with the majority of
these fragile systems showing neither evidence for an interaction, nor
distorted morphologies (Gallagher Han and Wyse 1997).

\subsection{Abell 2199}

Abell 2199, is the richest cluster in our sample, with a BM
richness class of I, and a radial velocity of 9063 \kms
(z$\approx$0.032; D $\approx$ 120 Mpc).  This cluster is dominated
by the cD cooling flow galaxy, NGC 6166.  NGC 6166 has been the
object of several detailed studies, and is the one of the best
candidates for cD galaxy cannibalism (Pritchet \& Harris 1990;
Lauer 1986; Bridges et al. 1996; Conselice \& Gallagher, 1998b).
The core of NGC 6166 contains several distinct nuclei.  These
objects are thought to be the remnants of merging elliptical
galaxies which formed the cD, or are projection effects caused by
eccentric orbits of isolated E galaxies (Lauer 1986).

Initial X-ray studies of Abell 2199 displayed a smooth distribution
of emission with no evidence for substructure (Forman \& Jones 1982).
More recent X-ray images of this cluster reveal an elongated shape,
as well as a significant cooling flow (Siddiqui, Stewart, and
Johnstone 1998).  Wang \& Ulmer (1997) have found a strong
correlation between X-ray elongation and the blue galaxy fraction.
Abell 2199 has a flattened central distribution in both X-rays and
optical light, and hence should have a high fraction of blue
galaxies.  Velocity dispersion work on this cluster has also shown
that there exists a large number of galaxies that have negative
peculiar velocities (Lucy et al. 1991).  A recent study combining
both X-ray and radio observations (Owen \& Eilek 1998) indicates that
the core of Abell 2199 is complex and cannot be described by a simple
spherical cooling flow model.  Maps of the overall velocity structure
of Abell 2199 (Zabludoff, Huchra \& Geller 1990), however, find no
significant subclumps, or velocity structures, indicating that the
core of Abell 2199 is viralized.

\subsection{AWM 3 \& AWM 5}

These two clusters are among the set of poor clusters first studied
by Morgan, Kayser \& White (1975) and  Albert, White \& Morgan
(1977).  These Morgan poor clusters were initially identified from
Palomar Sky-Survey prints as systems that contained large central D or
cD galaxies, but lack the high number of normal luminosity members as
seen in rich Abell clusters.  A typical criterion for a AWM or MKW poor
cluster is a system with 10 to 50 galaxies with magnitudes fainter than
m$_{3}$+2, where m$_{3}$ is the third ranked member of the cluster.
AWM 5 is the furthest cluster in our sample with a radial velocity of
10346 \kms (z$\approx$0.035), D $\approx$ 140 Mpc, and is centered
about the cD galaxy NGC 6269.  AWM 3 is the closest cluster in our
sample with a radially velocity of 4497 \kms (z$\approx$0.015) and a
distance of D $\approx$ 60 Mpc.

These morphologically defined clusters were later shown to be
physical systems with properties comparable to Abell clusters
(Bahcall 1980).  Additional X-ray studies (Kriss et al. 1983)
revealed extended X-ray emission in these clusters.  Galaxies in poor
clusters also contain neutral hydrogen (Williams \& Lynch 1991), as
well as evidence for H$\alpha$ emission (Beers et al. 1984).  These
systems are also comparable in age to richer clusters, and contain a
hot intracluster medium similar to those found in rich clusters
(Price et al. 1991).  In general,  these Morgan poor clusters in both
their X-ray and optical properties represent a class of objects that
form a continuation from the rich Abell clusters towards galaxy
groups; some having core densities comparable to those found in
richer Abell clusters as well as similar X-ray properties (Bahcall
1980; Beers et al. 1984; Price et al. 1991).

\section{THE CATALOGS}

We found 248 examples of fine-scale substructures in the images of our
five clusters.   We can place every galaxy  in our catalog into two
broad categories: galaxy associations, and disturbed galaxies, with a
third overlapping category, interacting galaxies, which are an
association with a disturbed morphology.   We can further divide these
two broad categories into the seven distinct and more descriptive
morphologies of line galaxies, distorted galaxies, multiple galaxies,
tailed galaxies, interacting galaxies and galaxy aggregates (see Fig
1).   The structures in our catalog are each assigned one of these
morphological types.  This initial approach is based purely on
morphologies.  As discussed in Section 5, some categories are likely to
be dominated by contamination from background galaxies; not all fine
scale structures seen towards a cluster are part of it.

Tables 6 through 10 contain our catalog of the fine-scale
substructures found in our five clusters: Coma, Perseus, Abell 2199,
AWM 3 and AWM 5.  Column one contains the name of the object,
starting with S (for structure) followed by the name of the cluster,
and ending in a number.  For the Coma cluster, we find a total of 116
structure, in Perseus 69, Abell 2199 30, AWM 5 with 23 and AWM 3 with
10 structures.  Column two and three list the R.A. (J2000) and Dec.
(J2000) for each structures.   For Coma, Abell 2199, and AWM 5,
columns four and five contain the R mag. and (B-R) color for each
structures (see Section 2 for further explanation).   The last two
columns of Tables 6-10 list the morphological type, sometimes given
with a '?' to indicate that the identification is in doubt.
Sometimes, the morphology changes between the R and B band images,
and is noted in the last column.

The distance range between the closest and most distant galaxy in
our sample is different by a factor of two, and it is possible that
in the more distant clusters, we could be mistaking some
morphological types for others.  This is due to the degraded
resolution seen in the more distant clusters.  This is discussed in
more detail in section 4.5.  Briefly, this is a long standing
problem for any morphological study, and can insert biases if not
address appropriately.  For this study, our resolution will only be
degraded by a factor of two, and in other morphological studies
concerning distant galaxies this is generally not enough to cause a
change in morphology (Conselice et al. 1999, in prep).    Additionally,
physical features such as size, and (B-R) color are similar for each
form of fine-scale substructure in each of our clusters, indicating
that we are not overly biased by resolution degrading.

The following sections describe the different morphologies listed in
the catalogs, giving possible field counterparts, and potential
origins for each type of structure. We emphasize that our data are
classifications of apparent galaxy structures, following the general
approach used in previous galaxy catalogs (e.g. Arp 1966, Lauberts
1982, Arp \& Madore 1987).  Physical interpretations in many cases
will require additional observations (see Sections 5 and 6).
Comparisons to various models for the existence of these fine-scale
substructures will be discussed in Section 7.

\subsection{Distorted Galaxies}

Distorted galaxies (DG), or peculiar galaxies have been observed in
the field for decades (e.g. Arp 1966); as well as in distant clusters
(e.g. Lavery \& Henry 1998, 1994).  The origin of these galaxies in
the field is almost always attributed to some form of low-velocity
galaxy-galaxy interaction.  Distorted galaxies in the field are
usually given the catch-all 'peculiar' classification.  When a
peculiar galaxy is isolated, it is either the result of a merger, or
is a starburst galaxy, which can also be the result of a galaxy
merger.    The presence of distorted galaxies in nearby clusters is
sometimes anecdotally noted, but has not been the subject of a
focused morphological study.  Nearly all of the disturbed
morphologies in our sample could be classified in general terms as
distorted normal galaxies, e.g. as expected from interactions.
However, for the present purposes, distorted galaxies are a catch-all
category that includes any isolated galaxy that has a peculiar
structure and does not fit into the other categories.  Galaxies that
have a disturbed morphology, and have a nearby apparent companion,
are classified as interacting galaxies, but may have a similar origin
to distorted galaxies.  For the purposes of this paper, we will
classify large ellipticals, including cD galaxies, that have multiple
nuclei (NGC 6166), or shells (NGC 5629) as distorted galaxies, with
the understanding that not all distorted galaxies are caused by the
same processes.

What could cause an apparently isolated galaxy in a cluster to be
distorted?  It is possible, although not likely, that an apparently
single object is in reality an interacting pair, or a recent
merger.  In our sample we have about twice as many distorted
galaxies as cadidate interacting galaxies, and it would be hard to
explain the existence of so many distorted galaxies by recent
mergers.  Many distorted galaxies could be products of past
interactions, a model consistent with fast collisions in clusters
where after effects linger (e.g. Moore et al. 1998).  Alternatively,
distorted galaxies could be interacting with the intracluster
medium, producing a starburst, or with the cluster potential (see
Section 7).  The tidal radii (Merritt, 1984) for some of our
distorted galaxies are consistent with stripping due to the
cluster's potential, but some galaxies have stripped material which
extends well beyond their gravitatonal tidal radii.  Also, we see
galaxies that are disturbed but are smaller than their tidal
radius.  However, by combining high-velocity galaxy interactions
interactions  with tidal forces from the cluster potential, as
predicted in galaxy harassment models (Moore et al. 1998), we may be
able to explain distorted cluster galaxies.   These models will be
further discussed in Section 7.

\subsection{Galaxy Aggregates}

In a previous paper we have described the morphology of galaxy
aggregates observed in the Coma cluster (Conselice \& Gallagher,
1998a).  A detailed description of these objects as seen in Coma is
presented in that paper.

Aggregates are systems of galaxies inside clusters which are
dominated by a central disk galaxy with smaller dwarf galaxies, or
star knots surrounding it in an asymmetrical pattern; generally all
on one side.   The colors of the smaller members tends to vary
according to the specific aggregate.  Some of the Coma aggregates have
aggregates which are redder, some bluer than the central galaxy.
Others have colors that span a range from red to blue.

In Conselice \& Gallagher (1998a) the criteria for an object to be
an aggregate is defined to be:

\begin{enumerate}

\item Primary galaxy must be disk galaxy which is not a dominate member
of the cluster.

\item The knots or dwarf galaxies must be asymmetrically distributed
and within 2-3 optical radii of the primary.

\item The number of dwarf galaxies, or knots surrounding the primary
member must be a statistical excess over the number of similar objects
in the cluster field. \\

\end{enumerate}

Interestingly, we find nine examples of possible galaxy aggregates
in the Coma cluster, including the three reported previously in
Conselice and Gallagher (1998); however, we find no examples in the
other clusters (with the exception of one possibility in AWM 5).
These aggregates could in some way be related to the clumping of
dwarf galaxies, which we have 10 examples of in Coma, and only four
total for the other clusters.   Additionally, we find the
aggregates to be spread out in the images; that is we do not find
an overabundance of this morphology in the densest parts of Coma.
If solely a projection effect, caused by the high density of
galaxies in this cluster, then we would expect to see more
aggregates in the densest part of Coma, however this is strictly
not true (see Table 2).  Perseus and Abell 2199 also contain an
abundant amount of dwarf galaxies, as well as being just as rich as
Coma, but have no examples of aggregates.  Our failure to see
aggregates in these clusters and the other poor AWM clusters
suggests that these objects are not mainly in the background.

In this paper we present what are a number of other possible
aggregates; these are not as prominent as the three we presented in
Conselice \& Gallagher (1998), but none the less morphologically
qualify as being aggregates.  Interestingly, the sizes of the
aggregates are all nearly the same, about 20 kpc, which is the size
of the tidal diameter for stripping in Coma (Merritt 1984), based on
a 100 \kms   velocity dispersion assumption for early type disk
galaxies (see Section 7).

\subsection {Multiple Galaxies}

In our clusters individual galaxies rather frequently appear to be
in association with other galaxies, with no apparent interaction
occurring.   Usually these are in the form of pairs, but
occasionally these types of galaxies are found to be in multiple
systems.  Classification as a multiple galaxy (MG) further requires
that the two or more members must be approximately the same size and
apparently close together, with no visible evidence for
interactions, in the form of tidal plumes or distortions.  Our
defintion of a multiple galaxy differs from those of Arp (1966) or 
Zwicky (1957)
in that we are not imply a physical association, just an apparent one.
Galaxies which are multiple, but show signatures of interactions are
classified as interacting galaxies (Section 4.4).   Dwarf galaxies
which appear to be congregating together are classified separately as
dwarf galaxy groups, which will be discussed in Section 4.7.  The MG
morphological category is the most susceptable to chance
superpositions.

Curiously, there is only a slight correlation between galaxy
richness and the number of multiple galaxies found.  This indicates
that some of these systems may be galaxies which are close to one
another but are not strongly interacting.  They could, for example,
be in the early phases of an interaction, before damage is done, or
moving at high relative velocities which weakens collisional
effects.  The color distribution of these systems is shown in
Figures 10, 11 and 12.  Some of the MG pairs show surprisingly blue
colors (B-R) $<$ 1, indicating that star formation is probably
occurring.  It is possible that we are missing some of the finer
features of an interaction and are thus misclassifying what should
be an interacting galaxy system; suggesting that some multiple
galaxies are physical systems.  However, the average (B-R) color for
multiple galaxies is one of the reddest morphological class (see
Table 5), consistent with chance superpositions of non-interacting
galaxies accounting for some of this class.

\subsection{Interacting Galaxies}

Interacting galaxies (IG) are systems which appear to have a
disturbed morphology and are in close proximity to another galaxy.
For our purposes, close proximety is defined to be two or more
galaxies of similar sizes that appear to be within 3 galaxy radii of
each other.  In some cases, the interaction is fairly obvious for
both galaxies; but for others, only one galaxy shows a significant
distortion.

If these systems are indeed real, then this would be direct proof of
low-velocity galaxy-galaxy interactions in clusters.  This category,
like the multiple galaxies (MG) is highly subject to chance
alignments, but the probability of superimposing two distorted
galaxies of similar sizes in a cluster is not high, although
possible.  Therefore, for a very few cases,  we are probably
observing a form of low-velocity galaxy-galaxy interaction.
Interacting galaxies are also the bluest form of structure, with an
average (B-R) = 1.59$\pm$0.49, with individual examples with colors
(B-R) $<$ 1.   The blue colors of IG gives a clue to their origins.
If these other structures are caused by an interaction with a
cluster's gravitational potential, and if any star formation is
occurring, it is at a reduced rate as compared with a low-velocity
galaxy-galaxy interaction.  This could be the result of the
mechanisms which produce the  structures being somehow 'weaker', but
may also be the result of gas stripping occuring prior to tidal
interactions, an idea which will be further discussed in Section 7.

\subsection{Tailed Galaxies}

Galaxies which have a luminous (probably tidal) tail, but no obvious
companion, are surprisingly common among our sample, and are likely
due in large part to background galaxies (see Section 5).   The
galaxies in this category have a wide range of sizes, but similar
surface brightnesses.  Some of these are  smaller than the average
sized cluster member, with faint tails, facilitating their anonymity
until now.

There are abundant examples of tailed galaxies in a variety of
different environments that are consistent  with several formation
scenarios.   Tails associated with field galaxies are very common,
and are almost always due to interactions between  galaxies (e.g.
Chromey et al. 1998; Elmegreen et al. 1998).  Tidal tails are also
locations where dwarf galaxies can form (e.g. Hunsberger et al.
1996), and therefore may be related to the galaxy aggregates.
Alternatively, tailed galaxies might be the result of gas 
stripping in clusters.

Calibrating with our blank field image (Section 2), we find that a
significant fraction of the tailed galaxies seen in these clusters
are probably due to projection of background galaxies.   The galaxies
seen in the Hubble Deep Field image are, however, very tiny, and
cannot explain the larger galaxies that we find in the clusters that
have tails.  These tailed galaxies could be interacting with the
global potential of the cluster, or could be an interaction with
companions which are too small to be detected.

An interaction between two field disk galaxies, which would appear
as simply a tailed galaxy when the system is very distant, is not
unusual.   The interacting galaxies NGC 4038, `The Antennae', and
NGC 5996, both consists of a large spiral with one of its arms
extended out towards a companion.  If very far away, these two
galaxies would blend together, and the appearance of these galaxies
would be similar to the tailed galaxies seen in our clusters.
These galaxies are also possibly  morphological similar to the
"tadpole" type galaxy found in relative abundance in the Hubble Deep
Field (van den Bergh et al. 1996).   These ``tadpole'' galaxies are
also isolated and appear to be the result of a distant merger, or
from some other dynamical effect.  A local analogy of these types of
galaxies is possibly NGC 3991\footnote{NED describes this galaxy as
a Magellanic irregular which is involved in an interaction with NGC
3994 and NGC 3995,  accounting for its peculiar morphology}, first
described by Morgan (1958).

Radio tails attached to galaxies in nearby clusters, including Coma
and Perseus are well-known (Feretti et al. 1990; Owen and Eilek
1998), and may have origins related to the optical tailed galaxies
found here.  Radio head-tail galaxies are generally found near the
core of their clusters, and are thought to originate from
ram-pressure stripping with an interaction between the stripped
plasma and the ambient gas (Begelman et al. 1979).  Tailed galaxies
are  most common in Perseus, with 0.2 tailed galaxies per
arcmin$^2$.  Interestingly, Perseus is also contains more examples
of radio head-tail galaxies than any other cluster (Owen and Eilek
1998).  This is another indication that the tail galaxies likely
include representatives in clusters, and may point towards an
intracluster, ram-pressure event to account for their morphology.

\subsubsection{Debris Arcs}

Debris arcs, like that discovered by Trentham \& Mobasher (1998) and
the one reported here, are likely to be related to the phenomenon of
tailed galaxies.  These are luminous bands, thought to consist of
stellar tidal debris from galaxies. The debris arc reported in this
paper (SA1656-029), has a size of about 20 kpc.   These objects, only
a few of which are known, are only seen in the Coma cluster.  The
possible origin of this interesting form of fine-scale substructure
is from an interaction with the potential of a cluster, or by some
form of galaxy harassment, an idea discussed more in Section 7.

\subsection{Line Galaxies}

Line galaxies with red colors are probably the most unusual, and
interesting class of galaxy, which are not known to exist in the
field.    These are exactly what their morphology indicates.  They
are extremely thin galaxies which have lengths that are at least 5
times longer  than their width.  In the field these superthin
galaxies probably are extreme late type galaxies seen edge on (e.g.
Goad \& Roberts 1981; Karachentsev et al. 1993) in which case they
are dynamically cold disk systems (Karachentsev \& Xu 1991; Matthews
\& Gallagher 1998).   If all line galaxies were superthin extreme
late-type systems, then  the presence of an excess density of line
galaxies towards clusters would be surprising; clusters are
generally deficient in late-type spirals.  This type of galaxy is
also present in our blank field, and a sizable portion of the
smaller line galaxies are probably in the background.  The bluer
line galaxies are likely to be small extreme late type galaxies,
Scd-Sdm, seen edge on.  Red line galaxies are an oddity, and
probably have a different origin.

Of our seven morphological types, line galaxies are the ones that are
most likely a normal type of galaxy seen in projection.   Possibly,
we are seeing a previously unrecognized LSB extreme late-type galaxy
component in clusters, which are detected only when they are edge-on
and therefore have a higher surface brightness (Bergvall \&
R\"onnbeck, 1995, Dalcanton \& Schectman 1996).   They could also be a
projection effect from a long tailed galaxy, or possible pieces of an
arc (Section 4.5.1) seen edge on.  Line galaxies, however, present no
significant curvature and are much more abundant than the very few
examples of arcs as seen in Coma.  Also, by far the most abundant
location for finding line galaxies is in the Perseus cluster, which
has no known examples of galaxy arcs.

A problem with this scenario is the extremely red colors of the
line galaxies in Coma have an average color of (B-R) = 2.18, a
higher average by 0.3 magnitudes than any other fine-scale
substructure.   Red colors are inconsistant with normal extreme
late-type spirals (e.g. Matthews \& Gallagher 1997).  Because of
their red colors, these galaxies might be related to early-type
objects within the clusters, or background galaxies seen in
projection (Secker et al. 1997).  If these are background systems,
they are likely to be very thin galaxies seen edge-on.  However, we
see a significant overabundance of these objects in Perseus, which
also has the highest density of tailed galaxies, as compared with
the other clusters.    Additionally, we find that for Abell 2199
and AWM 5 the line galaxies have average colors near (B-R) =
1.79$\pm$0.0 and (B-R) = 1.42$\pm$0.62, respectively.  The blue
colors of line galaxies in AWM 5, and the wide range in colors as
seen in Coma (see Figure 16), with some line galaxies having (B-R)
$\approx$ 1, indicates that a portion of these objects are probably
located within these clusters.  Redshifts of line galaxies are
required to properly resolve this issue.

\subsection{Dwarf Galaxy Groups}

Excess concentrations of dwarf galaxies in one region define a
'dwarf galaxy group'.    These objects usually consist of more than
five, mostly small, low surface brightness galaxies in the same
region.  This type of structure is almost exclusively found in the
Coma cluster (Table 3), its origin could be related to the galaxy
aggregates, which are also almost exclusively found in the Coma
cluster.   The tendency for dwarf galaxies to clump together in Coma
was originally noticed by Ulmer et al. (1994).  Along with multiple
galaxies, and galaxy aggregates,  the apparent relationship between
these galaxies may be due to projection effects.    However, a strong
argument can be made that some of these structures are real, as they
are not found in clusters other than Coma.  Coma has a relatively low
dwarf galaxy to giant ratio (Secker and Harris, 1996) for its
richness, and other clusters like Perseus and Abell 2199 will have
the same, or higher dwarf galaxy to giant ratios.  Coma, therefore
does not have a higher fraction of dwarf galaxies than other rich
clusters, and hence there is no reason why we should see more
clumping of these dwarf galaxies, if the effect is solely one of
projection.  The fact that we see only one dwarf galaxy group in
Perseus, and only two in Abell 2199, as opposed to the 10 we see in
Coma indicates that these objects are most likely real
associations.    The number density of dwarfs in dwarf groups is
around 49 galaxies arcmin$^{2}$.  The average number density of
similar, or brighter galaxies in other areas of Coma is 5 galaxies
arcmin$^{2}$. These systems have densities 20$\sigma$ above the mean
cluster level, and hence are almost certainly real physical systems.

\section{FIELD GALAXY CONTAMINATION}

When studying any group of astronomical objects where distances to
the objects cannot be known with certainty, one must consider, and
correct for, the possibility that some of the objects under study
are either in the foreground or background of the system you are
trying to study.  This is a long standing problem in galaxy cluster
research, and is particularly troublesome for the work presented in
this paper.  However, there are several tests we can perform to
determine whether or not what we are observing in these clusters are
statistically likely to be part of the cluster, or in the foreground
or background.

A large portion of the smaller galaxies in our sample could be the
result of projections of background galaxies mixing with and
contaminating our samples, and this is a likely origin for the
fainter, red objects.  Assuming that the Universe, and hence
background objects are isotropically distributed, then if our
fine-scale substructures are entirely of non-cluster origin, we
should see similar proportions in every field.   That we do not, and
the fact that we see more structures in the richer and closer
cluster fields implies that at least some of the features we are
seeing are real (Tables 2 \& 3).  To test this further, we can use
an image of a blank area of the sky to determine if any of the
structures we are seeing are present in an area which contains no
clusters.

To determine a lower limit to contaimination, we imaged the Hubble
Deep Field (HDF) at the same exposure times used for all of our
cluster images (600s). This area was chosen for its lack of any
bright stars or galaxies, or galaxy clusters (William et al. 1996)
to determine how many of our tiny background galaxies may be due to
contamination from distant and small field galaxies.  Performing a
similar type of analysis that we did on the cluster images, we find
four examples of what we would have classified as tailed galaxies,
one possible multiple galaxy, two line galaxies, and one distorted
galaxy.  Naturally, objects like multiple galaxies, and dwarf groups
are prone to chance superpositions among cluster members, and this
cannot be tested by using the HDF image, but we can compare the
non-associated structures.  We find no examples of interacting
galaxies in the HDF images, but on average we find three per WIYN
field in the cluster images.  The HDF images have one distorted
galaxy, whereas we find seven on average for our cluster fields;
distorted galaxies are  probably components of the clusters.

 However, we do not find this to be true for the line and tailed
 galaxies.  We find an average of 2.5 line galaxies per cluster
 field as compared to the two found in the blank HDF field.  We
 also find an average of 5.3 tailed galaxies per cluster field as
 compared to the four seen in the blank field.  This indicates that
 these galaxies might be solely background objects.  However, as
 will be seen, some of these objects are found in large abundance
 in certain clusters, as well as having peculiar trends in color
 (see Section 4.5), creating difficulty with a purely background
 explanation.  Additionally, the HDF tailed galaxies are very
 small, with barely discernible tails; some tailed galaxies in our
 sample have similar sizes, but others are very prominent. Future
 observations using HST images should clarify the morphologies of
 these galaxies.  The line galaxies in the HDF field likewise are
 morphologically the smallest and faintest.   No galaxy aggregates,
 dwarf groups, or interacting galaxies were observed in the blank
 field.

The best, and most accurate way to determine if a galaxy is in a
cluster is to know its redshift.  Unfortunately, we do not have
redshift information for the majority of the galaxies in our sample.
However, a few of the galaxies in our sample do have known published
redshifts, mostly in the Coma cluster.  Published  redshift studies
of the other clusters are disappointingly meager.  The galaxies for
which redshifts were available from NED\footnote{The NASA/IPAC
Extragalactic Database (NED) is operated by the Jet Propulsion
Laboratory, California Institute of Technology, under contract with
the National Aeronautics and Space Administration.} are shown in
Table 11.

We find 21 galaxies in the Coma cluster in our catalog can be
identified with a redshift, one with Perseus, and four for A2199,
and 1 with AWM3.  Most of these galaxies, the only ones that we can
confirm a redshift for,  have velocities which are consistent with
cluster membership; with most galaxies having velocities less than 1
$\sigma$ of the cluster velocity dispersion.  However, a few
exceptions must be noted and may give clues as to the nature of some
of the other galaxies in our catalogs.   In the Coma cluster the
galaxies, SA1656-018, SA1656-026, SA1656-030 have recession
velocities cz = 18300, 48108, and 35268 \kms  respectively (Table
11).  The average Coma velocity is 6917 \kms; these objects are in
the background and their morpholgies are DG, MG, and MG.  They also
appear very small in our images, similar to the galaxies in our
blank field image.   Note that SA1656-057, a tailed galaxy, has a
recessional velocity of 6629 \kms, only 288 \kms different from the
average Coma velocity, and certainly a member of the cluster.  Also,
the line galaxy SA1656-071 has cz = 8135, which is within 1 $\sigma$
of the mean cluster velocity, and hence also part of the cluster.
Therefore, we can be sure that at least one example of a tailed and
line galaxy exist in the Coma cluster, and are not background
objects.

We can further test membership of the galaxies by taking the averages
of the Coma galaxies in our sample that have known radial velocities,
excluding the three obvious background galaxies (SA1656-019,
SA1656-026, SA1656-030). The average velocity of these 18 Coma
galaxies is 7196$\pm$1726 \kms, which is close to the value
6917$\pm$47 \kms  found with a complete sample of 552 redshifts of
spectroscopically confirmed members (Colless \& Dunn 1996).  While
obviously not proof that the majority of the galaxies in our catalog
are part of their respective clusters, it does show that in Coma, the
fine-scale substructure is distributed in velocity space in a similar
manner as the average Coma galaxy.  The other six galaxies with
fine-scale substructures and velocity measurements in A2199, Perseus
and AWM3, all have velocities within 1$\sigma$ of the cluster mean.
Also, if our fine-scale substructures are purely background objects,
then these should be distributed nearly randomly on the sky.  We
however, do not see this type of behavior when we plot the positions
of our objects (Figures 2 and 3, see next section).

We therefore conclude that our sample probably contains a significant
fraction of objects in, or near their respective clusters.

\section{RESULTS}

\subsection{Spatial Distributions}

In each of our cluster fields, we find some forms of fine-scale
substructure.  The denser clusters, e.g. Coma and Perseus, have more
examples than the poorer AWM clusters.  One of the ways to determine
the nature and possible origin of these structures is to find how
they vary with local environmental conditions.  If we find that the
objects are evenly distributed, clumped, or only present in some
parts of the clusters, this can give us information concerning the
physical nature of these objects.

Figure 3 shows the location as a function of R.A. and Dec. (J2000)
for the fields IC 4051, NGC 4881, AWM5, AWM3, NGC 6166, and for the
Perseus cluster.  Figure 2 shows the six Coma fields (see Table 1).
It appears that for most of the fields a form of clumping is
occurring, although it is not a significant or outstanding feature.
It is rare, even in the low density fields, to find a structure that
does not have another one nearby.  Also, for some fields,
particularly Coma 3, Abell 2199, and Perseus there appear to be
voids where no fine-scale substructures are found.  This is also
hinted at in the various other cluster fields.   These void areas do
not always, but sometimes contain the larger ellipticals in the
cluster.

There are a few more important points worth mentioning concerning
the distribution of these structures.  In general, the structures
seem to be more abundant in richer fields (e.g. Coma, Perseus) than
in the poorer clusters (AWM 3 \& 5), showing a possible fine-scale
substructure density relation.  Our structures also tend to avoid
the areas around the central galaxies. However, this could be the
result of selection bias.  Since these structures are not isolated
but tend to clump together, this may indicate that in-fall events
are occurring into the cluster not as isolated galaxies, but as
groups (Henriksen \& Byrd 1996).

\subsection{Relative Frequencies of Structures}

We can learn more about fine-scale substructures by examing the
relative numbers of each structure in our five clusters. Tables 2 and
3 show the number of different morphologies as found in each of our
five clusters.  We can then use this information to investigate the
morphological trends in our clusters.  Figures 4 - 6 show the
relative distributions of the structures in the six Coma fields
(Figure 4), the other two Coma fields surrounded by NGC 4881, and IC
4051, and the fields for Perseus, Abell 2199, AWM 3 and AWM 5 are
shown in Figure 5.  The total Coma distributions is shown in Figure
5.

To study the salient features of the distributions we can categorize
everything into two broad subgroups: galaxy associations (includes
AG, MG, dwG) and structures that show either a distortion or
interaction (includes DG, IG, TG, and LG).  We place every fine-scale
substructure into one of these categories, and define $\Gamma$ to be
the ratio of the number of disturbed structures to the number of
associations.  When we do this, we find that our clusters fall into
two broad types: clusters that have their fine-scale substructure
dominated by associations, and ones that do not.  A high value of
$\Gamma$, which in our case, would be $\Gamma$ $\approx$ 10, would
signify a cluster dominated by associations, while a $\Gamma$
$\approx$ 1 would be a cluster dominated by disturbed structures.
Table 4 shows the values of $\Gamma$ along with the number of
association and disturbed structures.  Coma, Abell 2199, and AWM 5
all have $\Gamma$ values near 1, while Perseus and AWM 3 have
$\Gamma$ values near 10.

Why would some clusters appear to be overabundant in  associations,
while others are not?  There is no obvious answer to this, but we
propose that the value of $\Gamma$ gives a rough indication of the
recent dynamical history of a cluster.

We can further see this subdivision of cluster by examining the
histograms for the different clusters.  The clusters that have a low
$\Gamma$ (Coma, Abell 2199, and AWM 5) all have similar distributions
of types of structures (Figures 5 \& 6).  Multiple galaxies are the
most common type of structure in all three of these clusters, with
almost identical histograms for A2199 and Coma.  In addition, all but
one of the dwarf galaxy groups are in these three clusters.  On the
other hand, for our $\Gamma$ $\approx$ 10 clusters, we have very few
multiple galaxies (AWM 3 \& Perseus; see Figure 6).  Both clusters
are also dominated by distorted galaxies and tailed galaxies.
Therefore, the two types of clusters is our sample  have both similar
values for $\Gamma$ as well as similar distributions of structures.

\subsection{Magnitudes and Colors}

A powerful method of deciphering the origin of these fine-scale
substructures is to investigate their colors and their
magnitudes.   We are able to perform photometry to obtain the
colors and magnitudes of the fine-scale substructures in the three
clusters: Coma, Abell 2199, and AWM 5.  The Perseus images were
taken with the WIYN CCD in the fall of 1996 when the CCD was not
linear, and we only have one R-band image for AWM 3.   Photometry
is performed on the R and B images, using only an aperture size of
a few pixels at the center of our galaxy.  The magnitude and color
information presented here is only for the core of each object.
Typically the colors of the fainter features, such as tidal tails
and distorted parts of the galaxies that are of low surface
brightness, cannot be computed with any accuracy with our
relatively shallow survey imaging.  When a particular morphological
structure has more than one galaxy, such as multiple galaxies,
dwarf groups and galaxy aggregates, we measure the brightest
member.

Figure 7 shows the color-magnitude digram for all fine-scale
substructure object in the Coma cluster.  While this plot has a large
scatter, we can still pick up some trends.  The color-magnitude
diagram shows a very slight trend towards bluer colors at fainter
magnitudes with a large scatter.   In clusters, it is not surprising
to find a color-magnitude diagram with this type of weak
correlation.  Secker et al. (1997) find a similar type of
relationship for Coma dwarf elliptical galaxies.   This diagram
includes both the large galaxies in our sample, as well as the
fainter dwarf galaxies which constitute a part of the sample.  It
also contains the inevitable background galaxies.    Overall,
however, there is a pattern at the faint and bright ends of the
diagram;  objects which have a magnitude brighter than R = 18.0 have
colors greater than (B-R) = 2.  However, for the objects with R
fainter than 23, nearly every object is bluer than (B-R) $\approx$
2.

We do not see similar color-magnitude diagrams for the clusters
Abell 2199, and AWM 5 (Figure 8 \& 9).  In these two clusters, most
of the objects have similar colors, which do not show a dependence on
the magnitude.  It is possible that we do not pick up the same slight
dependence on magnitude that we see in Coma because we do not have
enough structures in those clusters to detect any correlations.

By looking at the colors of the individual fine-scale substructures
in our clusters, we might be able to better understand the origin of
these objects.  Figures 10, 11, and 12 show (B-R) scatter plots as a
function of the morphological type for our three clusters with color
information.   For Coma, the average colors for the galaxy types
are, from bluest to reddest (see Table 5): Interacting galaxies
(B-R) = 1.62$\pm$0.42, tailed galaxies (B-R) = 1.67$\pm$0.65,
distorted galaxies (B-R) = 1.68$\pm$0.76, galaxy aggregates (B-R) =
1.72$\pm$0.51, dwarf galaxy groups (B-R) = 1.73$\pm$0.45, multiple
galaxies (B-R) = 1.90$\pm$0.59, and line galaxies (B-R) =
2.18$\pm$1.4.  With the other examples from Abell 2199 and AWM 5, we
find that the average colors of these structures throughout all
three clusters with photometric information are:  Interacting
galaxies (B-R) = 1.59$\pm$0.49, Tailed galaxies (B-R) =
1.69$\pm$0.97, distorted galaxies (B-R) = 1.71$\pm$0.73, galaxy
aggregates (B-R) = 1.73$0.43$, dwarf galaxy groups (B-R) =
2.10$\pm$0.61, multiple galaxies (B-R) = 1.90$\pm$0.61, and line
galaxies (B-R) = 1.91$\pm$1.0 (see Table 5).

The bluest objects are the interacting galaxies and tailed
galaxies.  For interacting galaxies, this is not surprising and is
what we would expect if new stars were being formed from the
disturbed galaxies undergoing an interaction.  However, this is not
an extremely blue color, and as can be seen from from Figures 10, 11
and 12, a large spread of colors exists for distorted galaxies;
some have a color (B-R)$<$ 1, while others have (B-R)$>$ 2.5.   It
is impossible to tell from our photometry what the origin of such a
wide scatter is for these and the other fine-scale substructures.
Several possible scenarios (see Section 7), various degrees of
H$\alpha$ emission, and likely background contaminations could be
producing this wide scatter.

The next bluest type of fine-scale substructure are the tailed
galaxies.  These objects are less prone to projection effects,
since the objects consist of only one galaxy, but contaminating
background galaxies are likely.  A (B-R) color of 1.69 is not very
blue, although like the distorted galaxies, the scatter ranges from
blue colors (B-R) $\approx$ 0, to very red ones (B-R) $\approx$
3.5.  If tailed galaxies originate from an intracluster medium
interaction followed by a tidal interaction, star formation does
not necessarily have to occcur.  When ram-pressure stripping occurs
in a galaxy in-falling into the cluster, it will be stripped of all
its gas.  Later, when gravitational, or other perturbing forces act
on the galaxy, the material that we see as a tail could be old
stellar material, with no star formation occurring due to a lack of
gas.    However, we do have some examples of extremely blue,  and
also extremely red tails in our sample.  These objects may be the
result of a form of star fomation.  Similar arguments can be made
for the other structures as well, all of which have a wide scatter
in their (B-R) colors (see Figures 16, 17 and 18 for histograms).

As expected, if the structures we are seeing are real, the average
colors of the associated objects (AG, dwG, and MG) are all redder
than the disturbed structures (IG, TG, DG) with the interesting
exception of the line galaxies (LG).  The line galaxies are the
reddest objects in the Coma cluster (but the bluest in AWM 5), and
have the second highest (B-R) color.   The color of these objects in
Coma are red enough that they could be considered to be background
objects in the Coma cluster (Secker et al. 1997).  However, the
moderate and bluer colors of these objects in Abell 2199 and AWM 5
suggests that this simple interpretation might not fully explain the
Coma line galaxies (see Section 4.5, for a full discussion).

Figures 13, 14 and 15 show the distributions of magnitudes for our
sample as a function of morphology.  There is a wide scatter of
magnitudes, with some interesting features.  The line galaxies in
all three clusters have the faintest magnitudes.  In Coma, the
brightest line galaxy has a central magnitude of about 20.5, which
for most of the other structures, the brightest central magnitude is
around 16.5.  This is an indication of the low surface brightness
nature of these objects, as well as the violation of the
``fainter-bluer'' relation found by Secker et al. (1997) for Coma
dwarf elliptical galaxies.  Figures 19, 20 and 21 show histograms of
the magnitudes of the fine-scale substructures found in our three
clusters.  We do not see a rise in the number of galaxies, for any
morphology at fainter magnitudes.   Additionally, the dwarf galaxy
groups (dwG) have similar magnitudes and much less of a spread as
compared with the other features.  The features with the largest
spreads in magnitudes are the distorted galaxies and the multiple
galaxies.  The other features with low surface brightness are the
interacting galaxies and tailed galaxies; however there are examples
within these classes that have high surface brightness features.

\section{ORIGINS OF STRUCTURES}

The structures that we see in these clusters have a few possible
origins.  They could be due to superpositions of background galaxies
(Section 5), causing us to think that a particular galaxy is part of
the cluster.  This is certainly the case for some of our galaxies,
but our failure to observe similar abundances of structures in our
control field, and the relative increase in density of the structures
in the fields of denser clusters, as compared with the density in the
poorer clusters, as well as confirmed membership for 89 \%  of the
galaxies with known redshifts all show that a significant fraction of
our catalog are cluster members (see Section 5 for a more detailed
discussion).  Furthermore, with the caveat that some of our catalog is
certainly of background origin, we will hereafter assume that each of
our morphological classes (AG, TG, DG, MG, LG, IG, dwG) have
representatives in galaxy clusters.

The Coma cluster  contains a signifcant number of blue galaxies
(e.g. Bothun \& Dressler 1986; Caldwell et al. 1993) which resemble
Butcher-Oemler galaxies as  seen in distant clusters (Butcher \&
Oemler 1978, 1984).  We further confirm this observation, by finding
a significant fraction of galaxies in our sample with blue colors.
In fact, almost every morphological class has objects with colors
near (B-R) $\approx$ 1.   However, for the most part, the
structures we find have moderate to fairly red (B-R) colors, and
hence are dissimilar in this respect to distant B-O galaxies.  All
of the galaxies in our sample are similar to B-O galaxies in that
they have distroted morphologies.  Since some of our sample is blue,
with the rest red, and B-O galaxies are both distorted and blue,
this indicates that another mechanism is probably producing the red
galaxies.  However, based on previous observations of blue Coma
galaxies (e.g. Caldwell et al. 1993), and the results presented
here, processes that produce the blue galaxies in our sample are
probably similar to those occurring in the B-O clusters.

Four basic models predict how structures might have formed;
low-velocity galaxy-galaxy interactions, gas stripping,
and tidal interactions with the cluster potential.  The fourth
option, 'galaxy harassment' combines interactions with the
cluster's gravitational potential with collisions with  galaxies in
the cluster.   From our color information, we know that some of the
structures in our catalogs have colors that are consistent with
very little active star formation, so starbursts alone cannot
produce the observed effects, although they are a likely factor in
the bluer galaxies.  These blue galaxies have been observed
previously in Coma (Caldwell et al. 1993), and are interpreted as
star forming galaxies in the cluster.   Several models have  been
developed which predict how star formation within a galaxy will
behave during its in-fall into a rich cluster, which we will also
discuss.

\subsection{Gas Stripping}

If hierarchical structure formation is occurring in clusters, then
galaxies should be accreted by the cluster, slowly building up the
cluster's mass.  If these galaxies are field disk galaxies falling
into a cluster,  then a large portion of their gas may be stripped
into the cluster evironment by a variety of processes (Dressler \&
Gunn, 1986; Gunn \& Gott 1972; Valluri \& Jog 1990, 1991).   If this
is occurring, it will eventually lead to disk galaxies in the
cluster which are devoid of significant amounts of gas, potentially
converting spirals into S0 galaxies.  This model is supported by the
lack of H$\alpha$ emission from late-type spirals in the Coma
cluster (Moss \& Whittle 1997), and other observations of gas
deficiencies in cluster spirals (e.g. Haynes \& Giovannelli
1986).    If all the gas is stripped out of a disk while it enters
the cluster, then tidal forces will be unlikely to create a star
formation event, as the necessary raw mateirals are absent.

If ram-pressure is occurring, it could happen on a much faster time
scale than other mechanisms which produce tides and distortions in
an in-falling galaxy.  Soon after falling into a rich cluster, which
typically have intracluster  mediums extending significantly away
from the core of the cluster, gas will be removed from the galaxy
into the intracluster medium.  This may occur quickly before any any
other mechanism will have had time to produce an effect on the
in-falling galaxy.   During this stripping of gas, the galaxy will
start an orbit inside the cluster.   While on this orbit mechanisms
such as low-velocity galaxy-galaxy interactions (Section 7.2) or
galaxy harassment (Section 7.3) will begin to strip the galaxy
dynamically.  Usually when dynamical stripping of a galaxy occurs,
the gas inside is compressed and new stars are formed.  However,
since the in-falling galaxy has had its gas stripped by the
intracluster hot gas, no material is available to produce any new
stars.  Therefore, the stripped material only contains the old
stellar populations which existed before the galaxy was accreated by
the cluster.  Thus, galaxies which look distorted, but have red
colors are produced.

Ram-pressure stripping may also cause some forms of star formation
in an in-falling galaxy, however the specifics of such a process
remain somewhat uncertain (Henriksen \& Byrd 1996).

\subsection{Low-Velocity Galaxy-Galaxy Interactions}

Low-velocity galaxy-galaxy interactions are traditionally popular
mechanisms for explaining the B-O effect as seen in distant clusters
(Lavery \& Henry 1988, 1994).   In the field, it is fairly common to
observe low-velocity galaxy-galaxy interactions, most of which cause
some form of star formation in the galaxies undergoing an
interaction.  Galaxy interactions can also explain many other
features of unusual galaxies, including morphological changes
without associated star formation.

We also find in our nearby cluster images what appear to be examples
of galaxies which are interacting.   These morphological interacting
galaxies are, however, one of the rarer classes of structures in our
catalogs: we find about twice as many distorted galaxies - disturbed
galaxies without an obvious companion.   Additionally, some
morphologically selected interacting galaxies could be superpositions 
of two slightly abnormal galaxies.  Hence, low relative
velocity pairs with obvious signatures of interactions are rare.
However, for a few cases, noted in the catalogs, there are systems
undergoing a rather obvious interaction, which are also  members of
their respective clusters.  These galaxies typically have blue
colors.

The best way to determine if two galaxies are physically
interacting, is to know the radial velocity of each member of the
pair.  A high velocity difference would probably indicated that the
mechanism for producing the disturbed morphology is something other
than an interaction between the two galaxies, at least as envisioned
by Toomre and Toomre (1972).  One of our interacting pairs,
SA1656-065 (Markarian 60), has a velocity difference of ~4000 \kms.
This velocity difference is too high for this to be a standard
low-velocity galaxy-galaxy interaction.  However, other more
convincing cases than SA1656-065 are found, even though we do not
always have redshift information for these.

A likely candidate for a low-velocity galaxy-galaxy interaction is
in AWM 3, where the central galaxy NGC 5629 shows evidence of
ripples or shells, which point in the direction of an elongated
galaxy IC 1017.  The redshifts for these objects are 4498 \kms and
4382 \kms, respectively, a velocity difference of only 116 \kms.  This
is small enough to allow a strong interaction between these two
galaxies to occur.  Other possible examples exist, and can be confirmed
once their radial velocities are known.  We therefore conclude that
low-velocity galaxy-galaxy interactions in clusters probably do take
place, but are not a dominate effect in our sample (see Table 3).

Interacting galaxies are also the bluest form of structure in our
survey.    A low-velocity galaxy-galaxy interaction has a much
shorter time scale than an interaction between a galaxy and the
cluster potential, and is more likely to occur in a group in the
early stages of infall.  Therefore, this class of  interaction can
occur before ram-pressure stripping depletes the neighboring
galaxies of their gas new stars are created from.  Each low
relative-velocity, physically interacting galaxy pair thus is a
likely candidate for being in the the early phases of in-fall on the
periphery of the cluster.

\subsection{Tidal forces from cluster potential and Galaxy Harassment}

Tidal interactions between the cluster gravitational potential and a
member galaxy was considered in detail by Merritt (1983, 1984), and
applied to disk galaxies by Valluri (1993).  In the scenario
presented in these papers, a galaxy begins to undergo tidal stripping
when it falls into a cluster.   The outer portions of this in-falling
galaxy are dynamically heated and possibly stripped away from the
remainder of the galaxy, potentially producing a change in morphology
of the in-falling galaxy.  The radius of the in-falling galaxy to
where this stripping occurs, called the tidal radius, depends upon
the in-falling galaxies velocity dispersion $v_g$ and that of the
cluster $v_{cl}$. The {\it minimum} tidal radius $r_T$ of the in-falling
galaxy occurs for objects near the cluster core radius, R$_c$, given
by:

$$r_T=R_c\frac{1}{2}\frac{v_ g}{v_{\rm cl}}. $$

As shown in Conselice and Gallagher (1998), the star knots or dwarf
galaxies for RB 55 and RB 60 are located at about the tidal radius
for Coma.  The other aggregates found in Coma, are also of similar
size.  Therefore, at least these objects could be direct results of
tidal stripping caused by the cluster gravitational potential.

Other aspects of tidal stripping within rich clusters have been
considered as possible mechanisms for the production of active and
barred galaxies in clusters (Byrd \& Valtonen 1990;  Henriksen \&
Byrd).  These models predict that in-falling disk galaxies will
produce significant starbursts, especially near the cluster core;
this, of course, is based on the survival of suficient amounts of gas
in the inner disks.

Another method of producing disturbances in cluster galaxies is to
consider a combination of effects which lead to  'galaxy harassment.'
The models for galaxy harassment, developed by Moore and
collaborators (Moore, Lake \& Katz 1998; Moore et al. 1996a, Moore et
al. 1996b) present a natural explanation for some of our structures.
Galaxy harassment occurs when a galaxy experiences weak encounters from
bright cluster galaxy members, and is subject to tidal forces, as
discussed earlier.  The tailed galaxies and aggregates could be the
direct result of in-falling  galaxies which are currently undergoing a
form of ``morphological transformation'' via harassment that produces
the distorted structures.

In the model by Moore et al. (1998), disk galaxies orbiting a
viralized cluster are slowly transformed from disk galaxies, as
seen in higher z clusters, to dwarf spheroidal systems that are
abundant in nearby rich clusters (e.g. Secker \& Harris, 1996).  On the
way to becoming a spheroidal system the galaxies undergoing harassment
develop tidal tails composed of the material in the galaxy, both stars
and gas, and will contribute to the general intracluster medium.  One
particularly interesting effect of galaxy harassment, as shown
dramatically in the video accompanying Moore et al. (1998), is the
production of long tidal tails.  This may be the most natural
explanation for the occurrence of galaxy arcs and tidal tails as seen
in Coma, which are morphologically similar to the harassment
simulations.  Galaxies that are most most prone to undergoing this type
of transformation are low density, low mass galaxies that have orbits
that take them close to the cluster core.   If the galaxy in-fall
process is an ongoing one, then the tailed galaxies we see could be the
result of this in-fall occurring at the present time.    Additionally,
galaxy harassment allows for galaxies which are not in the core of the
cluster to become distorted, an effect that we see in our images, where
fine-scale substructures occur outside of the cores of the clusters.

Some of the galaxy aggregates (Conselice \& Gallagher, 1998) could
also be the result of galaxy harassment, or any mechanism that
produces tidal tails.  A disk galaxy falling into the core of a
rich cluster like Coma (where almost all galaxy aggregates are
found) will experience harassment, or some form of cluster
interaction, and as a result will developed tidal tails as proposed
by Moore et al. (1998), Byrd and Valtonen (1990) and Merritt
(1985).  If the tidal stripping is sufficient enough to propel
enough material from the disk galaxy, dwarf galaxies could in
principle form in the debris, creating a galaxy aggregate.

Models where dwarfs are created from galaxy interactions (Zwicky
1956; Barnes \& Hernquist 1992) and observations of dwarfs in tidal
tails (Yoshida et al. 1994; Hunsberger et al. 1996), support, in
principle, this possibility.  In addition the knots surrounding the
aggregates RB 55 and RB 60, both located in the Coma core, have
absolute magnitudes ranging from -12 to -14, which with fading would
overlap the magnitudes of local dwarf spheroidal systems (Conselice
\& Gallagher, 1998).   However, before this possibility could be
entertained further, the true physical nature of the aggregates based
on spectra, including redshfits, as well as further modeling, is
necessary.

Our data suggest that interactions with the gravitational potential of a 
cluster via harassment, or gas stripping are probably the dominate physical
mechanisms for the creation of fine-scale substructures in nearby
clusters.  The observations of isolated distorted and tailed
galaxies evidently are not explained by low-velocity galaxy-galaxy
interactions, and the few possible low-velocity galaxy-galaxy
interactions that we do see could result from pairs and
groups undergoing tidal compression as they enter the cluster.

\section{CONCLUSIONS}

We have performed a purely morphological and photometric study of
unusual fine-scale substructures in five nearby clusters (Coma,
Perseus, Abell 2199, AWM 5 and AWM 3); placing all observed
structures into seven different categories.  These morphological
categories are: Interacting galaxies (IG), distorted galaxies (DG),
tailed galaxies (TG), line galaxies (LG), multiple galaxies (MG) and
dwarf galaxy groups (dwG) (see Fig. 1).

Our photometric information indicates that for most of the galaxies
in our catalogs the central (B-R) colors have normal distributions
with some galaxies having extremely blue, (B-R) $\approx$ 0, or
extremely red colors (B-R) $\approx$ 3.5.  While a minority of
structures are associated with some form of a starburst, others are
probably quiescent, perhaps due to a lack gas, and still others are
of background objects.

By examining the relative distributions of structures, we classify
the galaxy clusters in our sample into two types:  those that are
dominated by apparent galaxy associations (dwarf groups, multiple
galaxies, galaxy aggregates), and those that have very few
associations, relative to distorted structures.  We define a factor
$\Gamma$ to be the ratio of distorted structures (interacting
galaxies, distorted galaxies, tailed galaxies, line galaxies) to
apparent galaxy associations (dwarf groups, multiple galaxies and
galaxy aggregates).  We find the value of  $\Gamma$ to be either
close to 1 for association dominated clusters (Coma, Abell 2199, and
AWM 5), and about 10 for non-association dominated clusters (Perseus
and AWM 3).  We further propose that $\Gamma$ may be related to the
viralization status, or is a measure of the recent galaxy in-fall
rate into a cluster.

We show that structures have a clumpy appearance and avoid the areas
occupied by the cD galaxies in the rich clusters.  Since distorted
structures tend to cluster together,  this may be an indication that
in-falling galaxies are entering galaxy clusters as members of
galaxy groups.   We further find that the bluest structures of our
seven categories are the distorted, interacting and tailed galaxies,
which is what one would expect if these features are cluster
members.

Based on models, and the overabundance of structures that appear to
be isolated from nearby companions, we conclude that the mechanisms
producing these fine-scale substructures are interactions through the
cluster potential, via galaxy harassment (Moore et al. 1998), or by
tidal stripping (Merritt 1984; Henriksen \& Byrd 1996).  The
implication is that tides and interactions in clusters are an
important component of nearby galaxy cluster evolution.  These
features are  fairly common in nearby clusters and are not solely the
providence of  distant clusters.  Furthermore, these features must be
considered and accounted for in any future attempts to understand the
structure and evolution of nearby clusters.

\acknowledgements
C.J.C. thanks members of the astronomical community for their
suggestions and encouragement during his poster presentation on this
topic at the January 1998 AAS meeting in Washington D.C. (Conselice
\& Gallagher, 1997). We thank F. Schweizer and K. Strobach for useful
comments and suggestion on a draft of this paper.  We also thank the
referee for their suggestions which improved and clarified several
important points.  We also thank B. Otte for help with the
observations.   This work was supported in part by the National
Science Foundation though grant AST-980318.

\newpage

\vspace{3cm}

FIGURE CAPTIONS:

Fig 1 --- Six examples of the morphological structures discussed in
this paper.   Aggregate images can be found in the paper Conselice \&
Gallagher (1998), and are not included here.  Each example is typical
for the entire sample of our catalog.

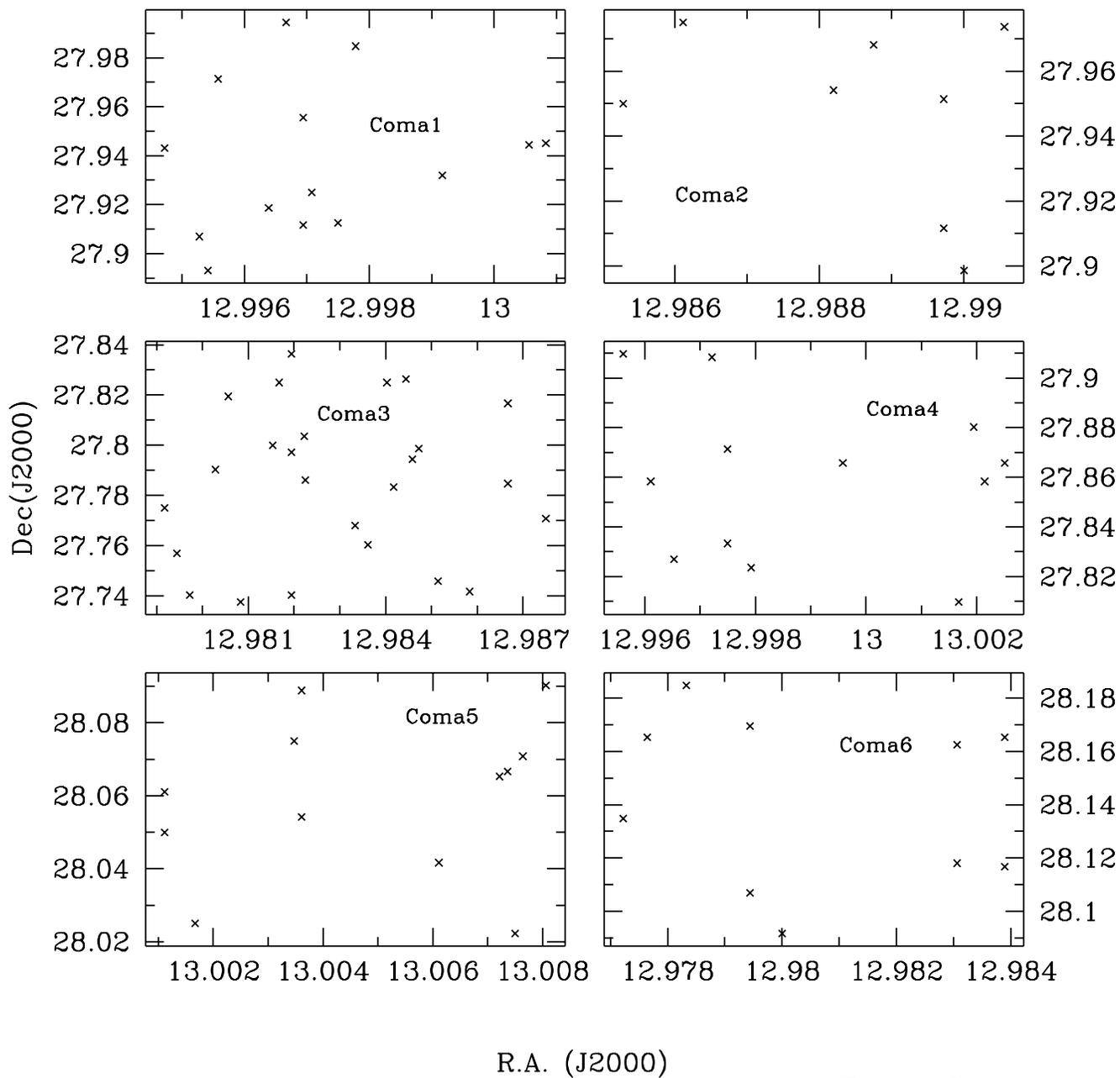
Fig 2 --- The distributions of fine-scale substructures in the Coma
cluster for the first six Coma fields (Table 2). The symbol x marks the
location of a structure.   In this figure and in Figure 3, it can be
seen that the structures tend to congregate together.

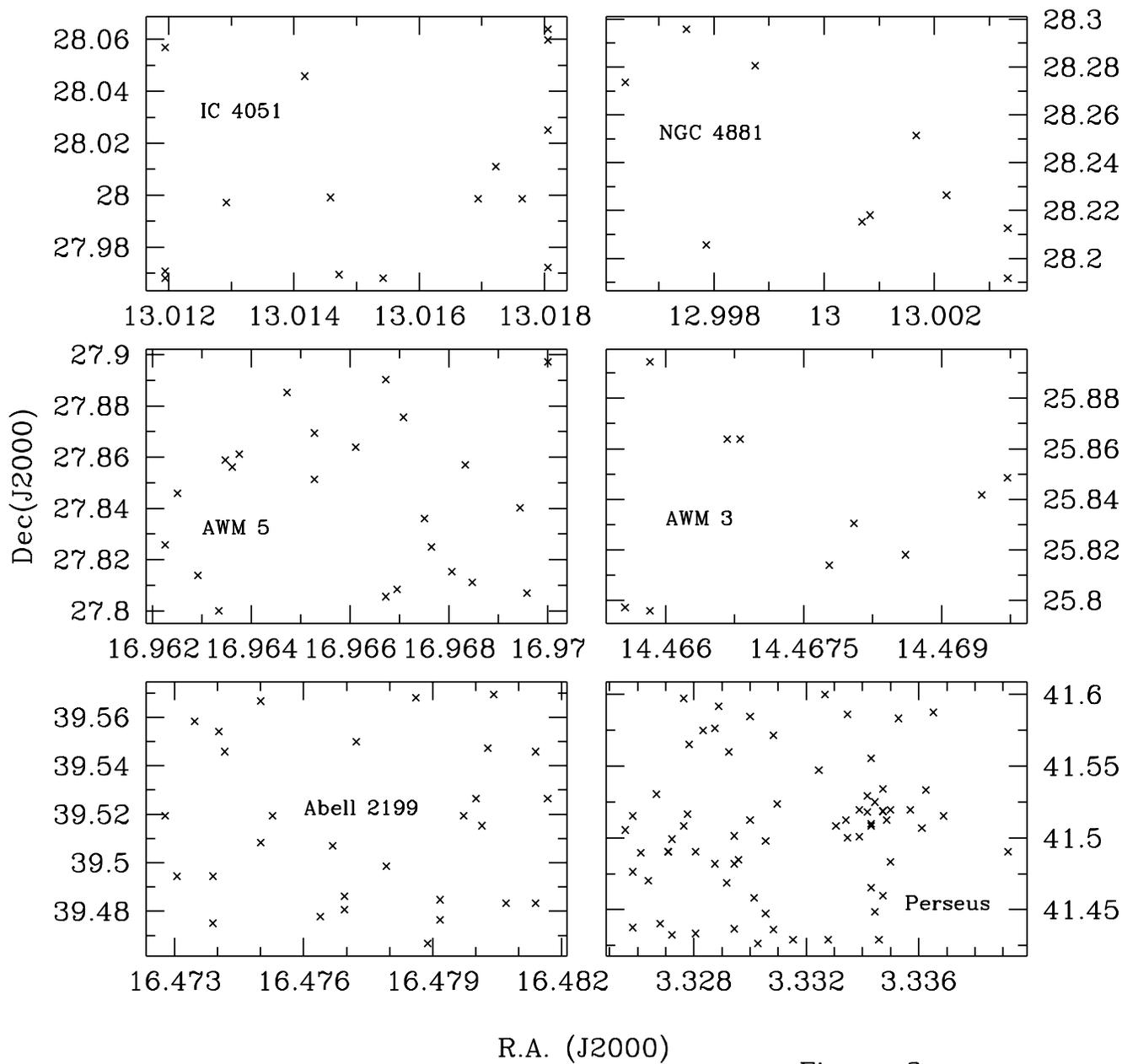
Fig 3 --- The distributions for the two Coma fields surrounded by NGC
4881 and IC 4051, as well as the distributions in the clusters:
Perseus, Abell 2199, AWM 3 and AWM 5.  We see similar effects occurring
as in the other Coma fields (Fig 2).  The structures tend to clump
together, with few structures appearing alone, and some in apparent
groupings.  We can also see 'bubble' shaped structures in these
diagrams, especially for the rich fields of Abell 2199, Perseus and
AWM5.

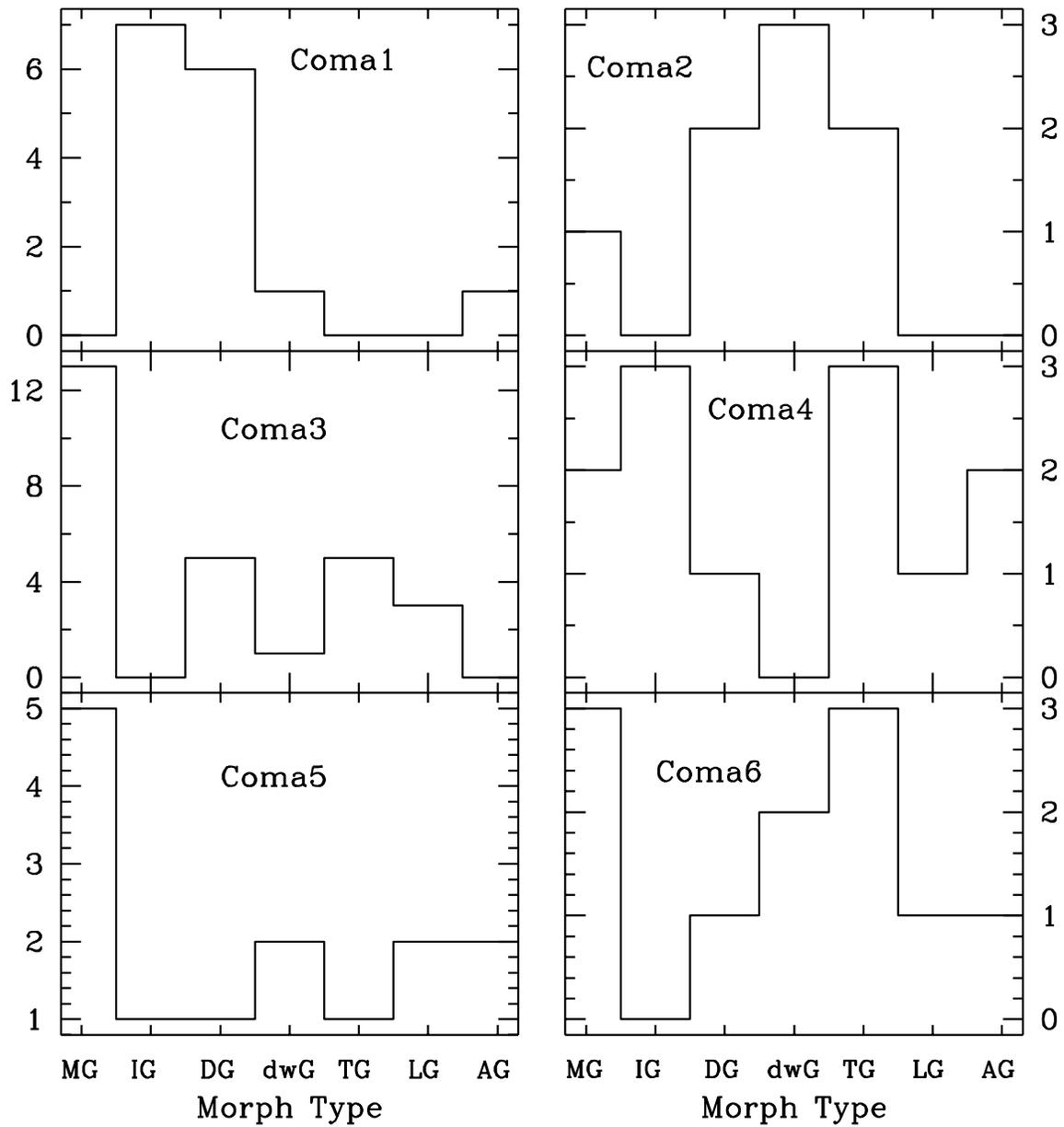
Fig 4 --- Histogram of structures as found in the six Coma fields.  For
most fields the dominate type of structure is either the multiple
galaxy, or dwarf galaxy group morphology.  There is also a dominance of
tailed galaxies in these fields of Coma.  Coma 1 field is unusual in
that it is dominated by interacting galaxies, and not multiple galaxies
and tailed galaxies.

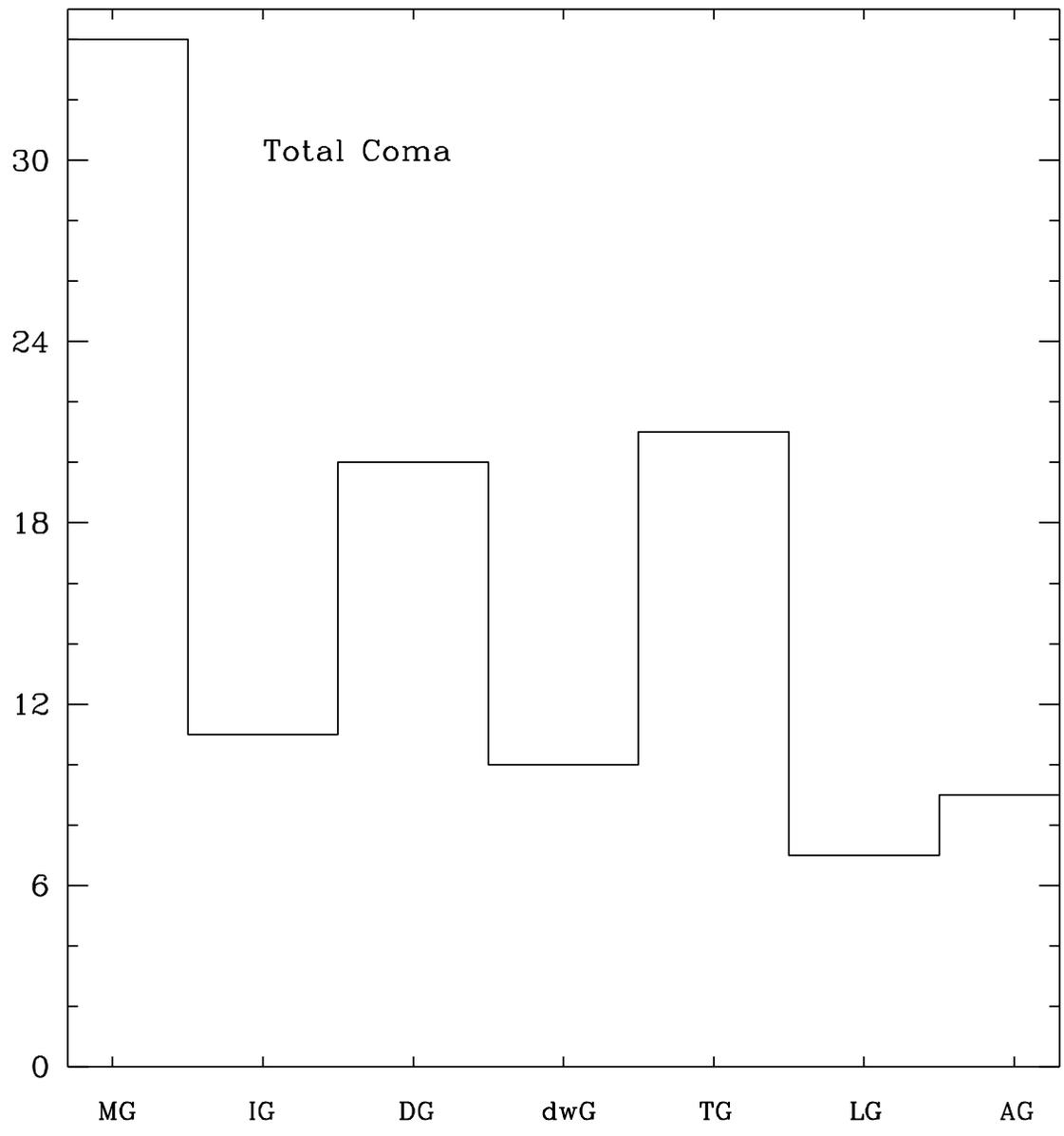
Fig 5 --- Total Coma histogram with all objects included showing a
similar dominance of multiple galaxies and tailed galaxies as seen in
Fig. 4.

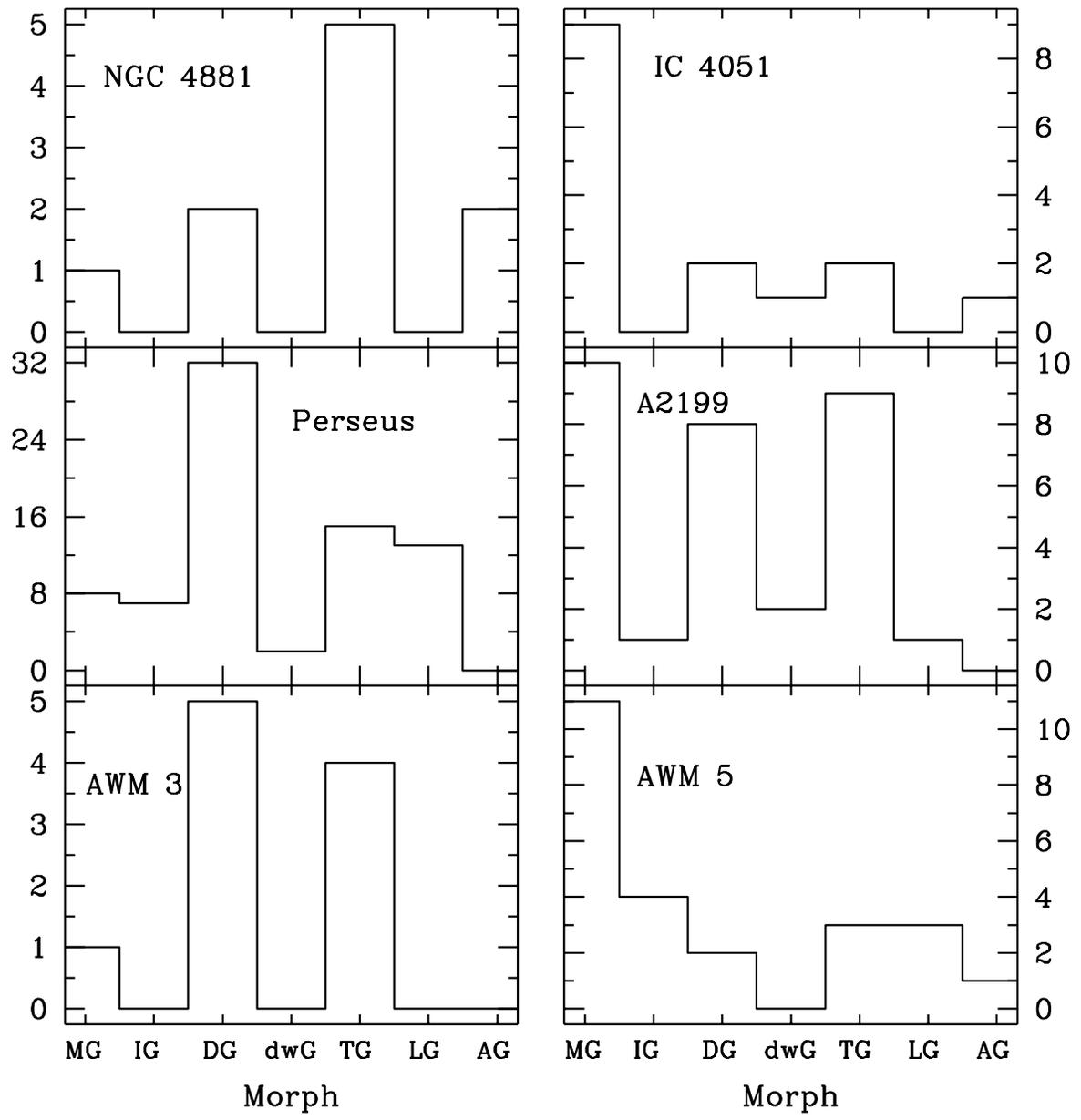
Fig 6 --- Histograms of structures for the two remaining Coma fields,
as well as the fields for the Perseus cluster, Abell 2199 and AWM 3 \&
5.  Here, we can see significant differences in the histogram trends
between these clusters and the Coma histogram (Fig 5).  The histograms
for Coma (Fig 5), Abell 2199, and AWM 5 are similar, and AWM 3 \&
Perseus are similar to each other.  The values for $\Gamma$ for these
clusters, also correlate (Table 4).  Coma, Abell 2199, and AWM 5 having
$\Gamma$ $\approx$ 1, while Perseus and AWM 3 have $\Gamma$ $\approx$
10.

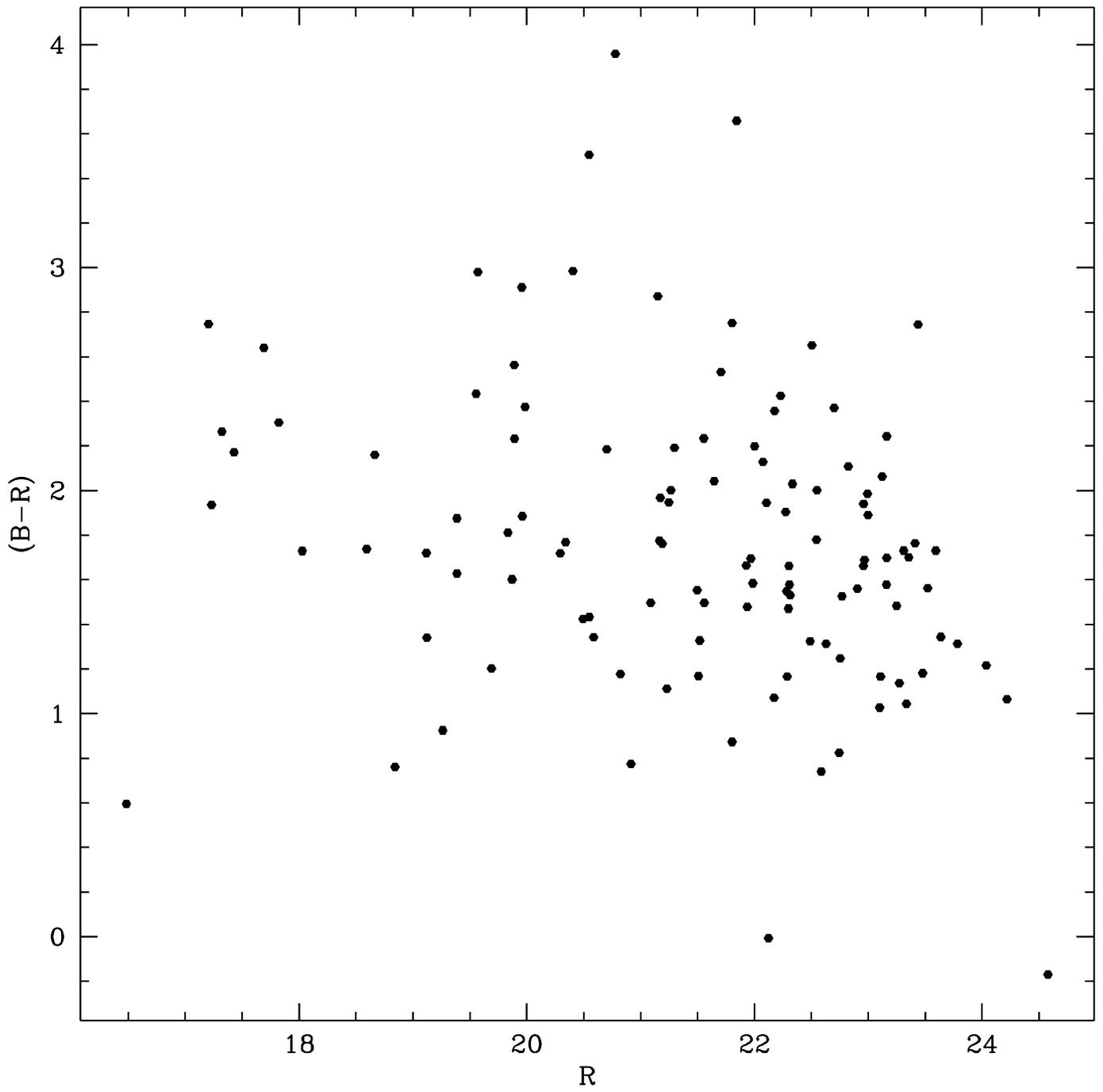
Fig 7 --- The color-magnitude diagram for all structures found in the
Coma cluster.  A very slight correlation between color and magnitude
can be seen, with fainter objects bluer.  This relation may have origin
similar to the one for dwarf ellipticals in Coma found by Secker et al.
(1997).

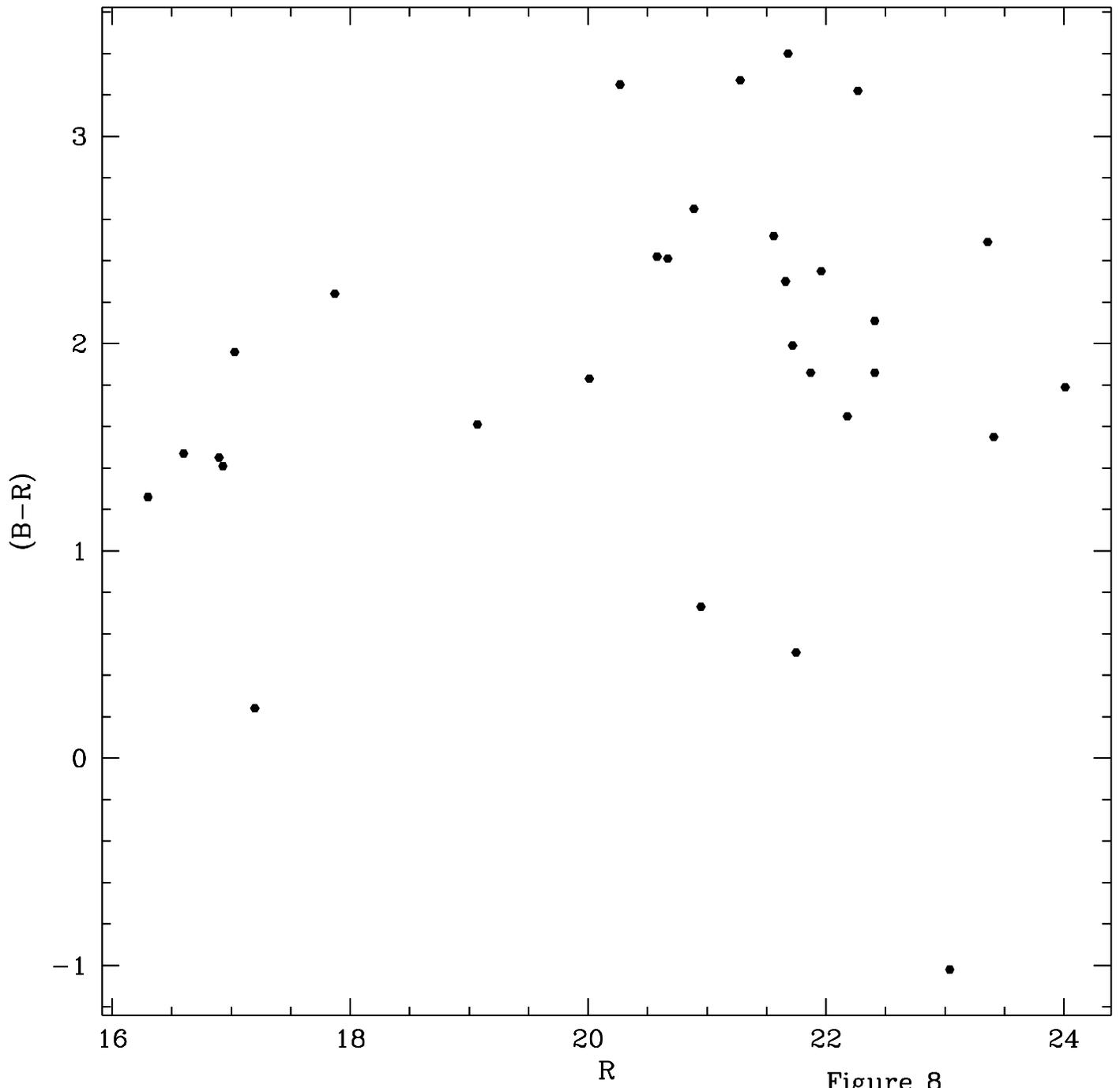
Fig 8 --- The color-magnitude diagram for Abell 2199, including all
structures.

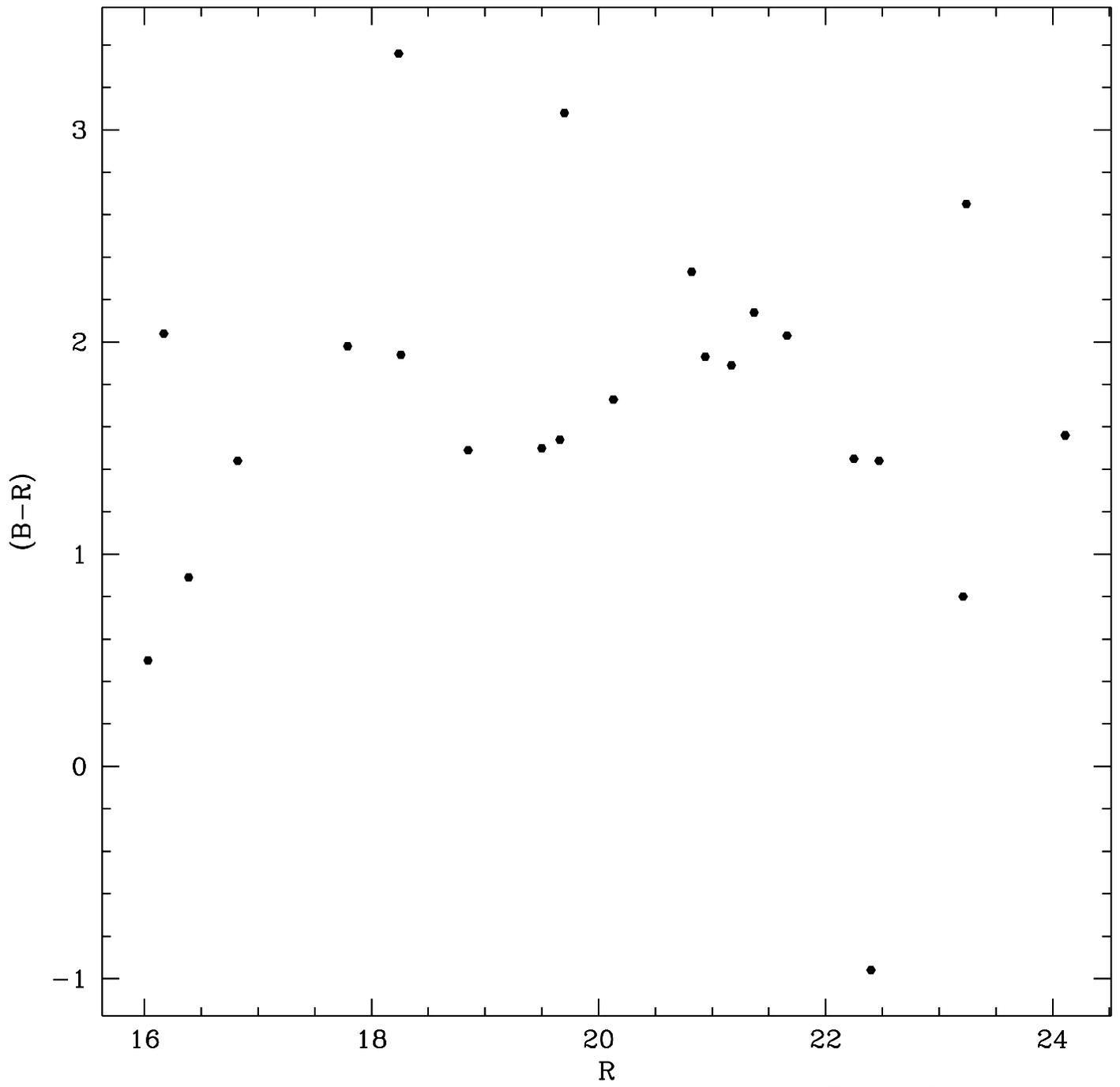
Fig 9 --- The color-magnitude diagram for AWM 5, including all
structures.

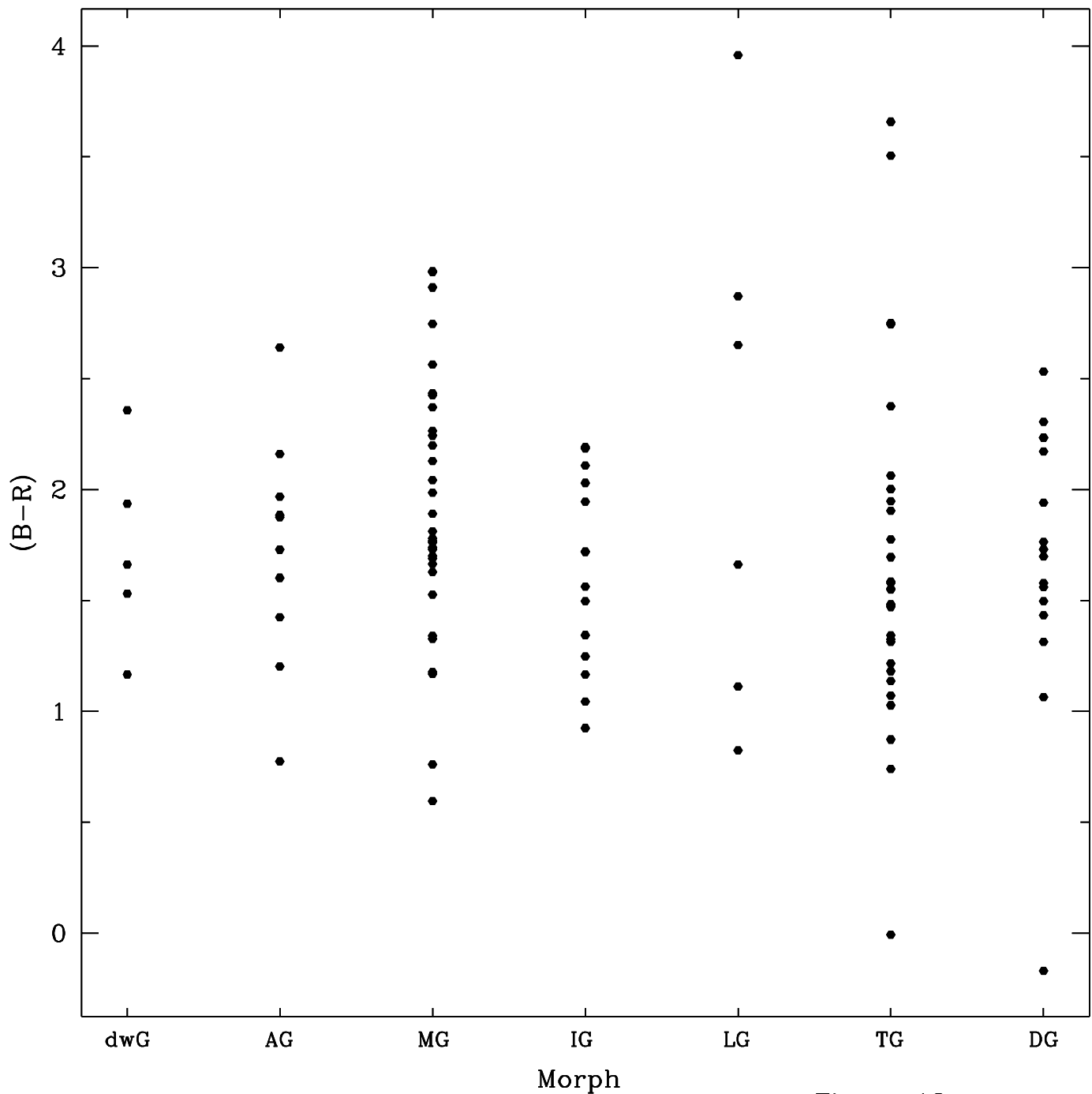
Fig 10 --- The color distributions for each of the morphology types as
found in the Coma cluster.  The average colors for each morphology is
shown in Table 5.  Each morphological type shows a wide scatter in
color, with most having objects as blue as (B-R) $\approx$ 1 and as red
as (B-R) $\approx$ 3.0.  The widest scatter is for the line and tailed
galaxies, possiblely indicating that the reddest members are background
galaxies.  Most of the features however have average colors which are
slightly bluer than the average Coma galaxy.

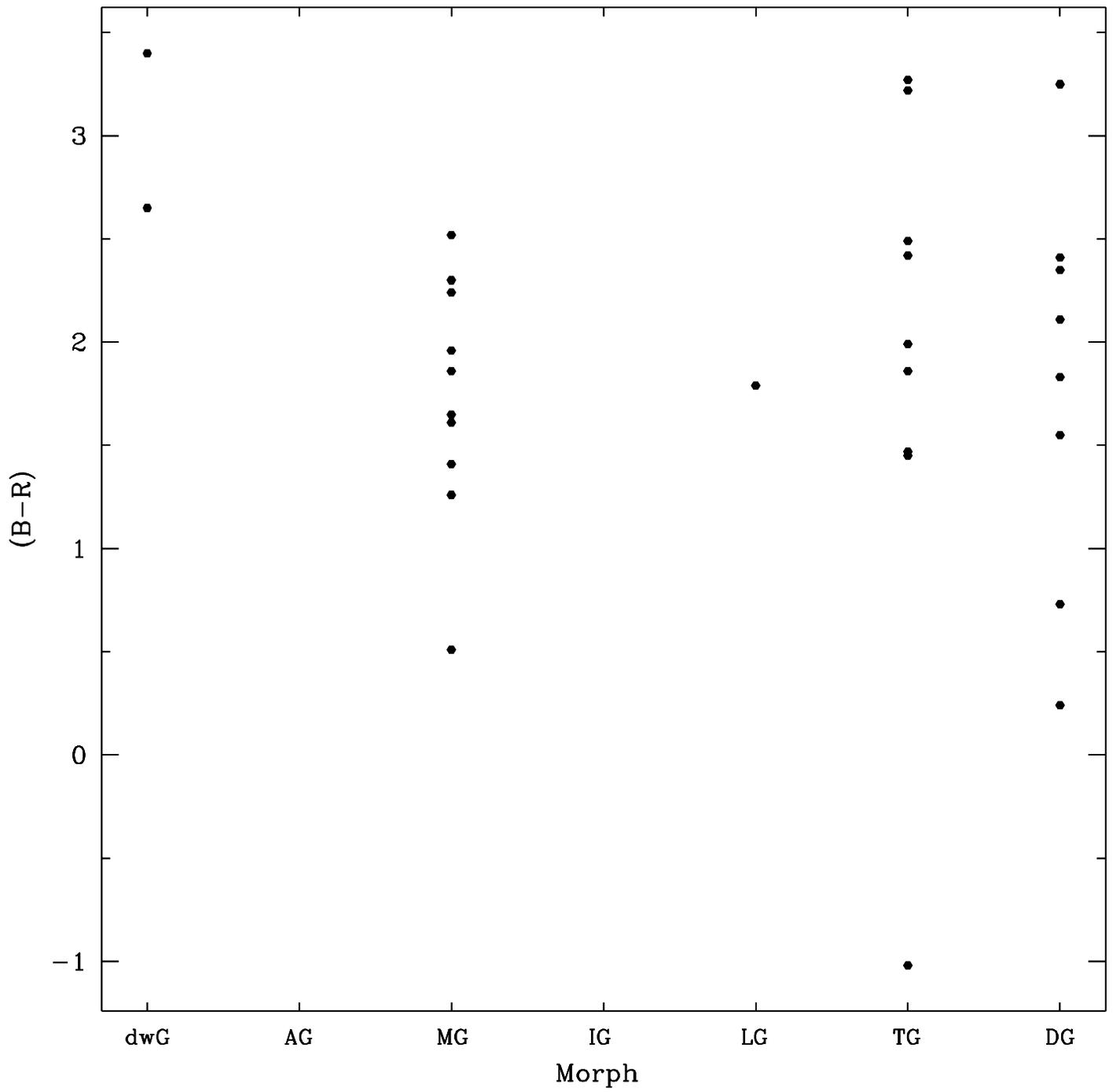
Fig 11 ---  The color distributions for structures in Abell 2199.  As
with the Coma cluster (Fig 10), we see a wide range in the colors for
the individual structures.

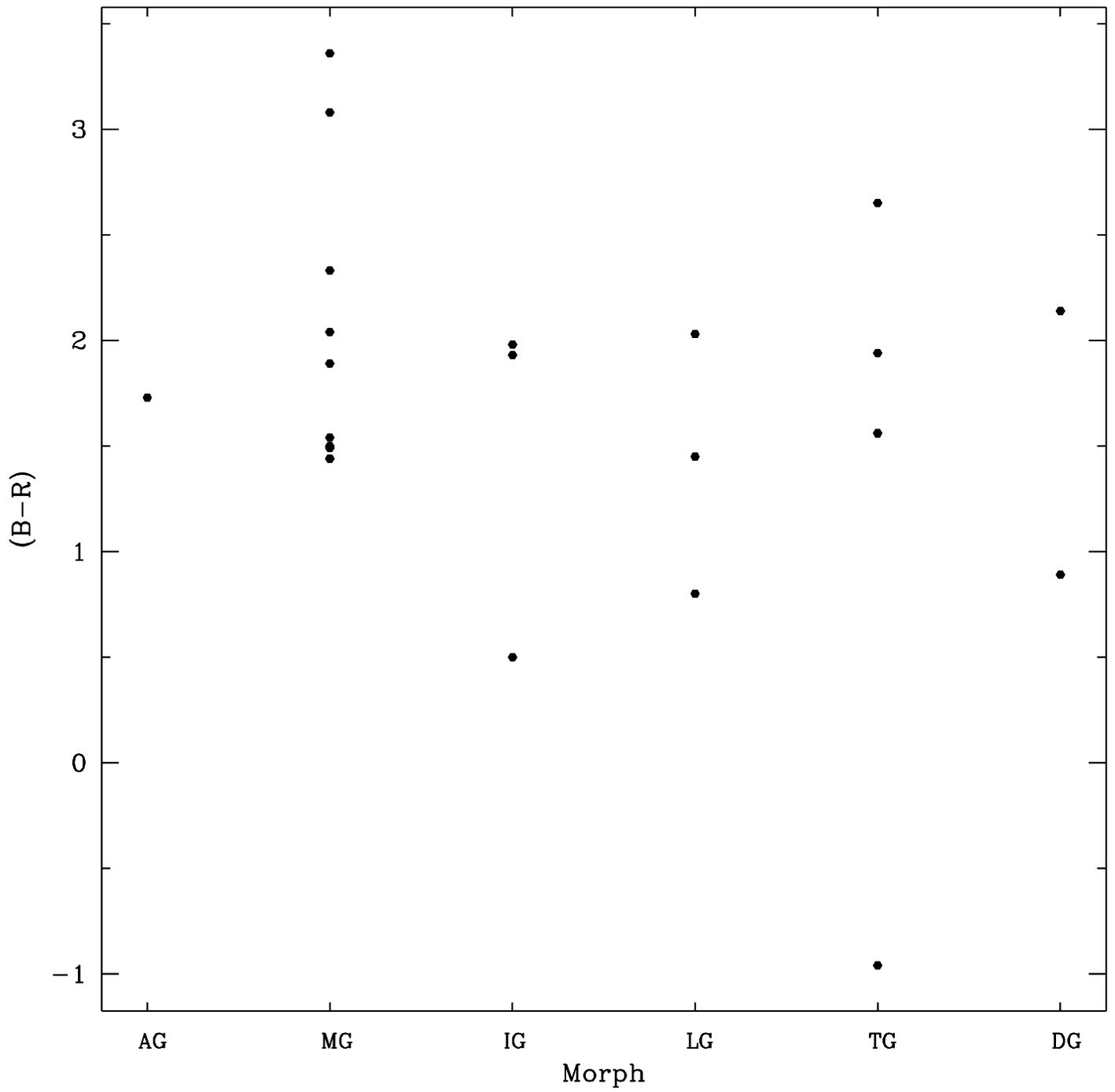
Fig 12 ---  The color distributions for the structures in AWM 5.  Like
Abell 2199 (Fig 11) and Coma (Fig 10), the distributions span a wide
range in color, indicating potentially different origins for some of
the structures we are seeing.

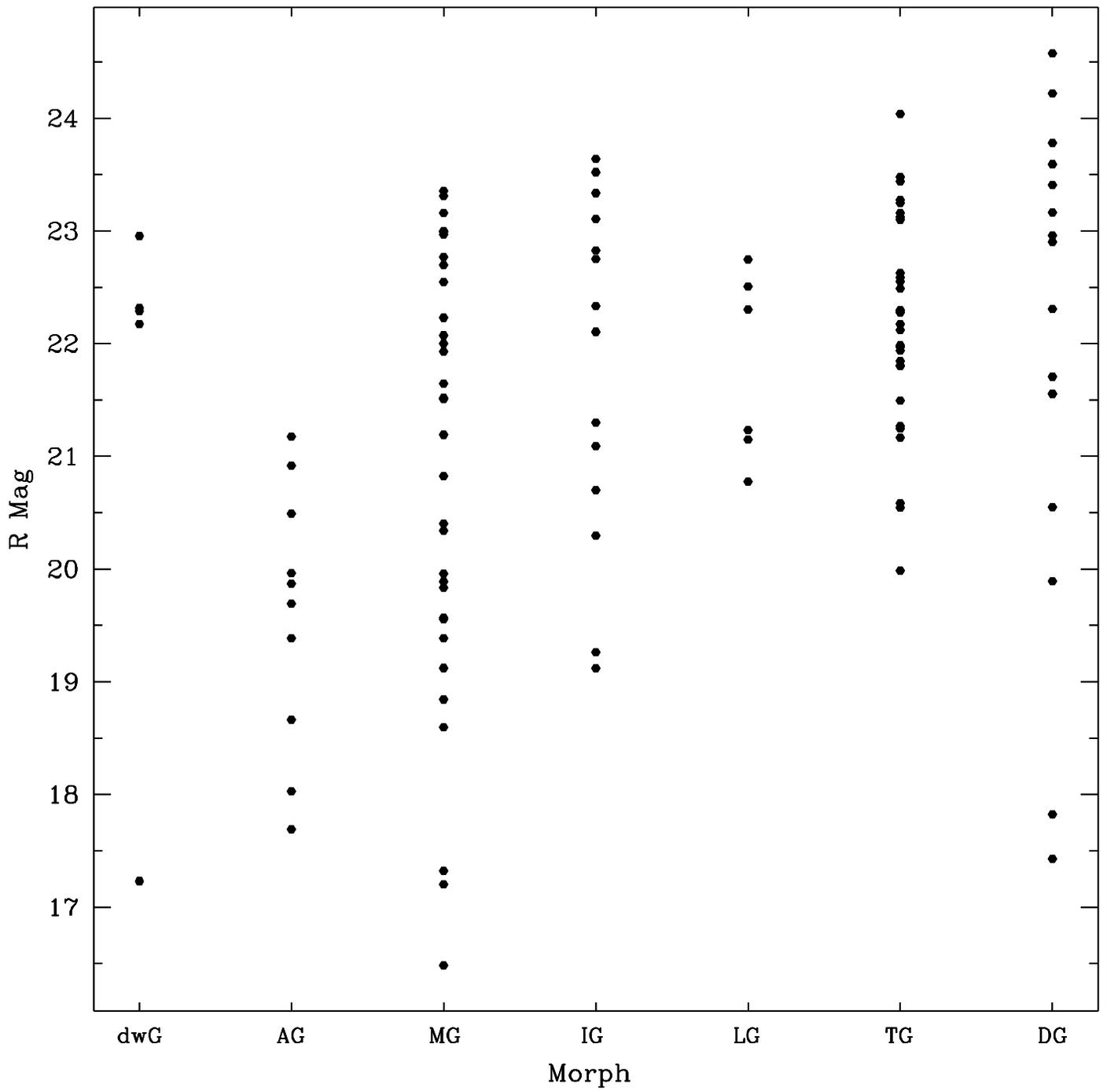
Fig 13 ---  The magnitude distributions for the structures found in the
Coma cluster.   Like the color distribution (Fig 10), there is a wide
range in magnitudes found for each morphology class.  Interestingly,
the line and tailed galaxies have the faintest magnitudes, and are also
the most likely to be of background origin based on blank field
comparisons (see text).  Galaxy aggregates have magnitudes which span
the smallest range in magnitudes with a range of only about 3.5
magnitudes.  The multiple galaxies and distorted galaxies span a wide
range from 16.5 to 24.5 central magnitudes.

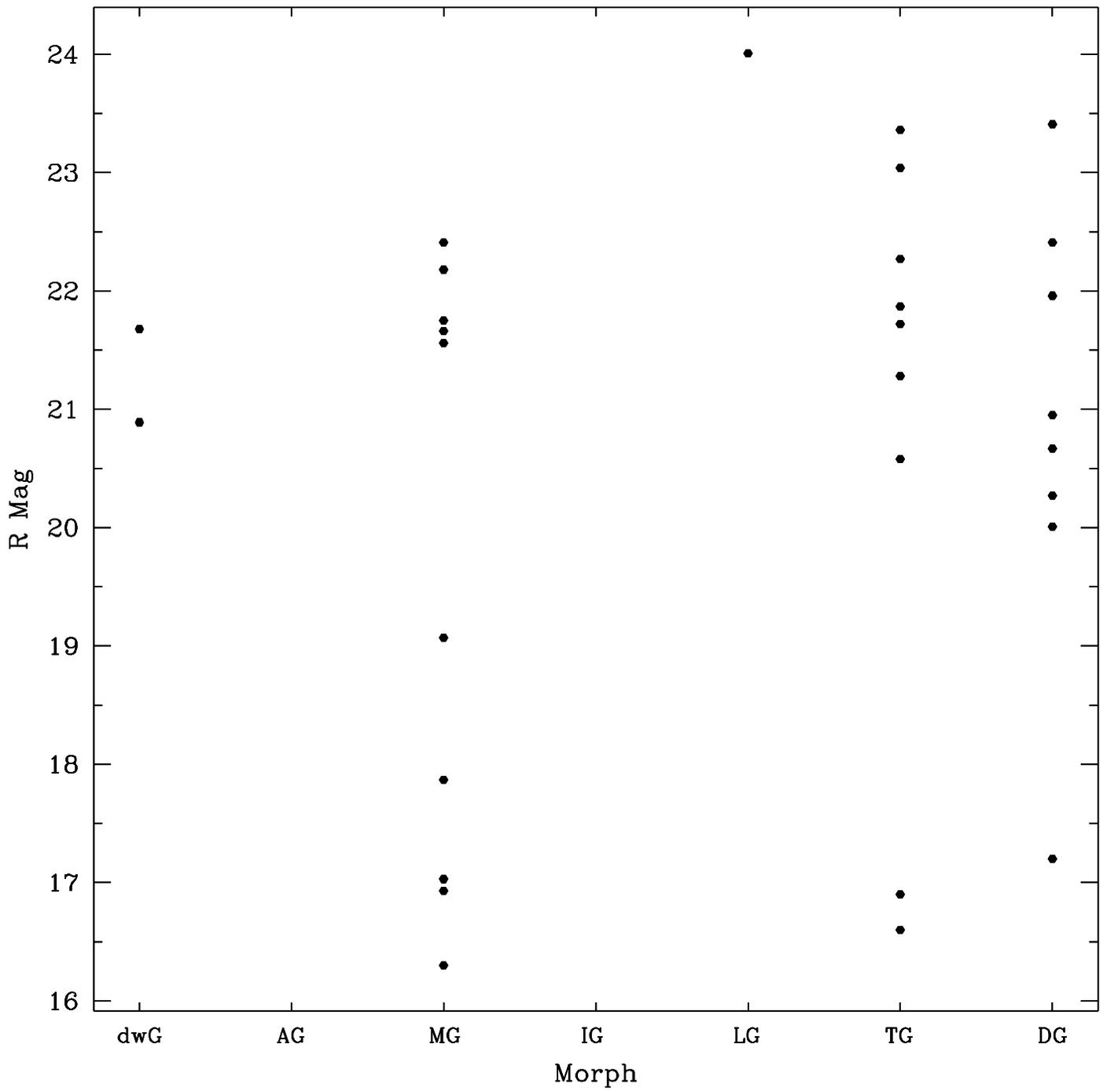
Fig 14 ---  The magnitude distributions for the structures found in
Abell 2199.  Like the distributions in Coma (Fig 13), the range for
each morphological class is large.  Some types, like aggregates and
interacting galaxies are not found in this cluster.

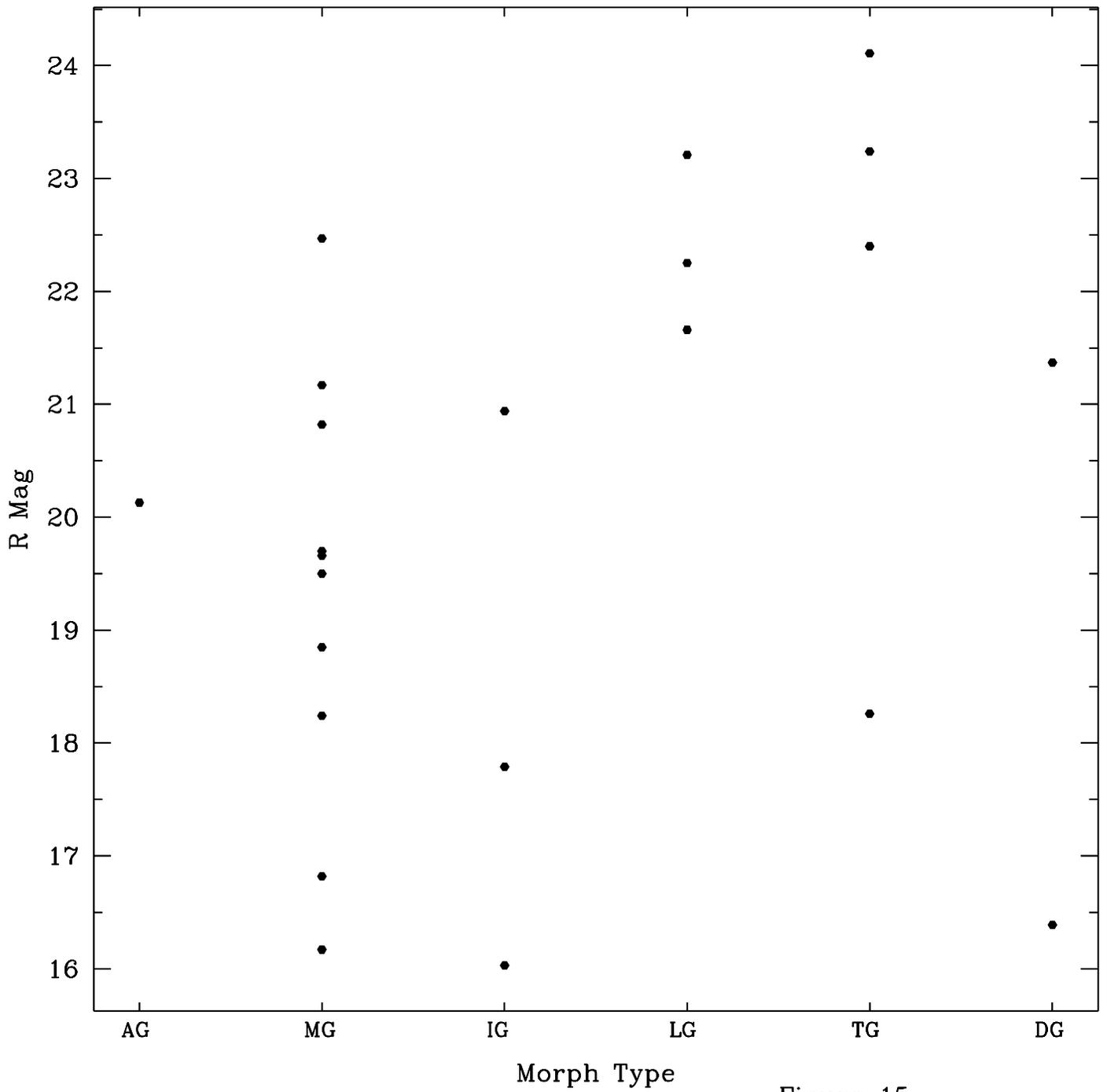
Fig 15 --- The magnitude distributions for the structures found in AWM
5.  As with Abell 2199 (Fig 14) and Coma (Fig 13), there is a wide
scatter in each morphological type.  With line galaxies and tailed
galaxies having the faintest magnitudes, a similar effect seen in the
Coma cluster.

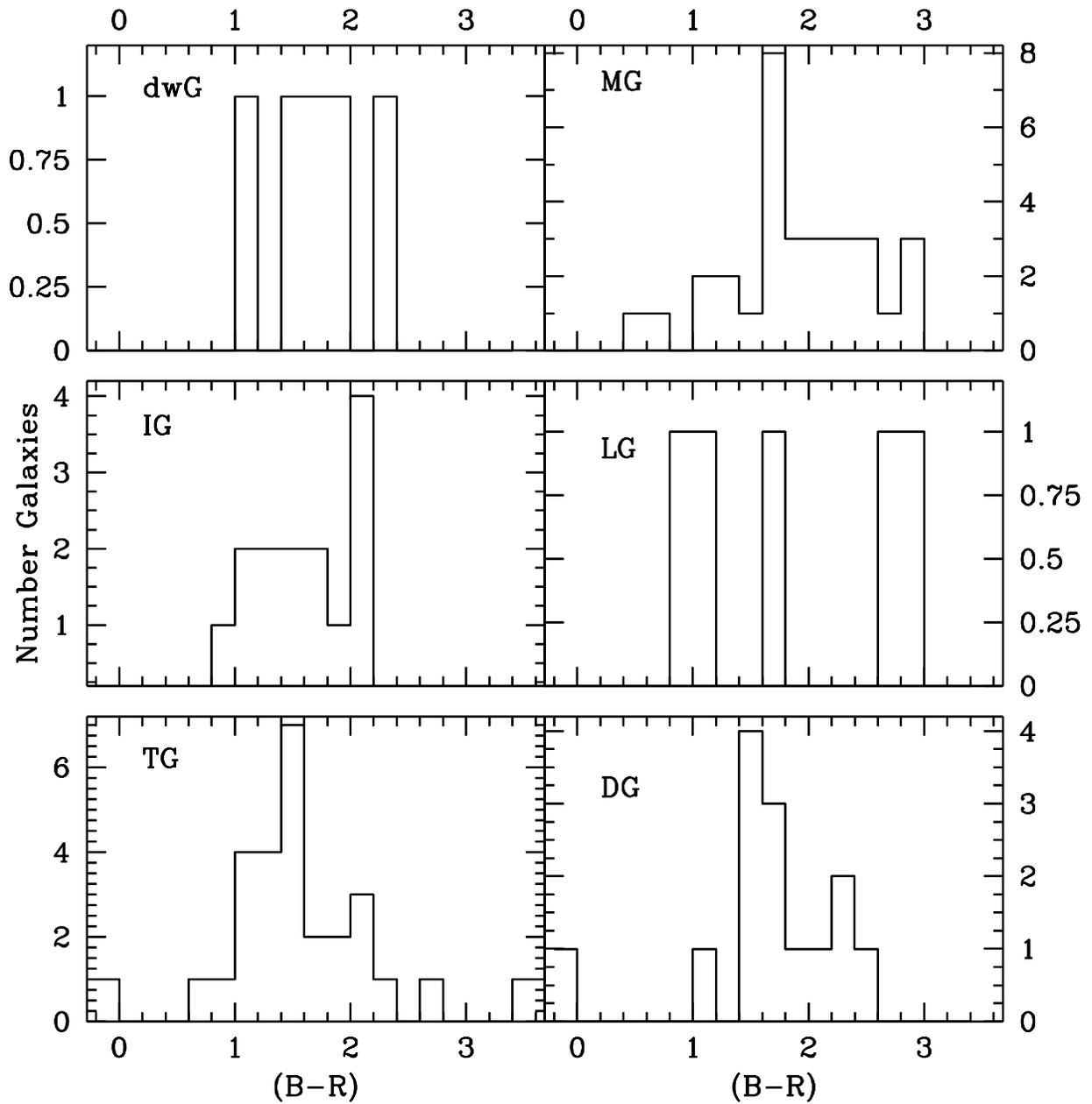
Fig 16 --- Color histograms for each morphological type as found in the
Coma cluster.  Most of the histograms have a Gaussian type shape, with
exception of the line galaxies.

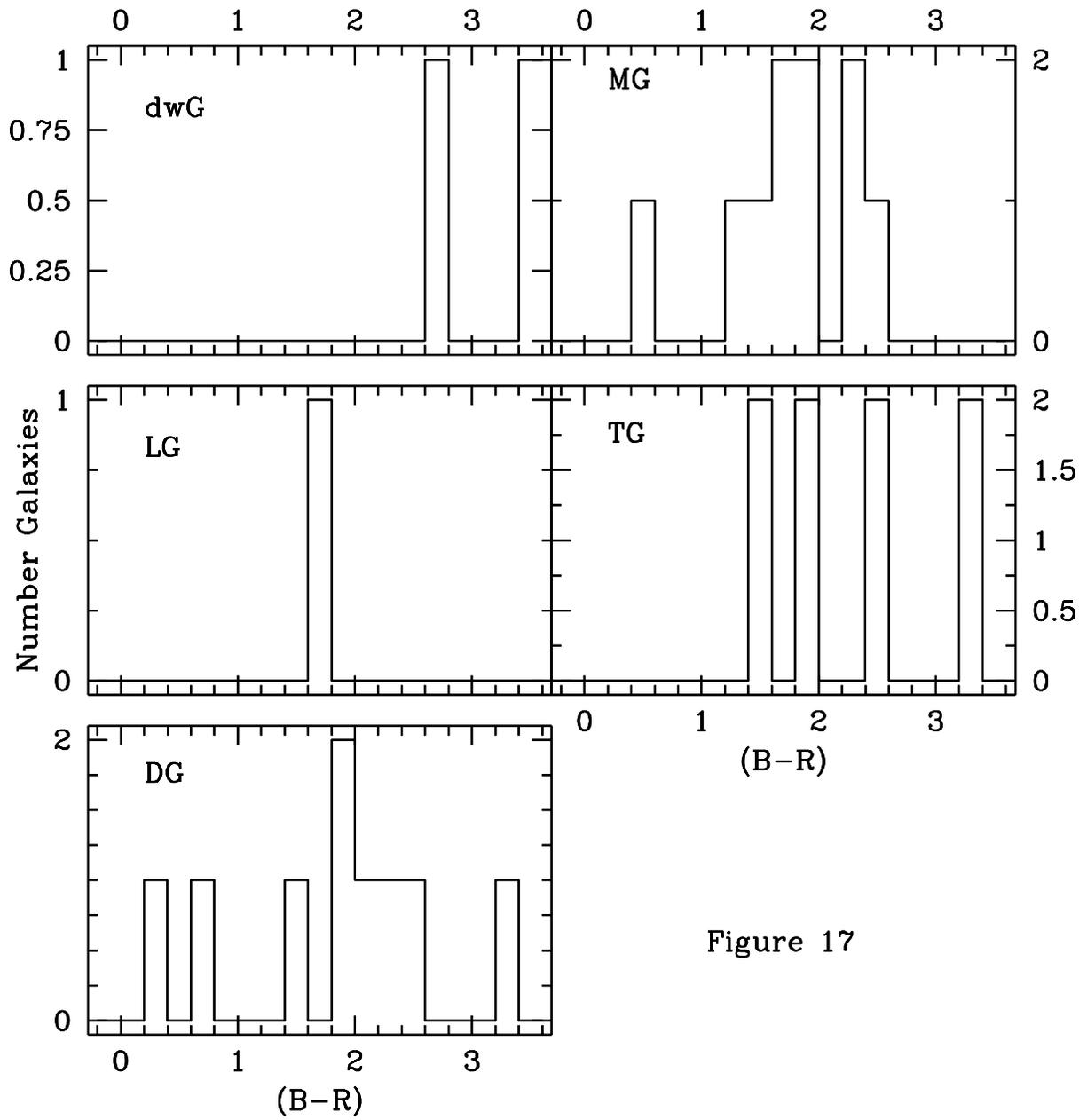
Fig 17 --- (B-R) color histogram for the structures as found in Abell
2199.

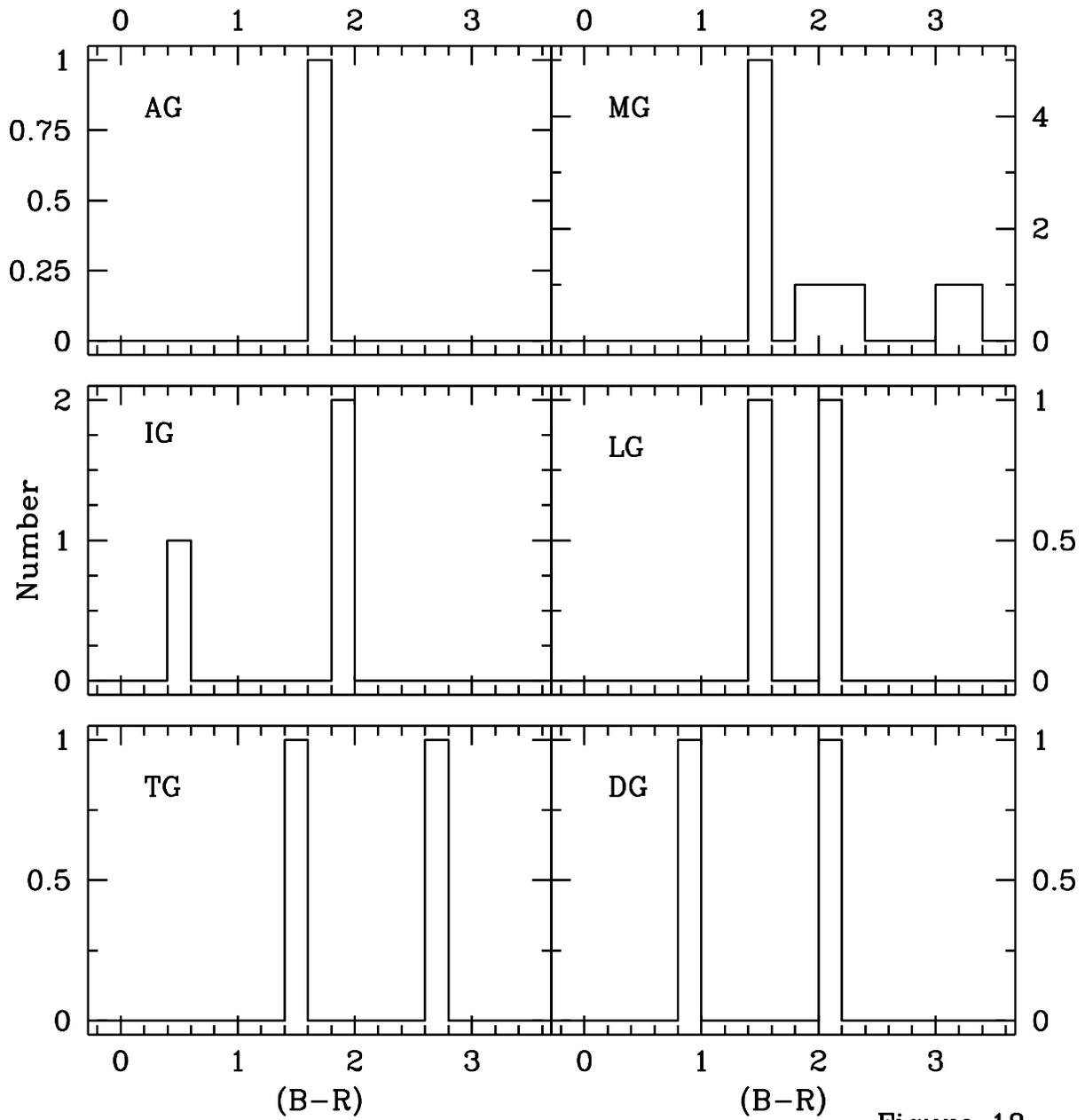
Fig 18 --- (B-R) color histogram for the structures found in AWM 5.

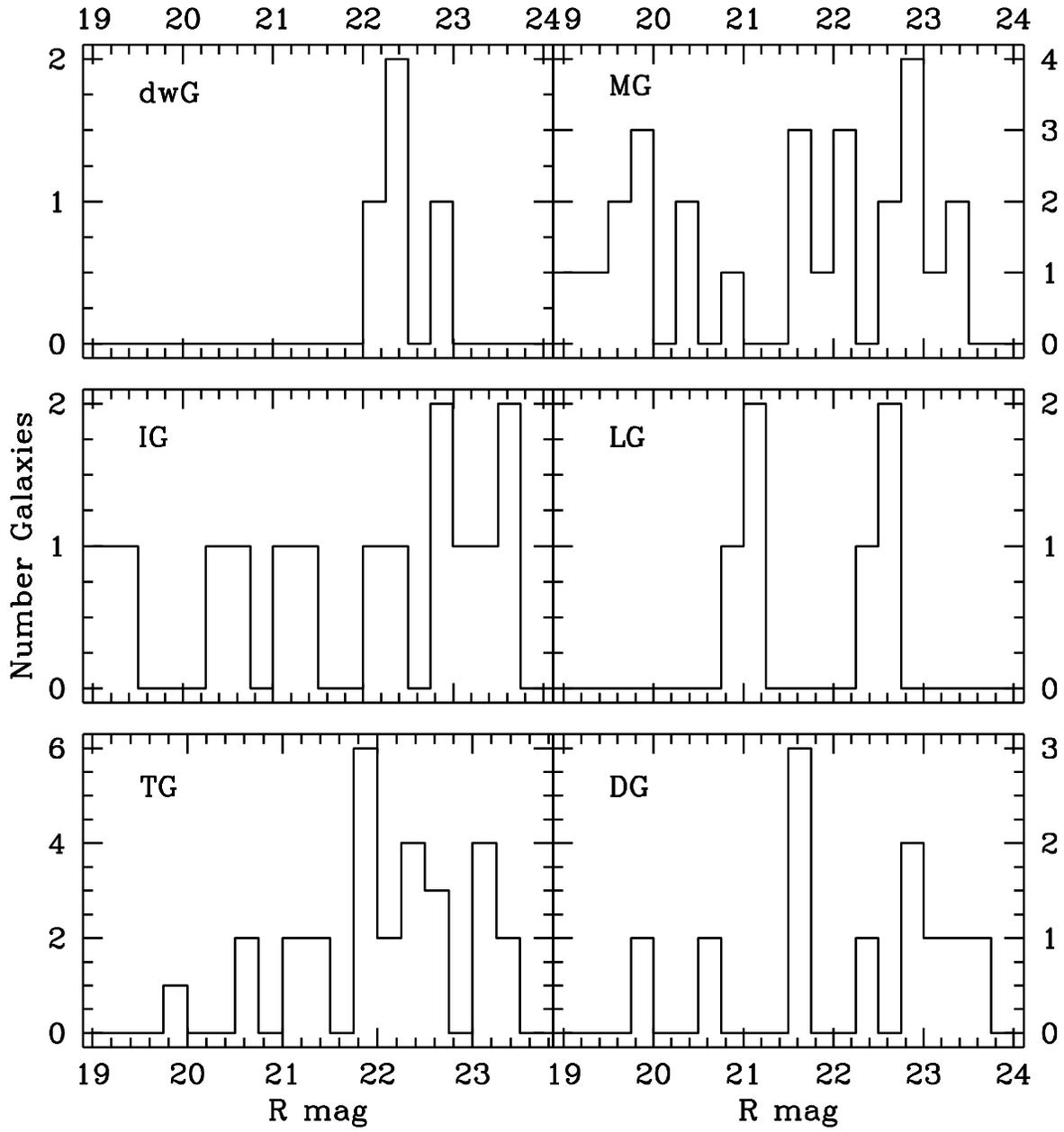
Fig 19 --- Magnitude histograms for the structures found in the Coma
cluster.  We do not see a Gaussian shape for the histograms, as we do
in the color histograms (Fig 16).  A wide spread in magnitudes for each
type is seen, with the exception of the dwarf groups.

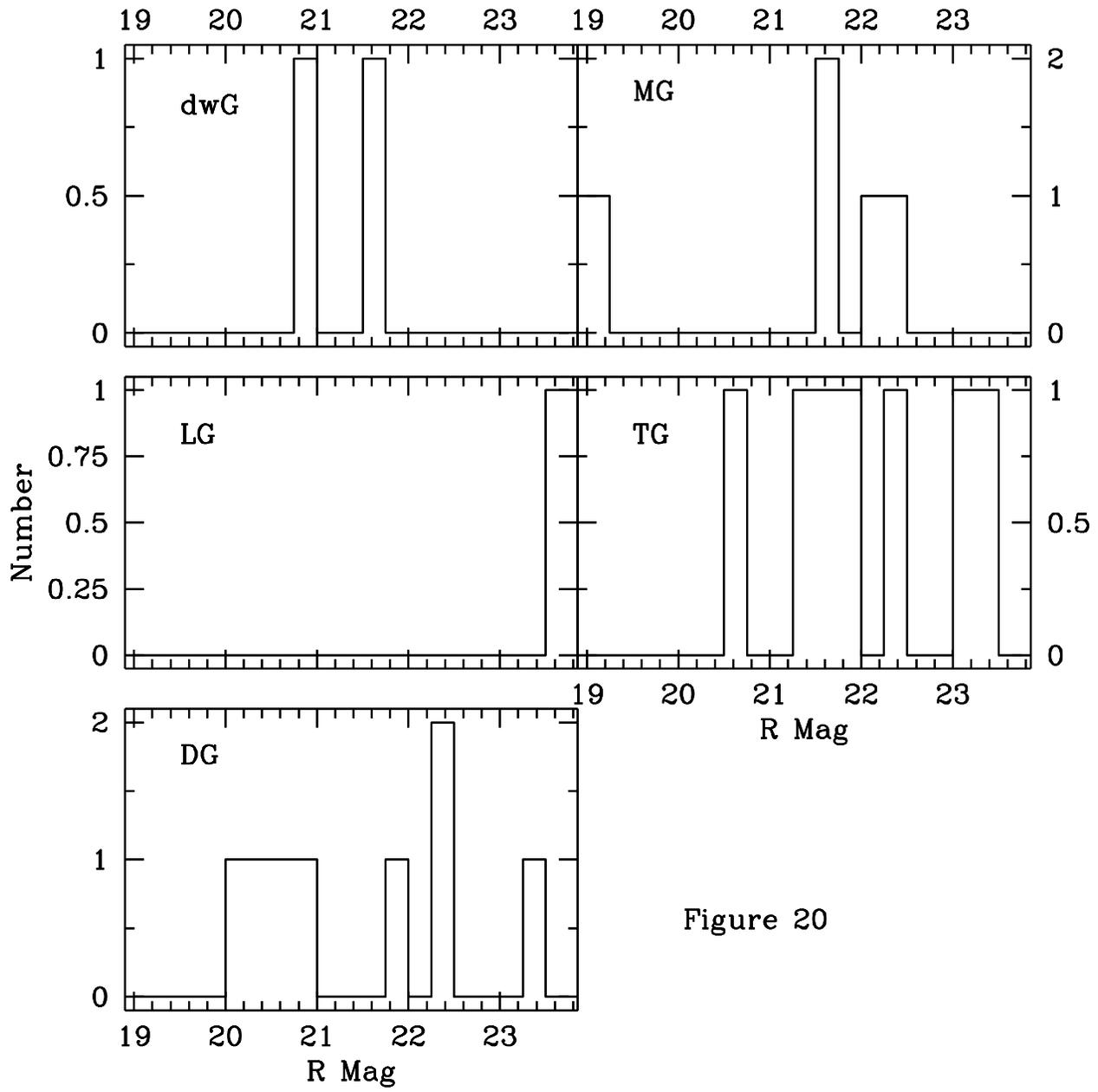
Fig 20 --- Magnitude histograms for the structures in Abell 2199,
showing a wide range in magnitudes, similar to Coma (Fig 19)

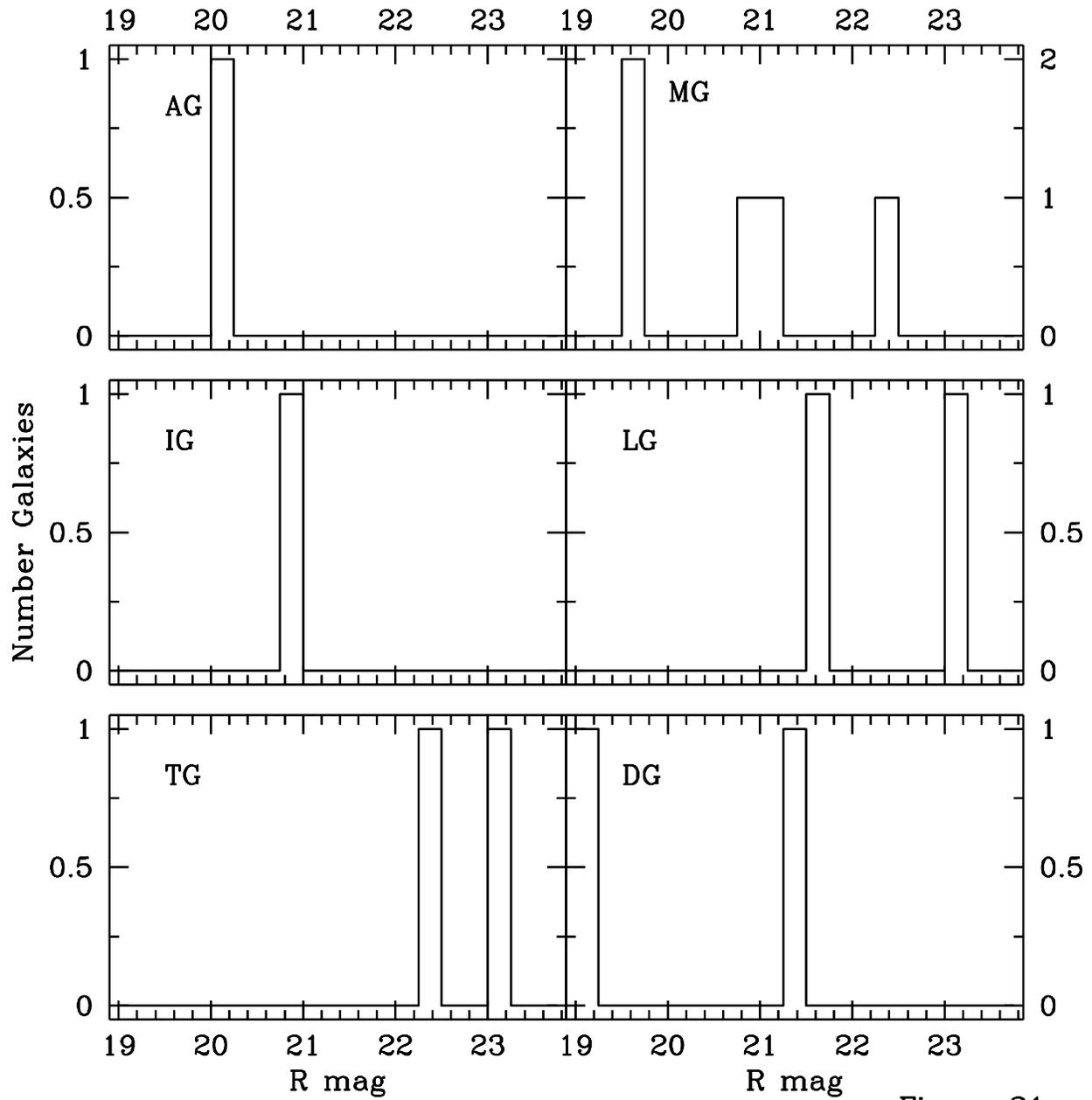
Fig 21 --- Magnitude histograms for structures found in AWM 5.

\begin{table}
\begin{center}

Table 1: Observing Log \\
\vspace{1ex}
\begin{tabular} {llllllll}

\hline
 ID & Cluster & R.A. (J2000) & Dec (J2000)  & filters &  Exp. Times & cz$_{cluster}$\\
\hline
\hline

Coma1 & Coma & 12 59 43.1 & 27 57 44.0 & R \& B & 600s, 900s & 6942 \kms\\
Coma2 & Coma & 12 59 08.0 & 27 57 49.0 & R \& B & 600s, 900s & 6942 \kms \\
Coma3 & Coma & 12 58 59.1 & 27 47 30.0 & R \& B & 600s, 900s & 6942 \kms\\
Coma4 & Coma & 12 59 56.0 & 27 51 60.0 & R \& B & 600s, 900s & 6942 \kms \\
Coma5 & Coma & 13 00 18.0 & 28 03 00.0 & R \& B & 600s, 900s & 6942 \kms\\
Coma6 & Coma & 12 58 50.0 & 28 08 00.0 & R \& B & 600s, 900s & 6942 \kms \\
NGC 4881 & Coma & 12 59 57.7 & 28 14 48.0 & R \& B & 600s, 900s & 6942 \kms\\
IC 4051 & Coma & 13 00 54.5 & 28 00 26.6 & R \& B & 600s, 900s & 6942 \kms\\
PerNR & Perseus & 03 19 53.7 & 41 33 46.0 & R \& B & 600s, 900s & 5486 \kms\\
PerS & Perseus & 03 19 47.7 & 41 28 46.0 & R \& B & 600s, 900s & 5486 \kms \\
NGC 6166 & Abell 2199 & 16 28 38.0 & 39 31 05.0 & R \& B & 600s, 900s & 9063 \kms\\
AWM 5 & AWM 5 & 16 55 58.0 & 27 55 47.0 & R \& B & 600s, 900s & 10346 \kms \\
AWM 3 & AWM 3 & 14 26 06.0 & 26 01 13.0 & R \& B & 600s, 900s & 4497 \kms \\

\hline

\end{tabular}
\end{center}
\end{table}

\begin{table}
\begin{center}

Table 2 : Structures in Coma Fields \\

\vspace{1ex}
\begin{tabular} {llllllllll}

\hline
 Field & Cluster & MG & IG & DG & dwG & TG & LG & AG \\
\hline
\hline

Coma 1 & Coma & 0 & 7 & 6 & 1 & 0 & 0 & 1 \\
Coma 2 & Coma & 1 & 0 & 2 & 3 & 2 & 0 & 0 \\
Coma 3 & Coma & 13 & 0 & 5 & 1 & 5 & 3 & 0 \\ 
Coma 4 & Coma & 2 & 3 & 1 & 0 & 3 & 1 & 2 \\
Coma 5 & Coma & 5 & 1 & 1 & 2 & 1 & 2 & 2  \\
Coma 6 & Coma & 3 & 0 & 1 & 2 & 3 & 1 & 1 \\
NGC 4881 & Coma & 1 & 0 & 2 & 0 & 5 & 0 & 2 \\
IC 4051 & Coma & 9 & 0 & 2 & 1 & 2 & 0 & 1 \\
\hline
Total & & 34 & 11 & 20 & 10 & 21 & 7 & 9 \\

\hline
\end{tabular}
\end{center}
\end{table}

\newpage

\begin{table}
\begin{center}

Table 3 : Number of Structures in Each Cluster (per field) \\
\vspace{1ex}
\begin{tabular} {lllllllll}

\hline
 Field & Cluster & MG & IG & DG & dwG & TG & LG & AG \\
\hline
\hline

Coma & Coma & 34 (4.3) & 11 (5.5) & 20 (2.5) & 10 (1.3) & 21 (2.7) & 7 (1) & 9 (1.2) \\
Perseus & Perseus & 8 (4) & 7 (3.5) & 32 (16) & 2 (1) & 15 (8) & 13 (7) & 0 (0) \\
NGC 6166 & A2199 & 10 & 1 & 8 & 2 & 9 & 1 & 0 \\
AWM 3 & AWM 3 & 1 & 0 & 5 & 0 & 4 & 0 & 0  \\
AWM 5 & AWM 5 & 11 & 4 & 2 & 0 & 3 & 3 &  1 \\

\hline
Total & & 64 & 23 & 67 & 14 & 52 & 24 & 10 \\
\hline
\end{tabular}
\end{center}
\end{table}

\begin{table}
\begin{center}

Table 4: Values of $\Gamma$ \\
\vspace{1ex}
\begin{tabular} {lcccc}

\hline
 Cluster & Disturbed Structures & Associations &$\Gamma$ \\
\hline
\hline

AWM 3 & 9 & 1 & 9.00 \\
Perseus & 67 & 10 & 6.70 \\
A2199 & 19 & 12 & 1.58 \\
Coma & 59 & 52 & 1.13 \\
AWM 5 & 12 & 12 & 1.00 \\

\hline
\end{tabular}
\end{center}
\end{table}

\begin{table}
\begin{center}

Table 5: Average Colors of Structures \\
\vspace{1ex}
\begin{tabular} {lcccc}

\hline
 Morph & Coma & Abell 2199  & AWM 5 & Total Ave\\
\hline
\hline

IG (Interacting) & 1.62$\pm$0.42 & --- & 1.47$\pm$0.84 & 1.59$\pm$0.49 \\
TG (Tailed) & 1.67$\pm$0.65 & 1.90$\pm$1.28 & 1.29$\pm$1.80 & 1.69$\pm$0.97 \\
DG (Distorted) & 1.68$\pm$0.76 & 1.80$\pm$0.90 & 1.51$\pm$0.79 & 1.71$\pm$0.73 \\
AG (Aggregates) & 1.72$\pm$0.51 & --- & 1.73$\pm$0.00  & 1.73$\pm$0.42 \\ 
dwG (Dwarf Groups) & 1.73$\pm$0.45 & 3.02$\pm$0.87 & --- & 2.10$\pm$0.76 \\
MG (Multiple) & 1.90$\pm$0.59 & 1.73$\pm$0.58 & 2.01$\pm$0.71 & 1.90$\pm$0.61 \\
LG (Line) & 2.18$\pm$1.40 & 1.79$\pm$0.00 & 1.42$\pm$0.62 & 1.91$\pm$1.00 \\
\hline

\end{tabular}
\end{center}
\end{table}

\begin{table}
\begin{center}

Table 6: Structures Found in the Coma Cluster \\
\vspace{1ex}
\begin{tabular} {rlllllll}

\hline
 Name & R.A.(J2000) & Dec (J2000) & R & B-R & Morphology & Notes \\
\hline
\hline

SA1656-001 &   12 59 43.5 &  27 53 35  & 23.16$\pm$0.30 & 2.24$\pm$0.40 & MG, IG    &    Sml, Wispy  \\ 

SA1656-002 &   12 59 43.0 &  27 54 25 & 23.16$\pm$0.20 & 1.58$\pm$0.24 &     TG & Possible Interaction\\
 
SA1656-003  &  12 59 41.0 &  27 56 35 & 22.31$\pm$0.10 & 1.58$\pm$0.12 &   DG & IG? \\

SA1656-004  &  12 59 44.1 &  27 58 17 & 22.96$\pm$0.19 & 1.663$\pm$0.22 &  dwG & 8 galaxies, all D \\

SA1656-005  &  12 59 48.0 &  27 59 40 & 22.99$\pm$0.26 & 1.986$\pm$0.28 &   MG, IG  &  one bright, one faint \\

SA1656-006  &  12 59 52.0 &  27 59 05 & 22.01$\pm$0.13 & 2.20$\pm$0.14 &  MG, IG   &  Same size and elong.\\

SA1656-007  &  12 59 49.0 &   27 57 20 & 21.71$\pm$0.13 & 2.53$\pm$0.14 &  DG &      Sprial w/material arnd. \\

SA1656-008  &  12 59 49.5 &  27 55 30  & 18.66$\pm$0.02 & 2.16$\pm$0.02 &  AG       &   RB 55 \\

SA1656-009  &  12 59 47.0 &  27 55 07  & 22.96$\pm$0.21 & 1.94$\pm$0.24 &   DG       & LSB \\
 
SA1656-010  &   12 59 49.0 &  27 54 42 & 21.30$\pm$0.08 & 2.19$\pm$0.90 &  IG or DG?  & \\

SA1656-011  &  12 59 57.0 & 27 55 55 & 19.12$\pm$0.03 & 1.72$\pm$0.04 &   IG  &   Sa and E, RB64 \\ 

SA1656-012  &  13 00 02.0 & 27 56 40 & 20.30$\pm$0.03 & 1.72$\pm$0.03 &  IG     &   RB 71 \\

SA1656-013  &  13 00 03.0 & 27 56 42 & 23.52$\pm$0.29 & 1.56$\pm$0.34 &  IG    & LSB, one D. \\

SA1656-014  & 12 59 07.0 & 27 57 00 & 22.18$\pm$0.20 & 2.36$\pm$0.20 &      dwG & \\

SA1656-015  &   12 59 10.0 &  27 58 30  & 22.30$\pm$0.10 & 1.47$\pm$0.11 &      TG & \\

SA1656-016 &   12 59 17.5 &  27 57 15  & 21.17$\pm$0.05 & 1.78$\pm$0.05 &  TG   & \\

SA1656-017  &  12 59 19.5 &  27 58 05 & 20.59$\pm$0.03 & 1.34$\pm$0.05 & TG   & Lrg. red tail\\ 

SA1656-018  &  12 59 24.0 &  27 53 55 & 20.55$\pm$0.03 & 1.43$\pm$0.03 &  DG    &     a bent S0 18 w/comp.  \\

SA1656-019  &  12 59 23.0 &  27 54 42 & 17.23$\pm$0.01 & 1.94$\pm$0.01 &  dwG     &   cntrd on  NGC4869 \\

SA1656-020 &   12 59 23.0 &  27 57 05 & 22.90$\pm$0.17 & 1.56$\pm$0.20 &  DG  &  possible dwf irregular \\

SA1656-021 &  12 59 26.0 &  27 58 25 & 18.60$\pm$0.02 & 1.74$\pm$0.02 &   MG   &     RB 141 \\

SA1656-022 &  12 59 21.0 & 27 57 25  & 22.28$\pm$0.10 & 1.55$\pm$0.12 &   TG & Slight Tail \\

SA1656-023 &  12 58 46.0 & 27 45 25 & 23.10$\pm$0.12 & 1.03$\pm$0.15 &   TG   & Blue Tail  \\

SA1656-024 &  12 58 47.0 &  27 44 25 & 20.70$\pm$0.04 & 2.19$\pm$0.04 &   IG  &   one b, one LSB \\
 
SA1656-025  &  12 58 51.0 &  27 44 15  & 24.22$\pm$0.32 & 1.07$\pm$0.42 &  DG & very faint\\           

\hline

\end{tabular}
\end{center}
\end{table}

\begin{table}
\begin{center}
Table 6 Cont.: Structures in Coma Cluster \\
\vspace{1ex}
\begin{tabular} {rlllllll}

\hline
 Name & R.A.(J2000) & Dec (J2000) & R & B-R & Morphology & Notes \\
\hline
\hline

SA1656-026  &   12 58 45.0 &  27 46 30 & 19.56$\pm$0.02 & 2.44$\pm$0.02 & MG   &       \\

SA1656-027  &  12 58 49.0 &  27 47 25  & 22.63$\pm$0.10 & 1.31$\pm$0.12 & MG, DG  &  3 gal, one wtail\\

SA1656-028  &   12 58 53.5 &  27 48 00 & 21.99$\pm$0.08 & 1.58$\pm$0.10 & TG & large gal.   \\

SA1656-029  &   12 58 55.0 &  27 50 11 & 21.56$\pm$0.05 & 1.50$\pm$0.06 &  DG  & debris arc \\

SA1656-030  &  12 58 50.0 &  27 49 10 & 19.83$\pm$0.02 & 1.81$\pm$0.02 &   MG  &    \\

SA1656-031  &  12 58 54.0 &  27 49 30 & 22.30$\pm$0.10 & 1.66$\pm$0.11 &   LG & \\

SA1656-032  &  12 58 55.0 &  27 47 50  & 19.39$\pm$0.01 & 1.63$\pm$0.02 &  MG &  RB 210 \\

SA1656-033 &    12 59 00.0 &  27 46 05  & 18.89$\pm$0.03 & 2.23$\pm$0.03 &  DG    &  spiral \\

SA1656-034 &   12 58 56.1 &  27 47 10   & 21.94$\pm$0.06 & 1.48$\pm$0.08 &  TG &  LG in R\\

SA1656-035 &    12 58 55.0 &  27 44 25   & 20.82$\pm$0.02 & 1.18$\pm$0.03 & MG & 3 gal.\\

SA1656-036 &    12 59 01.0 &  27 45 37   & 19.57$\pm$0.03 & 2.98$\pm$0.03 &  MG & 4 galaxies\\

SA1656-037 &   12 59 03.0 &  27 47 00   & 22.77$\pm$0.14 & 1.53$\pm$0.16 &  MG & 2 galaxies\\

SA1656-038 &    12 58 56.0 &  27 48 13  & 20.34$\pm$0.02 & 1.77$\pm$0.03 &  MG & One lrg. One sml.  \\

SA1656-039 &   12 59 04.0 &  27 49 35   & 21.27$\pm$0.06 & 2.00$\pm$0.06 & TG & object at tail's end\\

SA1656-040 &   12 59 02.5 &  27 49 30  & 22.75$\pm$0.08 & 0.82$\pm$0.11 &  LG & Near another gal.\\

SA1656-041  &  12 59 04.5 & 27 47 40 & 21.70$\pm$0.03 & 0.76$\pm$0.03 &  TG & \\

SA1656-042  &  12 59 05.0 &  27 47 55  &  19.89$\pm$0.03 & 2.56$\pm$0.03 &  MG & One LSB \\

SA1656-043  &  12 59 15.0 & 27 46 15  &  20.40$\pm$0.06 & 2.99$\pm$0.07 &  MG & \\

SA1656-044  &  12 59 06.5 & 27 44 45  & 22.33$\pm$0.15 & 2.03$\pm$0.16 & IG & DG in B \\

SA1656-045  &  12 59 09.0 & 27 44 30  & 21.93$\pm$0.07 & 1.67$\pm$0.08 &  MG & Very close\\

SA1656-046  &  12 59 12.0 & 27 47 05  & 22.28$\pm$0.12 & 1.90$\pm$0.13 &  TG  & LG in R\\

SA1656-047  &  12 59 12.0 & 27 49 55  & 22.55$\pm$0.13 & 1.78$\pm$0.14 &  MG & faint and close\\

SA1656-048 &   12 58 47.5 & 27 49 50 & 22.55$\pm$0.16 & 2.00$\pm$0.18 & TG & not vis. in B \\

SA1656-049 &  12 58 55.0 & 27 49 35 & 19.99$\pm$0.03 & 2.38$\pm$0.03 & TG & large S0 \\ 

SA1656-050 &  12 59 06.0 & 27 46 10 & 22.51$\pm$0.28 & 2.65$\pm$0.29 & LG & \\

\hline

\end{tabular}
\end{center}
\end{table}

\begin{table}
\begin{center}
Table 6 Cont.: Structures in Coma Cluster \\
\vspace{1ex}
\begin{tabular} {rlllllll}

\hline
 Name & R.A.(J2000) & Dec (J2000) & R & B-R & Morphology & Notes \\
\hline
\hline

SA1656-051 &  12 58 51.0 & 27 46 45 & 16.48$\pm$0.01 & 0.60$\pm$0.01 & MG & \\

SA1656-052 &  12 59 14.0 & 27 46 30 & 17.43$\pm$0.01 & 2.17$\pm$0.02 &DG & Large E with Shells \\

SA1656-053 &  12 59 10.0 & 27 47 10 & 21.37$\pm$0.03 & 1.18$\pm$0.03 & TG  &  \\

SA1656-054  &  12 59 51.0 &  27 50 00 & 19.39$\pm$0.05 & 1.88$\pm$0.07 & AG &  RB 60 \\

SA1656-055  &  12 59 46.0 &  27 51 30 & 17.69$\pm$0.03 & 2.64$\pm$0.04 &  AG? &  RB 49 \\

SA1656-056  &  12 59 47.5 &  27 49 37 & 22.49$\pm$0.11 & 1.33$\pm$0.14 &  TG & \\

SA1656-057  &  12 59 44.2 &  27 54 35 & 23.48$\pm$0.24 & 1.18$\pm$0.30 &  TG  & NGC 4876 \\

SA1656-058  &  12 59 50.0 &  27 54 30 & 21.09$\pm$0.06 & 1.50$\pm$0.08 &   MG, IG & one mem. a disk system \\

SA1656-059 &  12 59 51.0 &  27 52 17  & 22.99$\pm$0.28 & 1.89$\pm$0.30 &  MG  & f pair \\

SA1656-060 &  12 59 52.5 & 27 49 25  & 21.50$\pm$0.06 & 1.55$\pm$0.08 &  TG  &  pos. dwf companion \\

SA1656-061 &  12 59 58.5 &  27 51 57  & 23.11$\pm$0.19 & 1.17$\pm$0.24 & MG, IG  & \\

SA1656-062 &  13 00 07.0 &  27 52 49  & 22.11$\pm$0.14 & 1.946$\pm$0.15 & LG &  pos. sprl/edge on, bent arm \\

SA1656-063 &  13 00 07.7 &  27 51 30  & 23.16$\pm$0.27 & 1.70$\pm$0.31 &   DG &   \\

SA1656-064 &  13 00 09.0 &  27 51 57  & 19.26$\pm$0.05 & 0.93$\pm$0.08 &   IG &  diffuse debris nearby \\

SA1656-065 &  13 00 06.0 &  27 48 35  & 23.64$\pm$0.32 & 1.34$\pm$0.27 &   IG    &  CGCG 160-240 \\

SA1656-066 &  13 00 06.0 & 28 01 30  & 19.69$\pm$0.04 & 1.20$\pm$0.07 &  AG?   &   RB 74 \\

SA1656-067 &  13 00 04.0 & 28 03 00  & 23.34$\pm$0.17 & 1.04$\pm$0.21 &  dwG, IG \\

SA1656-068 &  13 00 04.0 & 28 03 40  & 23.36$\pm$0.30 & 1.70$\pm$0.34 &  MG & 3 gal\\

SA1656-069 &  13 00 12.5 & 28 04 30  & 18.03$\pm$0.03 & 1.73$\pm$0.04 &  AG  &  RB 87  \\

SA1656-070 &  13 00 13.0 &  28 05 20  & 21.55$\pm$0.17 & 2.23$\pm$0.20 &  DG  & slight \\

SA1656-071 &  13 00 13.0 &  28 03 15  & 21.23$\pm$0.06 & 1.11$\pm$0.10 & LG  &   very flat, and bent \\

SA1656-072 &  13 00 22.0 &  28 02 30  & 22.59$\pm$0.07 & 0.74$\pm$0.10 &  TG &  \\

SA1656-073 &   13 00 26.5 &  28 04 00 & 21.15$\pm$0.10 & 2.87$\pm$0.11 &  LG & possible tail in R \\

SA1656-074 &   13 00 29.0 &  28 05 25 & 23.31$\pm$0.26 & 1.73$\pm$0.30 & MG & \\

SA1656-075 &   13 00 27.0 &  28 04 30 & 22.75$\pm$0.12 & 1.25$\pm$0.14 & IG & two tails joining \\

\hline

\end{tabular}
\end{center}
\end{table}

\begin{table}
\begin{center}
Table 6 Cont.: Structures in Coma Cluster \\
\vspace{1ex}
\begin{tabular} {rlllllll}

\hline
 Name & R.A.(J2000) & Dec (J2000) & R & B-R & Morphology & Notes \\
\hline
\hline

SA1656-076 &   13 00 27.5 &  28 04 15 & 22.07$\pm$0.15 & 2.13$\pm$0.17 & MG & 4 gal\\

SA1656-077 &   13 00 26.0 &  28 03 55 & 22.29$\pm$0.08 & 1.17$\pm$0.11 & dwG & main gal. w/tail\\

SA1656-078  &  13 00 27.0 &  28 01 20 & 23.12$\pm$0.34 & 2.06$\pm$0.40 & TG, MG & possible arc \\

SA1656-079  &  12 58 38.0 &  28 08 05 & 17.32$\pm$0.01 & 2.27$\pm$0.01 & dwG, MG & \\

SA1656-080  &  12 58 39.5 &  28 09 55 & 21.80$\pm$0.18 & 2.75$\pm$0.19 & TG & \\

SA1656-081  &  12 58 42.0 &  28 11 05 & 21.17$\pm$0.10 & 1.97$\pm$0.12 &  AG  &  LSB material, debris \\

SA1656-082  &  12 58 46.0 &  28 06 25  & 22.23$\pm$0.19 & 2.43$\pm$0.20 & MG & DG in B\\

SA1656-083  &  12 58 48.0 & 28 05 30  & 20.55$\pm$0.10 & 3.51$\pm$0.11 & TG & knots in R\\  

SA1656-084  &  12 58 59.0 &  28 07 05 & 23.59$\pm$0.40 & 1.73$\pm$0.44 &  DG & very distored, two red knots in R nearby  \\

SA1656-085  &  12 59 02.0 &  28 07 00 & 19.87$\pm$0.09 & 1.60$\pm$0.12 & AG, MG  &  N4858 \& N4860  \\

SA1656-086  &  12 58 59.0 &  28 09 45  & 22.83$\pm$0.24 & 2.11$\pm$0.26 & LG & Tailed in R\\

SA1656-087  &  12 59 02.0 &  28 09 55  & 19.96$\pm$0.05 & 2.91$\pm$0.05 &  MG  & triple, l cntr \\

SA1656-088  &  12 58 46.0 &  28 10 10  & 21.84$\pm$0.38 & 3.66$\pm$0.40 & TG  & 2 red knots\\

SA1656-089  &  12 59 52.3 &  28 12 20  & 21.80$\pm$0.05 & 0.87$\pm$0.06 &  TG  & \\

SA1656-090 &   12 59 47.0 &  28 16 25  & 19.96$\pm$0.04 & 1.89$\pm$0.04 &  AG? &   mat. arnd gal. w/bridge \\

SA1656-091 &   12 59 55.5 &  28 16 50  & 21.25$\pm$0.08 & 1.95$\pm$0.09 & TG & \\

SA1656-092 &   12 59 51.0 &  28 17 45  & 20.92$\pm$0.03 & 0.78$\pm$0.04 &  AG & \\  

SA1656-093 &   13 00 03.0 &  28 13 05  & 17.20$\pm$0.01 & 2.75$\pm$0.02 &  MG & 3 members\\

SA1656-094 &   13 00 02.5 &  28 12 55  & 23.27$\pm$0.21 & 1.14$\pm$0.26 &  TG & not obvious in R\\

SA1656-095 &   13 00 06.0 &  28 15 05  & 24.58$\pm$0.36 & -0.17$\pm$0.95 &  DG  & lrg & very Distorted \\

SA1656-096 &   13 00 08.0 &  28 13 35  & 23.41$\pm$0.43 & 1.76$\pm$0.48 &  DG  & \\

SA1656-097 &   13 00 12.0 &  28 12 45  & 22.17$\pm$0.08 & 1.07$\pm$0.09 & TG & \\

SA1656-098 &   13 00 12.0 &  28 11 30  & 23.25$\pm$0.30 & 1.48$\pm$0.32 &  TG & f \\ 

SA1656-099  &  13 00 01.0  & 28 15 05   & 22.12$\pm$0.04 & -0.01$\pm$0.08 &  TG & Blue Tail\\

SA1656-100  &  12 59 59.0  & 28 14 05   & 20.78$\pm$0.32 & 3.96$\pm$0.34 &  LG & Large \\

\hline

\end{tabular}
\end{center}
\end{table}

\begin{table}
\begin{center}
Table 6 Cont.: Structures in Coma Cluster \\
\vspace{1ex}
\begin{tabular} {rlllllll}

\hline
 Name & R.A.(J2000) & Dec (J2000) & R & B-R & Morphology & Notes \\
\hline
\hline

SA1656-101  &  13 00 04.0  & 28 15 35   & 21.97$\pm$0.15 & 1.70$\pm$0.20 &  TG & Curved Tail \\

SA1656-102 &   13 00 43.0 &  27 58 05 & 17.82$\pm$0.05 & 2.31$\pm$0.07 &  MG &  RB113 \& IC4042 \\

SA1656-103 &   13 00 43.0 & 27 58 15  & 17.82$\pm$0.05 & 2.31$\pm$0.07 & DG  & SA0/a w/shells \\

SA1656-104 &   13 00 46.5 & 27 59 50  & 21.65$\pm$0.17 & 2.04$\pm$0.18 &  MG   & s and l, blends in B \\

SA1656-105 &   13 00 43.0 & 28 03 25  & 22.32$\pm$0.17 & 1.53$\pm$0.20 &  dwG & line of galaxies\\

SA1656-106 &   13 00 51.0 &  28 02 45 & 18.84$\pm$0.09 & 0.76$\pm$0.11 &  MG &  N4908 \\

SA1656-107 &   13 00 52.5 &  27 59 57 & 24.04$\pm$0.56 & 1.22$\pm$0.64 &  TG & debris srndg\\

SA1656-108  &  13 00 53.0 &  27 58 10 & 21.520$\pm$0.10 & 1.33$\pm$0.13 &  MG & \\

SA1656-109 &   13 00 55.5 & 27 58 05  & 22.97$\pm$0.32 & 1.69$\pm$0.35 &  MG  &   s \\
 
SA1656-110  &  13 01 01.0 & 27 59 55  & 22.70$\pm$0.43 & 2.37$\pm$0.46 &  MG & 4 gal. in arc\\

SA1656-111  &  13 01 02.0 & 28 00 40  & 23.44$\pm$1.20 & 2.75$\pm$1.30 & TG & \\

SA1656-112  &  13 01 05.0 & 28 03 50  & 20.49$\pm$0.06 & 1.43$\pm$0.08 &  AG &  \\

SA1656-113 &   13 01 05.0 & 28 03 35 & 23.78$\pm$0.42 & 1.31$\pm$0.50 &  DG  & bits of material \\

SA1656-114 &   13 01 05.0 & 28 01 30 & 19.12$\pm$0.02 & 1.34$\pm$0.02 & MG & \\

SA1656-115 &   13 01 03.5 & 27 59 55 & 21.51$\pm$0.06 & 1.17$\pm$0.08 & MG  & \\

SA1656-116 &   13 01 05.0 & 27 58 20  & 21.19$\pm$0.08 &  1.76$\pm$0.09 &  MG   & s, l \\
\hline
\end{tabular}
\end{center}
\end{table}

\begin{table}
\begin{center}

Table 7: Structures Found in Perseus Cluster (Abell 0426) \\
\vspace{1ex}
\begin{tabular} {rlllll}

\hline
 Name & R.A.(J2000) & Dec (J2000) & Morphology & Notes \\
\hline
\hline

SA0426-01  &  03 19 48.0 & 41 30 45  &     DG	&    Perseus A \\

SA0426-02  &  03 19 40.0 & 41 31 00  &     DG  &    butterfly Shaped\\

SA0426-03  &  03 19 45.3 & 41 33 36  &     TG  \\

SA0426-04  &  03 19 40.2 & 41 33 55  &     TG  \\

SA0426-05  &  03 19 42.0 & 41 34 30  &     MG  &        Sprls\\

SA0426-06  &  03 19 43.5 & 41 34 35  &     TG\\

SA0426-07  &  03 19 39.5 & 41 35 50  &     DG  &        twisted Gal\\

SA0426-08  &  03 19 48.0 & 41 35 05  &     TG\\

SA0426-09  &  03 19 44.0 & 41 35 30  &     TG     &     s\\

SA0426-10  &  03 19 57.6 & 41 36 00  &     DG or MG &im    peanut Shp\\

SA0426-11  &  03 19 51.0 & 41 34 18  &     DG     &     flat btm\\

SA0426-12  &  03 19 56.8 & 41 32 50  &     DG     &     a LSB Sprl\\

SA0426-13  &  03 20 06.0 & 41 31 10  &     MG, LG\\

SA0426-14  &  03 20 12.8 & 41 30 55  &     DG     &     LSB\\

SA0426-15  &  03 20 03.5 & 41 30 35  &     DG     &     diffuse\\

SA0426-16  &  03 20 08.5 & 41 31 10  &     DG     &     Peanut Shp\\

SA0426-17  &  03 20 10.0 & 41 30 25  &     IG     &     sprl \\

SA0426-18  &  03 20 05.0 & 41 32 03  &     DG, TG &    very disturbed\\

SA0426-19  &  03 20 03.0 & 41 31 45  &     DG, TG &     extented halo mat.\\

SA0426-20  &  03 20 10.5 & 41 32 00  &     TG\\

SA0426-21  &  03 20 05.0 & 41 31 07  &     MG\\

SA0426-22  &  03 20 07.0 & 41 35 00  &     TG\\

SA0426-23  &  03 20 00.5 & 41 35 10  &     DG    &      batman Galaxy\\

SA0426-24  &  03 20 11.5 & 41 35 15  &     DG    &      looks blown up\\

SA0426-25  &  03 20 02.0 & 41 31 10  &     IG    &      faint bridge\\

\hline

\end{tabular}
\end{center}
\end{table}

\begin{table}
\begin{center}
Table 7 Cont.: Structures in Perseus Cluster \\
\vspace{1ex}
\begin{tabular} {rlllll}

\hline
 Name & R.A.(J2000) & Dec (J2000) & Morphology & Notes \\
\hline
\hline

SA0426-26  &  03 19 33.0 & 41 26 15   &    LG\\

SA0426-27  &  03 19 36.5 & 41 26 25   &    MG, IG\\

SA0426-28  &  03 19 41.0 & 41 26 00   &   MG or DG \\

SA0426-29  &  03 19 38.0 & 41 25 57   &    LG   &       near an E \\

SA0426-30  &  03 19 43.5 & 41 28 55   &    LG  &        very flat\\

SA0426-31  &  03 19 41.0 & 41 29 25   &    LG\\

SA0426-32  &  03 19 33.0 & 41 28 35   &    TG\\

SA0426-33  &  03 19 37.5 & 41 29 25   &    DG      &    LSB \\

SA0426-34  &  03 19 34.0 & 41 29 23   &    MG or DG\\

SA0426-35  &  03 19 35.0 & 41 28 13   &    IG   &       smudgy \\

SA0426-36  &  03 19 33.0 & 41 30 55   &    DG \\

SA0426-37  &  03 19 39.5 & 41 30 30   &    DG\\

SA0426-38  &  03 19 38.0 & 41 29 57   &    TG, LG\\

SA0426-39  &  03 19 32.0 & 41 30 20   &    dwG\\

SA0426-40  &  03 19 36.0 & 41 31 50   &    DG    &      LSB \\

SA0426-41  &  03 19 51.5 & 41 31 25   &    TG   &       thick \\

SA0426-42  &  03 19 46.0 & 41 30 05   &    LG \\

SA0426-43  &  03 19 50.0 & 41 29 52   &    DG  &        smuge \\

SA0426-44  &  03 19 46.5 & 41 29 05   &    LG\\

SA0426-45  &  03 19 46.0 & 41 28 55   &    IG, dwG  &   bits of material \\

SAO426-46  &  03 19 47.0 & 41 28 50   &    DG    &      LSB \\

SA0426-47  &  03 19 45.0 & 41 28 07   &    DG   &       smuge \\

SA0426-48  &  03 19 48.5 & 41 27 30   &    MG \\

SA0426-49  &  03 19 50.0 & 41 26 50   &    LG    &      LSB  \\

SA0426-50  &  03 19 51.0 & 41 26 10   &    IG    &      box Like \\

\hline

\end{tabular}
\end{center}
\end{table}

\begin{table}
\begin{center}
Table 7 Cont.: Structures in Perseus Cluster \\
\vspace{1ex}
\begin{tabular} {rlllll}

\hline
 Name & R.A.(J2000) & Dec (J2000) & Morphology & Notes \\
\hline
\hline

SA0426-51 &   03 19 46.0 & 41 26 12   &    MG \\

SA0426-52 &   03 19 49.0 & 41 25 35   &    DG     &     smudge \\

SA0426-53 &   03 19 53.5 & 41 25 45   &    DG    &      boxy, smudge, LSB \\

SA0426-54 &   03 19 58.0 & 41 25 45   &    DG   &       ring Gal. \\

SA0426-55 &   03 20 04.5 & 41 25 45   &    DG  &        LSB\\

SA0426-56 &   03 20 04.0 & 41 26 55   &    TG\\

SA0426-57 &   03 20 03.5 & 41 27 55   &    LG\\

SA0426-58 &   03 20 05.0 & 41 27 35   &    TG &         bent\\

SA0426-59 &   03 20 06.0 & 41 29 00   &    LG\\

SA0426-60 &   03 20 57.0 & 41 29 25   &    DG    &      smudge\\

SA0426-61 &   03 20 00.5 & 41 30 00   &    LG   &       flattened\\

SA0426-62 &   03 20 05.0 & 41 31 07   &    MG \\

SA0426-63 &   03 20 00.3 & 41 30 45   &    TG \\

SA0426-64 &   03 20 03.0 & 41 31 05   &    LG   &       bright core/halo\\

SA0426-65 &   03 20 03.5 & 41 30 30   &    DG \\

SA0426-66 &   03 20 02.0 & 41 30 03   &    DG   &       diffuse\\

SA0426-67 &   03 20 04.0 & 41 31 30   &    DG \\

SA0426-68 &   03 20 05.5 & 41 30 45   &    LG \\

SA0426-69 &   03 20 03.5 & 41 03 20   &    DG      &    boxy\\

\hline
\end{tabular}
\end{center}
\end{table}

\begin{table}
\begin{center}

Table 8: Structures Found in Abell 2199 \\
\vspace{1ex}
\begin{tabular} {rlllllll}

\hline
 Name & R.A.(J2000) & Dec (J2000) & R & (B-R) & Morphology & Notes \\
\hline
\hline

SA2199-01  &   16 28 26.0 &  39 28 30 & 16.90$\pm$0.01 & 1.45$\pm$0.01 & TG & slight \\

SA2199-02  &   16 28 23.0 &  39 29 40 & 22.18$\pm$0.10 & 1.65$\pm$0.11 & MG \\

SA2199-03  &   16 28 30.0 & 39 30 30  & 21.66$\pm$0.11 & 2.30$\pm$0.12 & MG \\

SA2199-04  &   16 28 22.0 & 39 31 10  & 20.89$\pm$0.07 & 2.65$\pm$0.08 & dwG \\

SA2199-05  &   16 28 26.0 & 39 29 40  & 21.68$\pm$0.27 & 3.40$\pm$0.28 & dwG \\

SA2199-06  &   16 28 24.5 & 39 33 30  & 16.30$\pm$0.01 & 1.26$\pm$0.01 & MG \\

SA2199-07  &   16 28 27.0 & 39 32 45  & 16.93$\pm$0.01 & 1.41$\pm$0.02 & MG & s comp \\

SA2199-08  &   16 28 26.5 & 39 33 15  & 21.96$\pm$0.16 & 2.35$\pm$0.17 & DG \\

SA2199-09  &   16 28 30.0 & 39 34 00  & 24.01$\pm$0.54 & 1.79$\pm$0.60 & LG \\

SA2199-10  &   16 28 38.0 & 39 33 00  & 20.95$\pm$0.18 & 0.73$\pm$0.41 & DG & NGC 6166 \\

SA2199-11  &   16 28 31.0 & 39 31 10  & 17.03$\pm$0.02 & 1.96$\pm$0.03 & MG \\

SA2199-12  &   16 28 40.5 & 39 29 55  & 21.56$\pm$0.11 & 2.52$\pm$0.12 & MG & ringed \\

SA2199-13  &   16 28 36.0 & 39 30 25  & 23.41$\pm$0.24 & 1.55$\pm$0.28 & DG \\

SA2199-14  &   16 28 37.0 & 39 29 10  & 21.72$\pm$0.12 & 1.99$\pm$0.14 & TG\\

SA2199-15  &   16 28 35.0 & 39 28 40  & 21.87$\pm$0.08 & 1.86$\pm$0.09 & TG\\

SA2199-16  &   16 28 37.0 & 39 28 50  & 23.04$\pm$0.03 & -1.02$\pm$0.08 & TG\\

SA2199-17  &   16 28 45.0 & 39 28 35  & 20.01$\pm$0.07 & 1.83$\pm$0.09 & DG &  with arm\\

SA2199-18  &   16 28 45.0 & 39 29 05  & 17.87$\pm$0.02 & 2.24$\pm$0.03 & MG &  sprl \\

SA2199-19  &   16 28 44.0 & 39 28 00  & 20.67$\pm$0.08 & 2.41$\pm$0.10 & DG \\

SA2199-20  &   16 28 48.5 & 39 30 55  & 20.58$\pm$0.05 & 2.42$\pm$0.05 & TG?\\

SA2199-21  &   16 28 47.0 & 39 31 10  & 22.41$\pm$0.16 & 2.11$\pm$0.17 & DG\\

SA2199-22  &   16 28 48.0 & 39 31 35  & 22.41$\pm$0.14 & 1.86$\pm$0.16 & MG\\

SA2199-23  &   16 28 49.5 & 39 34 10  & 19.07$\pm$0.02 & 1.61$\pm$0.02 & MG\\

SA2199-24  &   16 28 43.0 & 39 34 05  & 21.28$\pm$0.15 & 3.27$\pm$0.15 & TG\\

SA2199-25  &   16 28 49.0 & 39 32 50  & 17.20$\pm$0.01 & 0.24$\pm$0.01 & DG\\

\hline

\end{tabular}
\end{center}
\end{table}

\begin{table}
\begin{center}
Table 8 Cont.: Structures in Cluster Abell 2199 \\
\vspace{1ex}
\begin{tabular} {rlllllll}

\hline
 Name & R.A.(J2000) & Dec (J2000) & R & (B-R) & Morphology & Notes \\
\hline
\hline

SA2199-26  &   16 28 54.0 & 39 31 35  & 20.27$\pm$0.07 & 3.25$\pm$0.07 & DG\\

SA2199-27  &   16 28 53.0 & 39 32 45  & 23.36$\pm$0.48 & 2.49$\pm$0.51 &  TG \\

SA2199-28  &   16 28 53.0 & 39 29 00  & 16.60$\pm$0.01 & 1.47$\pm$0.02 & TG, IG \\

SA2199-29  &   16 28 50.5 & 39 29 30  & 22.27$\pm$0.36 & 3.22$\pm$0.37 & TG \\  
SA2199-30  &   16 28 50.5 & 39 29 00  & 21.75$\pm$0.03 & 0.51$\pm$0.04 & MG \\

\hline
\end{tabular}
\end{center}
\end{table}

\begin{table}
\begin{center}

Table 9: Structures Found in Cluster AWM 5 \\
\vspace{1ex}
\begin{tabular} {rlllllll}

\hline
 Name & R.A.(J2000) & Dec (J2000) & R & (B-R) & Morphology & Notes \\
\hline
\hline

SAWM5-01 & 16 57 48.0 & 27 48 00 & 16.03$\pm$0.01 & 0.50$\pm$0.02 &  IG & one a disk gal. \\

SAWM5-02 & 16 57 46.5 & 27 48 50 & 20.82$\pm$0.05 & 2.33$\pm$0.05 & MG &  a LSB and HSB Es\\

SAWM5-03 & 16 57 44.1 & 27 49 33 & 18.85$\pm$0.01 & 1.49$\pm$0.02 &  MG &  l and s\\

SAWM5-04 & 16 57 45.0 & 27 50 45 & 18.24$\pm$0.02 & 3.36$\pm$0.02 &  MG &   equal b\\

SAWM5-05 & 16 57 49.0 & 27 51 22 & 22.25$\pm$0.09 & 1.45$\pm$0.11 & LG & one end larger \\

SAWM5-06 & 16 57 48.5 & 27 51 32 & 16.82$\pm$0.01 & 1.44$\pm$0.02 & MG & f and b\\

SAWM5-07 & 16 57 49.5 & 27 51 40 & 21.66$\pm$0.11 & 2.03$\pm$0.12 &  LG\\

SAWM5-08 & 16 57 53.0 & 27 53 07 & 21.17$\pm$0.05 & 1.89$\pm$0.06 & MG\\

SAWM5-09 & 16 57 55.0 & 27 52 10 & 17.79$\pm$0.02 & 1.98$\pm$0.02 & IG & large galaxies\\

SAWM5-10 & 16 57 58.0 & 27 51 50  & 23.24$\pm$0.50 & 2.65$\pm$0.60 & TG\\

SAWM5-11 & 16 57 55.0 & 27 51 05  & 21.37$\pm$0.10 & 2.14$\pm$0.11 & DG & bubble shaped\\

SAWM5-12 & 16 58 00.2 & 27 48 20  & 20.94$\pm$0.05 & 1.93$\pm$0.06 & IG \\

SAWM5-13 & 16 58 06.5 & 27 48 40  & 19.50$\pm$0.02 & 1.50$\pm$0.02 & MG & 3 gal. w/similar b.\\

SAWM5-14 & 16 58 03.5 & 27 49 30  & 23.21$\pm$0.11 & 0.80$\pm$0.18 & LG & two LSB thin lines\\

SAWM5-15 & 16 58 01.0 & 27 48 30  & 22.47$\pm$0.12 & 1.44$\pm$0.15 & MG    &   faint\\

SAWM5-16 & 16 58 03.0 & 27 50 10  & 19.66$\pm$0.02 & 1.54$\pm$0.02 & MG \\

SAWM5-17 & 16 58 10.0 & 27 50 25  & 20.13$\pm$0.04 & 1.73$\pm$0.06 & AG? & knots distant\\

SAWM5-18 & 16 58 01.5 & 27 52 32  & 16.37$\pm$0.01 & 0.89$\pm$0.01 & DG & s. gal. nearby \\  

SAWM5-19 & 16 58 00.2 & 27 53 25  & 18.26$\pm$0.02 & 1.94$\pm$0.02 & TG & near another gal.\\

SAWM5-20 & 16 58 12.0 & 27 53 50  & 19.70$\pm$0.04 & 3.08$\pm$0.04 & MG\\

SAWM5-21 & 16 58 06.0 & 27 51 25  & 22.40$\pm$0.03 & -0.96$\pm$0.06 & TG & near another gal., peanut shaped\\

SAWM5-22 & 16 58 05.0 & 27 48 55  & 24.11$\pm$0.50 & 1.56$\pm$0.60 & TG\\

SAWM5-23 & 16 58 10.5 & 27 48 25  & 16.17$\pm$0.01 & 2.04$\pm$0.01 & MG & l. gal. w/two comp.\\

\hline
\end{tabular}
\end{center}
\end{table}

\begin{table}
\begin{center}

Table 10: Structures Found in Cluster AWM 3 \\
\vspace{1ex}
\begin{tabular} {rlllll}

\hline
 Name & R.A.(J2000) & Dec (J2000) & Morphology & Notes \\
\hline
\hline

SAWM3-01 & 14 28 04.0 & 25 48 50   &   DG  &      LSB\\

SAWM3-02 & 14 27 57.0 & 25 47 45   &   TG\\

SAWM3-03 & 14 27 56.0 & 25 47 50   &   TG\\

SAWM3-04 & 14 28 00.0 & 25 51 50   &   DG  &      slt halo\\

SAWM3-05 & 14 28 00.5 & 25 51 50   &   DG\\

SAWM3-06 & 14 27 57.0 & 25 53 40   &   TG\\

SAWM3-07 & 14 28 05.0 & 25 49 50   &   MG\\

SAWM3-08 & 14 28 10.0 & 25 50 30   &   DG  &      twisted\\

SAWM3-09 & 14 28 07.0 & 25 49 05   &   TG  &      faint\\

SAWM3-10 & 14 28 11.0 & 25 50 55   &   DG  &  has shells, cD NGC 5629\\

\hline

\end{tabular}
\end{center}
\end{table}

\begin{table}
\begin{center}

Table 11: Objects identified with Redshifts \\
\vspace{1ex}
\begin{tabular} {rlllllll}

\hline
 ID & R & B-R & morph &  cz & Cluster$_{cz}$ & $\delta$(cz) \\
\hline
\hline

SA1656-008  & 18.66$\pm$0.02 & 2.16$\pm$0.02 &  AG  &    9833  &  6917 &   2,916 \\
SA1656-011  & 20.30$\pm$0.03 & 1.72$\pm$0.03 &  IG  &    7904    &  6917 &  987 \\
SA1656-018  & 20.55$\pm$0.03 & 1.43$\pm$0.03 &  DG  &   18300  &  6917 &  11383 \\
SA1656-019  & 17.23$\pm$0.01 & 1.94$\pm$0.01 &  dwG &    6788  &  6917 &  -129 \\
SA1656-026  & 19.56$\pm$0.02 & 2.44$\pm$0.02 &  MG  &   48108  &  6917 & 41191  \\   
SA1656-030  & 19.83$\pm$0.02 & 1.81$\pm$0.02 &  MG  &   35268  &  6917 & 28351  \\
SA1656-032  & 19.39$\pm$0.01 & 1.63$\pm$0.02 &  MG  &    6448  &  6917 & -469 \\
SA1656-038  & 20.34$\pm$0.02 & 1.77$\pm$0.03 &  MG &    7864  &  6917 & 947 \\

SA1656-055  & 17.69$\pm$0.03 & 2.64$\pm$0.04 &   AG &    8009  &  6917 & 1092 \\
SA1656-057  & 23.48$\pm$0.24 & 1.18$\pm$0.30 &  TG  &    6629  &  6917 & -288 \\
SA1656-064  & 19.26$\pm$0.05 & 0.93$\pm$0.08 &  IG  &    5128  &  6917 & -1789 \\
SA1656-065  & 23.64$\pm$0.32 & 1.34$\pm$0.27 &  IG  &    6568  &  6917 & -349\\
SA1656-066  & 19.69$\pm$0.04 & 1.20$\pm$0.07 &  AG? &    5922  &  6917 & -995 \\
SA1656-069  & 18.03$\pm$0.03 & 1.73$\pm$0.04 &  AG  &    7493  &  6917 & 576 \\
SA1656-071  & 21.23$\pm$0.06 & 1.11$\pm$0.10 &  LG  &    8135  &  6917 & 1218 \\ 
SA1656-085  & 19.87$\pm$0.09 & 1.60$\pm$0.12 &  AG, MG & 9436  &  6917 & 2519  \\
SA1656-095  & 24.58$\pm$0.36 & -0.17$\pm$0.95 & DG  &    7569  &  6917 & 652 \\
SA1656-103  & 17.82$\pm$0.05 & 2.31$\pm$0.07 &   MG  &    8366  &  6917 & 1449 \\
SA1656-104  & 21.65$\pm$0.17 & 2.04$\pm$0.18 &  DG  &    6363  &  6917 & -554 \\
SA1656-106  & 18.84$\pm$0.09 & 0.76$\pm$0.11 &  MG  &    8784  &  6917 & 1867 \\
SA1656-112  & 20.49$\pm$0.06 & 1.43$\pm$0.08 &   AG  &    2289  &  6917 & -4628 \\

SA0426-011   & --- & --- &  DG  &    4982  &  5486 & -504 \\

SA2199-010   & 20.95$\pm$0.18 & 0.73$\pm$0.41 &  DG  &    9324  &  9063 & 261 \\
SA2199-011   & 17.03$\pm$0.02 & 1.96$\pm$0.03 &  MG  &    8957  &  9063 & -106 \\
SA2199-017   & 20.01$\pm$0.07 & 1.83$\pm$0.09 &  DG  &    8156  &  9063 & -907 \\
SA2199-018   & 17.87$\pm$0.02 & 2.24$\pm$0.03 &  MG  &   10163  &  9063 & 1100 \\
SAWM3-010    & ---    &  --- &    DG  &    4489  &  4497 & -8 \\

\hline

\end{tabular}
\end{center}
\end{table}


\begin{references}

\reference{} Albert C.E., White R.A. \& Morgan W.W. 1977, ApJ, 211, 309
\reference{} Arp H. 1966, ApJS, 14, 1
\reference{} Arp H. \& Madore B.F. 1987, 'A Catalogue of Southern 
Peculiar Galaxies and Associations, Vol. I Positions and Descriptions', 
Cambridge: Cambridge University Press)
\reference{} Bahcall N.A. 1980, 238, 117
\reference{} Barnes J.E. \& Hernquist, L. 1992 ARA\&A, 30, 705
\reference{} -------- 1992, Nature, 360, 715
\reference{} Barnes J.E. 1997 in Galaxy Interactions at Low and High Redshifts, 
IAU Symp. 186, ed. D.B. Sanders, p.36
\reference{} Bautz L.P. \& Morgan W.W. 1970,  162, 149
\reference{} Beers T.C. \& Geller M.J. 1983, ApJ, 274, 491
\reference{} Beers T.C., Geller M.J., Huchra J.P., Latham D.W., \& Davis R.J. 
1984, ApJ, 283, 33
\reference{} Begelman M.C., Rees M.J., \& Blandford R.D. 1979, Nature, 279, 770
\reference{} Bergvall N. \& R\"onnbeck J. 1995, MNRAS, 273, 603
\reference{} Bothun G.D. \& Dressler A. 1986, ApJ, 301, 57
\reference{} Bridges T.J., Carter D., Harris W.E. \& Pritchet C.J. 1996, 
MNRAS, 281, 129
\reference{} Briel U.G., Henry J.P., \& Bohringer H. 1992, A \&A  259, L31 
\reference{} Butcher H. \& Oemler A, Jr. 1978, ApJ, 219, 18 
\reference{} -------- 1984, ApJ, 285, 426
\reference{} Byrd G. \& Valtonen M. 1990, ApJ, 350,89
\reference{} Caldwell N., Rose J.A., Sharpless R.M., Ellis R.S., \& Bower, R.G. 
1993, AJ, 106, 473
\reference{} Chromey F.R., Elmegreen D.M., Mandell A., \& McDermott J. 1998, 
AJ, 115, 2331
\reference{} Colless M. \& Dunn A.M. 1996, ApJ, 458, 435
\reference{} Conselice C.J. 1997, PASP, 109, 1251
\reference{} Conselice C.J. \& Gallagher J.S. III 1997, BAAS 191, 106.9
\reference{} -------- 1998a 1998, MNRAS, 297, L34
\reference{} -------- 1998b, in prep 
\reference{} Couch W.J., Ellis R.S., Sharples R., \& Smail I. 1994, ApJ, 
430, 121
\reference{} Crone M.M., Evard A.E. \& Richstone D.O. 1996, ApJ, 467,489
\reference{} Currie M.J. 1983 in Clusters and Groups of Galaxies, ed. F. Mardirossian, G. Giuricin., M. Mezzetti (D. Reidel Publishing: Dordrecht)
\reference{} Dalcanton J.J. \& Shectman S.A. 1996, ApJ, 465, L9
\reference{} Davis D.S. \& Mushotzky R.F. 1993, AJ, 105, 409
\reference{} Dressler A. \& Gunn J.E. 1983, ApJ, 270, 7
\reference{} -------- 1990, in Evolution of the Universe of Galaxies, ed. R. 
Kron (New York: PASP), 200
\reference{} Dressler A., Oemler A. Jr., Butcher H., \& Gunn J.E. 1994, ApJ, 430,107
\reference{} Dressler A., Oemler A. Jr., Couch W.J., Smail I., Ellis R.S., 
Barger A., Butcher H., Poggianti B.M., \& Sharples R.M. 1997, ApJ, 490, 577
\reference{} Dutta S.N. 1995, MNRAS, 276, 11
\reference{} Elmegreen D.M., Chromey F.R., Knowles B.D., \& Wittenmyer R.A. 
1998, AJ, 115, 1433
\reference{} Feretti L., Dallacasa D., Giovannini G., \& Venturi T. 1990, 
A\&A, 232, 337
\reference{} Fisher D., Illingworth G., \& Franx M. 1994, AJ, 107, 160
\reference{} Fitchett M. \& Webster R. 1987, 317, 653
\reference{} Forman W. \& Jones C.J. 1982, ARA\&A, 20, 547
\reference{} Gaetz T.J., Salpeter E.E. \& Shaviv G. 1987, ApJ, 316, 530
\reference{} Gallagher J.S., Han M. \& Wyse R.F.G. 1997, in Dark and Visible 
Matter in Galaxies, PASP Conf Ser 117, eds. M. Persic \& P. Salucci, p66 
\reference{} Gallego J., Zamorano J., Rego M., Alonso O., \& Vitores A.G. 1996, 
A\&AS, 120, 323 
\reference{} Girardi M., Escalera E., Fadda D., Giuricin G., Mardirossian F., \&
Mezzetti M. 1997, ApJ, 482, 41
\reference{} Gunn J.E. 1987, in Nearly Normal Galaxies, ed. S.M. Faber 
(New York:Springer Verlag), p. 455
\reference{} Gunn J.E. \& Gott J. R. 1972, ApJ, 176, 1
\reference{} Haynes M.P., Giovanelli R. 1986, ApJ, 306, 466
\reference{} Henriksen M.J. \& Byrd G.,  1996, ApJ, 459, 82
\reference{} Henriksen M.J. 1985, Ph.D. thesis, University of Maryland
\reference{} Hoessel J.G. 1980, ApJ, 241, 493
\reference{} Hunsberger S.D., Charlton J.C., \& Zaritsky D. 1996, ApJ, 462,50
\reference{} Icke V. 1985, A\&A, 144,115
\reference{} Johnson M.W., Cruddace R.G., Fritz G., Schulman S., \& Friedman H. 
1978, ApJL, 231, L45
\reference{} Karachentsev I.D., Karachenstseva V.E., \&  Parnovskij S.L 1993, 
AN, 314, 97
\reference{} Karachentsev I.D. \& Xu Z. 1991, PAZh, 17, 321
\reference{} Kauffmann G. 1995, MNRAS, 274, 153
\reference{} Kenney J.D.P., Koopmann R.A., Rubin V.C. \& Young J.S. 1996, AJ, 
111, 152
\reference{} Kenney J.D.P., Rubin V.C., Planesas P., \& Young J.S. 1995, 
ApJ, 438, 135
\reference{} Kent S.M. \& Gunn J.E. 1982, AJ, 87, 945
\reference{} Koopman R.A. \& Kenney J.D.P., ApJL, 497, 75
\reference{} Kriss G.A., Cioffi D.F. \& Canizares C.R. 1993, ApJ, 272, 439 
\reference{} Lake G. \& Dressler A. 1986, ApJ, 310, 605
\reference{} Landolt A.R. 1992, AJ, 104, 372
\reference{} Lauberts A. 1982, 'The ESO/Uppsala Survey of the ESO(B) Atlas', 
(Garching:ESO)
\reference{} Lauer T.R. 1986, ApJL, 311, 34
\reference{} Lavery R.J. \& Henry J.P. 1988, ApJ, 330, 596
\reference{} -------- 1994, ApJL, 426, 524
\reference{} Lucy J.R., Gray P.M., Carter D., \& Terlevich R.J. 1991, 
MNRAS, 248, 804
\reference{} Malphrus B.K., Simpson C.E., Gottesman S.T., \& Hawarden T.G. 
1997, AJ, 114, 1427
\reference{} Matthews L.D. \& Gallagher J.S.III. 1997, AJ, 114, 1899 
\reference{} Mellier Y., Mathez G., Mazure A., Chauvinau B., \& Proust D. 
1988, A\&A 199, 67
\reference{} Merritt D. 1984, ApJ, 276,26
\reference{} Merritt D. 1983, ApJ, 264, 24
\reference{} Mohr J.J., Fabricant D.G., \& Geller M.J. 1994, ApJ, 413, 492
\reference{} Moore B., Lake G., \&  Katz N. 1998, ApJ, 495, 139
\reference{} Moore B., Katz N., Lake G., Dressler A., \& Oemler A. 1996, 
Nature, 379, 613
\reference{} Moore B., Katz N., \& Lake G. 1996, ApJ, 457, 455
\reference{} Morgan, W.W. 1958, 70, 364
\reference{} Morgan, W.W., Kayser, S., \& White, R.A. 1975, ApJ 199,545
\reference{} Nulsen P.E.J. 1982, MNRAS 198, 1007
\reference{} Oemler A, Jr., Dressler A., \& Butcher H.R., 474, 561
\reference{} Ostriker J.P. \& Tremaine S.D. 1975, ApJL, 202, 1
\reference{} Price B., Burns J.O., Duric N., \& Newberry M. 1991, AJ, 102,14
\reference{} Pritchet C.J. \& Harris W.E. 1990, ApJ, 355, 410
\reference{} Ramella, M., Geller, M.J., \& Huchra, J.P. 1989, ApJ, 344, 57
\reference{} Rood H.J. \& Baum, W.A. 1968,  AJ, 72, 398
\reference{} Sarazin C.L. 1986, Rev. Mod. Phys., 58,1
\reference{} Schneider D.P., Gunn J.E., \& Hoessel J.G. 1983, ApJ, 268, 476
\reference{} Secker J. 1998, Private Communication
\reference{} Secker J., Harris, W.E., \& Plummer J.D. 1997, PASP, 109, 1377
\reference{} Secker J. \& Harris, W.E. 1996, ApJ, 496, 623
\reference{} Siddiqui H., Stewart G.C., \& Johnstone R.M. 1998, A\&A, 334,71
\reference{} Sijbring D. \& DeBruyn A.G. 1998, A \& A, 331, 901
\reference{} Slezak E., Durret F. \& Gerbal D. 1994, AJ, 108 1996
\reference{} Strom K.M \& Strom S.E. 1978,  AJ, 83, 1293
\reference{} Toomre A. \& Toomre J. 1972, ApJ, 178,623
\reference{} Toomre A. 1977, in The Evolution of Galaxies and Stellar 
Populations ed. B.M. Tinsley and R.B. Larson (New Haven: Yale Univ. 
Observatory), p. 401
\reference{} Toomre A. 1978, in Large Scale Structure of the Universe, ed. M.S. 
Longair \& J. Einasto (Reidel, Dordrecht), p. 109.
\reference{} Trentham N. \& Mobasher B. 1998, MNRAS (In Press)
\reference{} Ulmer M.P., Nichol R.C., Martin D.R., Wipf S., Bernstein G., 
Whittman D., \& Tyson J.A. 1994, in ESO/OHP Workshop - Dwarf Galaxies, ed. 
G Meylan, P Prugniel, 121
\reference{} van den Bergh S. 1982, PASP, 94, 459
\reference{} van den Bergh S., Abraham R.G., Ellis R.S., Tanvir N.R., \&
Santiago B.X. 1996, 112, 359
\reference{} Valluri M. 1993, ApJ, 408, 57
\reference{} Valluri M., \& Jog C. 1990, ApJ, 357, 367
\reference{} Vigroux L., Lachieze-Roy M., Thuan T.X., \& Vader, J.P. 1986, AJ, 91, 70
\reference{} Vikhlinin A., Forman W. \& Jones C. 1997, ApJL, 474, L7
\reference{} Wang D.Q. \& Ulmer M. 1997, MNRAS, 292, 920
\reference{} West M. J. 1994, in Clusters of Galaxies, ed. F. Durret, A. Mazure, J. Tran Thanh Van (Gif-sur-Yvette: Editions Frontires), 23
\reference{} White S.D.M., Briel U.G., \& Henry J.P. 1993, MNRAS, 261, L8
\reference{} Williams R.E., Blacker B., Dickinson M., Dixon W. V., Ferguson H.C.,
Fruchter A.S., Giavalisco M., Gilliland R.L., Heyer I., Katsanis R., Levay Z., 
Lucas R.A., McElroy, D.B., Metro L., \& Postman M. 1996, AJ, 112, 1335 
\reference{} Williams B.A., \& Lynch J.R. 1991 AJ, 101 1969
\reference{} Yoshida M., Taniguchi, Y., \& Murayama, T. 1994, PASJ 195, 46
\reference{} Zabludoff A.I., Huchra J.P., \& Geller M.J. 1990 ApJS, 74, 1
\reference{} Zwicky F. 1956, Ergebnisse d. exakten Naturw., 29, 344
\reference{} Zwicky, F. 1957, Morphological Astronomy (Berlin: Springer)
\end{references}
\end{document}